\documentclass[twocolumn]{aastex62}

\usepackage{graphicx}
\usepackage{ulem}
\usepackage{url}
\usepackage{threeparttable}
\usepackage{xcolor}
\usepackage{soul}
\usepackage{booktabs}
\usepackage{natbib}
\usepackage{amsmath}
\usepackage{mathrsfs}
\hypersetup{colorlinks, linkcolor=black, anchorcolor=green, citecolor=blue}
\usepackage{txfonts}
\usepackage{url}

\usepackage{rotating}
\usepackage{longtable, needspace}
\usepackage{overpic,color}
\usepackage{multirow}
\usepackage[]{subfigure}

\usepackage{graphbox} %loads graphicx package

\def\HII {H{\sc ii}}

\received{}
\revised{}
\accepted{}
\submitjournal{ApJ}

\shorttitle{70 micron dark CO depletion}
\shortauthors{Feng et al.}
\begin{document}

\title{
The chemical structure of young high-mass star-forming clumps: (II) parsec-scale CO depletion and deuterium fraction of $\rm HCO^+$ \footnote{}}

\correspondingauthor{Siyi Feng}
\email{siyi.s.feng@gmail.com}

\author[0000-0002-4707-8409]{S. Feng}
\affil{National Astronomical Observatories, Chinese Academy of Science, Beijing 100101, People's Republic of China}
\affil{Academia Sinica Institute of Astronomy and Astrophysics, No.1, Section 4, Roosevelt Road, Taipei 10617, Taiwan, Republic of China} 
%\affil{CAS Key Laboratory of FAST, National Astronomical Observatories, Chinese Academy of Sciences, Beijing 100101, People's Republic of China}
\affil{National Astronomical Observatory of Japan, National Institutes of Natural Sciences, 2-21-1 Osawa, Mitaka, Tokyo 181-8588, Japan}

\author{D. Li}
\affil{National Astronomical Observatories, Chinese Academy of Science, Beijing 100101, People's Republic of China}
%\affil{CAS Key Laboratory of FAST, National Astronomical Observatories, Chinese Academy of Sciences, Beijing 100101, People's Republic of China}
\affil{NAOC-UKZN Computational Astrophysics Centre, University of KwaZulu-Natal, Durban 4000, South Africa}

\author{P. Caselli}
\affil{Max-Planck-Institut f\"ur Extraterrestrische Physik, Gie{\ss}enbachstra{\ss}e 1,  D-85748,  Garching bei M\"unchen, Germany}

\author{F. Du}
\affil{Purple Mountain Observatory and Key Laboratory of Radio Astronomy, Chinese Academy of Sciences, Nanjing 210023, People's Republic of China}
\affil{School of Astronomy and Space Science, University of Science and Technology of China, Hefei 230026, People's Republic of China}

\author{Y. Lin}
\affil{Max Planck Institute for Radio Astronomy, Auf dem H\"ugel 69, D-53121 Bonn, Germany}

\author{O. Sipil\"a}
\affil{Max-Planck-Institut f\"ur Extraterrestrische Physik, Gie{\ss}enbachstra{\ss}e 1,  D-85748,  Garching bei M\"unchen, Germany}

\author{H. Beuther}
\affil{Max-Planck-Institut f\"ur Astronomie, K\"onigstuhl 17,  D-69117,  Heidelberg, Germany}

\author{Patricio Sanhueza}
\affil{National Astronomical Observatory of Japan, National Institutes of Natural Sciences, 2-21-1 Osawa, Mitaka, Tokyo 181-8588, Japan}

\author{K. Tatematsu}
\affil{Nobeyama Radio Observatory, National Astronomical Observatory of Japan, National Institutes of Natural Sciences, Nobeyama, Minamimaki, Minamisaku, Nagano 384-1305, Japan}
\affil{Department of Astronomical Science, SOKENDAI (The Graduate University for Advanced Studies), 2-21-1 Osawa, Mitaka, Tokyo, 181-8588, Japan}

\author{S. Y. Liu}
\affil{Academia Sinica Institute of Astronomy and Astrophysics, No.1, Section 4, Roosevelt Road, Taipei 10617, Taiwan, Republic of China}

\author{Q. Zhang}
\affil{Center for Astrophysics $|$ Harvard and Smithsonian, 60 Garden Street, Cambridge, MA 02138, USA}

\author{Y. Wang}
\affil{Max-Planck-Institut f\"ur Astronomie, K\"onigstuhl 17,  D-69117,  Heidelberg, Germany}

\author{T. Hogge}
\affil{Institute for Astrophysical Research, Boston University, Boston, MA 02215, USA}

\author{I. Jimenez-Serra}
\affil{Centro de Astrobiolog\'{\i}a (CSIC, INTA), Ctra. de Torrej\'{o}n a Ajalvir km. 4, Torrej\'{o}n de Ardoz, E-28850 Madrid, Spain}
 
\author{X. Lu}
\affil{National Astronomical Observatory of Japan, National Institutes of Natural Sciences, 2-21-1 Osawa, Mitaka, Tokyo 181-8588, Japan}

\author{T. Liu}
\affil{Shanghai Astronomical Observatory, Chinese Academy of Sciences, 80 Nandan Road, Shanghai 200030, PeopleÕs Republic of China}

\author{K. Wang}
\affil{Kavli Institute for Astronomy and Astrophysics, Peking University, 5 Yiheyuan Road, Haidian District, Beijing 100871, People's Republic of China}

\author{Z. Y. Zhang }
\affil{School of Astronomy and Space Science, Nanjing University, 163 Xianlin Avenue, Nanjing 210023, People's Republic of China}

\author{S. Zahorecz}
\affil{National Astronomical Observatory of Japan, National Institutes of Natural Sciences, 2-21-1 Osawa, Mitaka, Tokyo 181-8588, Japan}
\affil{Department of Physical Science, Graduate School of Science, Osaka Prefecture University, 1-1 Gakuen-cho,Naka-ku, Sakai, Osaka 599-8531, Japan}

\author{G. Li}
\affil{South-Western Institute for Astronomy Research, Yunnan University, Kunming, 650500 Yunnan, People's Republic of China}

\author{H. B. Liu}
\affil{Academia Sinica Institute of Astronomy and Astrophysics, No.1, Section 4, Roosevelt Road, Taipei 10617, Taiwan, Republic of China}

\author{J. Yuan}
\affil{National Astronomical Observatories, Chinese Academy of Science, Beijing 100101, People's Republic of China}



\begin{abstract}
The physical and chemical properties of cold and dense molecular clouds are key to understanding how stars form. 
Using the IRAM 30\,m and NRO 45\,m telescopes, we  carried out a Multiwavelength line-Imaging survey of the
70\,$\mu$m dark and bright clOuds (MIAO). 
At a linear resolution of 0.1--0.5\,pc,
this work presents a detailed study of parsec-scale CO depletion and $\rm HCO^+$ deuterium (D-) fractionation toward four sources (G\,11.38+0.81, G\,15.22-0.43,  G\,14.49-0.13, and G\,34.74-0.12) included in our full sample. In each source with $\rm T<20$\,K and $n_{\rm H}\rm\sim10^4$--$\rm 10^5\,cm^{-3}$, we compared pairs of neighboring 70\,$\mu$m bright and dark clumps and found that
(1) the $\rm H_2$ column density and  dust temperature of each source show strong spatial anticorrelation;
(2) the spatial distribution of CO isotopologue lines and dense gas tracers, such as 1--0 lines of $\rm H^{13}CO^+$ and $\rm DCO^+$, are anticorrelated; 
(3) the abundance ratio between $\rm C^{18}O$ and $\rm DCO^+$ shows a strong correlation with the source temperature;
(4) both the $\rm C^{18}O$ depletion factor and D-fraction of $\rm HCO^+$ show a robust decrease from  younger clumps to  more evolved clumps by a factor of more than 3;
and (5) preliminary chemical modeling indicates chemical ages of our sources are ${\sim}8\times10^4$ yr, which is comparable to their free-fall timescales and smaller than their contraction timescales, indicating that our sources are likely { dynamically and chemically young.} %{\color{blue}at the onset of global collapse.}

\end{abstract}



%______________________________________________________________
\section{Introduction}
The initial conditions of high-mass star  formation (HMSF) are still  under debate \citep[e.g.,][]{beuther07b,tan14,motte17,sanhueza19}. 
For example, how different are the kinematics and chemical evolution during the formation of  high-mass star clusters with respect to their low-mass analogs?
In particular, 
what is the chemical environment of these high-mass clumps \citep[e.g.,][]{sanhueza12,feng16a,tatematsu17}? How do gas motions  (e.g., infall, outflow) link the  parental clouds and the descendant high-mass clumps during star formation \citep[e.g.,][]{wangk14,beuther15,zhang15,sanhueza17,contreras18,lu18}?  Two steps are essential to address these questions   \citep[e.g.,][]{zhang09}: (1)identifying the ``initial" environments that have the potential to form  high-mass stars and (2) precisely characterizing the chemical and kinematic properties of these ``initial" environments from the observations. 

The dense  ($n\rm >10^3$--$\rm 10^5\,cm^{-3}$; \citealt{rathborne06}), cold ($\rm T<20\,K$; \citealt{wangk12}), and less luminous infrared-dark molecular clouds (IRDCs) are of particular interest \citep[e.g., ][]{tan13,sanhueza13}. In particular, the 70\,$\mu$m dark \citep{dunham08} high-mass clumps, with bolometric luminosity ($\rm L_{bol}$)-to-mass ($\rm M_c$) ratio less than $\rm 1\,L_\odot/M_\odot$ \citep{molinari16}, are prime targets for studying initial conditions.
These regions may contain clusters of low-mass young stellar objects or be prestellar and thus future sites of  intermediate-/high-mass protostellar objects.
{\color{black}Therefore, they are excellent space laboratories to test not only the chemical processes in the cold and dense environment, but also different kinematic  scenarios of HMSF (\citealp[e.g., competitive accretion,][]{bonnell04,bonnell06}, or \citealp[monolithic collapse, ][]{mckee03,krumholz05,krumholz09}).

Previous multiwavelength dust continuum surveys have provided several catalogs of initial HMSF clump  candidates \citep[e.g.,][]{ragan12,guzman15,svoboda16,yuan17}.
 However, observations of the dust continuum can discern  neither the kinematic nor chemical properties of these candidates. 
 Since these properties are crucial to understanding the high-/intermediate-/low-mass star formation in high-mass clumps,  spectroscopic images with a high spatial dynamic range (0.01\,pc--1\,pc) and fine velocity resolution are essential for these properties.

{To characterize the chemical processes  and gas motions in the early phase of high-mass clumps, we designed and carried out the Multiwavelength line-Imaging survey of the
70\,$\mu$m dark and bright clOuds (MIAO\footnote{``MIAO" shares a pronunciation with three Chinese characters: a noun (``the seed or something in the initial condition"), an adjective (``wonderful"), and a verb (``to draw the profile of something"). }) project (see Section~\ref{sec:project}).
Given that the data  collected for this project have a broad range of spatial and spectral coverage, we plan to carry out a series of analyses on the detailed chemistry  (this work; {\it Paper  IV--V}) and kinematics ({\it Paper III}) of our source sample. 
}

{In the present work},  we focus on two crucial chemical processes in  early star-forming  environments, when molecular clouds are  cold ($\rm T < 20$\,K) and dense ($n\rm > 10^4\,cm^{-3}$), namely, freeze-out and deuterium (D-) fractionation  \citep[e.g., ][]{caselli02a}.

Freeze-out is the process that allows gaseous species, including elements heavier than He, to adsorb on the surface of dust grains \citep[e.g., ][]{aikawa13}. 
The D-fractionation of gas-phase species is a process that starts by unlocking atomic deuterium from HD through cosmic ray-driven ion-molecule chemistry
\citep[e.g., ][]{millar89,ceccarelli14}.
Isotope exchange reactions take place via $\rm H_3^+$ yielding  $\rm H_2D^+$,  $\rm D_2H^+$, and $\rm D_3^+$ \citep[e.g., ][]{caselli02b,crapsi05,vastel06,chen10}, which then react with more abundant species, such as CO and $\rm N_2$, to produce species such as $\rm DCO^+$ and $\rm N_2D^+$.

Observationally, CO freeze-out (also called CO depletion) is measured as the ratio of the expected CO canonical abundance with respect to its observed gaseous abundance. Depletion of CO  has been widely detected toward cold and dense starless clumps and cores \citep[e.g., ][]{willacy98,kramer99,caselli99,bergin02,bacmann03,fontani12}, where the CO depletion peaks show spatial coincidence with the D-fractionation peaks of gas-forming  species, such as $\rm  N_2H^+$ and $\rm HCO^+$. In most cases, such a spatial coincidence appears at a subparsec spatial scale \citep[e.g., ][]{caselli99}.
Recent observations have revealed  parsec-scale CO depletion \citep{hernandez11,giannetti14,sabatini19}, associated with high D-fractionation \citep[e.g., ][]{barnes16,feng19a}.
However, cases of parsec-scale CO depletion are much more rarely reported than subparsecc scale cases. 
One reason is that the history of studying IRDCs is relatively short. In particular, IRDCs that are 70\,$\mu$m dark are mostly, if not entirely, identified with the {\it Herschel} Space Observatory which was launched only about a decade ago. Due to a lack of candidates, the chance of witnessing parsec-scale CO depletion toward the star-forming regions at the extreme young stage (dense and with a short timescale) are small.
Another reason is that, to identify CO depletion, adequate linear resolution is crucial, for high-depletion zones are localized in relatively small, low-temperature and high-density regions.  Millimeter/submillimeter interferometers offer sufficient resolutions, but CO data are hampered by missing fluxes due to their large spatial extent. Although observations from single-dish telescopes are not affected by missing fluxes, many of them offer too limited angular resolutions to probe the densest region. Moreover, imaging a large field at adequately high spatial resolution and high spectral sensitivity required good weather conditions and was
very time-consuming.

By taking advantage of new, high-sensitivity  observational instrumentation, we carried out a  line-imaging survey project on a large sample of sources. 
In Section~\ref{obs}, we introduce our MIAO survey project  and summarize the observations and the data quality. 
We present the maps of continuum and molecular line emission toward a pilot sample of four regions in Section~\ref{distribution}, and characterize their physical structures in Section~\ref{calculation}. 
In Section~\ref{discussion}, we discuss  the possible spatial relation between the CO depletion factor, D-fraction of $\rm HCO^+$, source temperature and density  toward each region, as well as fit our chemical model to the observations. 
Finally, a summary of our main results can be found in Section~\ref{conclusion}.

 %%%%%%%%%%%%%%%
\section{Survey Design and Observations}\label{obs}

\subsection{MIAO}\label{sec:project}

During 2016--2017, 
we carried out a pilot line-imaging survey toward the filamentary IRDC G\,28.34+0.06 \citep[e.g., ][]{wangk18}. Using the Institut de Radio Astronomie Millim\'etrique 30\,m telescope (IRAM 30\,m),  we comparatively observed G\,28.34 P1-S, a pair of 70\,$\mu$m bright and dark dense clumps 
separated at subparsec distance in this IRDC, at 1\,mm--4\,mm wavelength. On the one hand, we unveiled varying degrees of high-mass star-forming activities from prestellar objects to protostellar objects, such as parsec-scale infall signature \citep[][]{feng16a} and dynamically extremely young outflows \citep[$\rm \sim10^4$\,yr,][]{feng16b,tan16,kong18}. On the other hand, we also revealed the chemical variations  in the framework of evolutionary stages of star formation, such as, parsec-scale CO depletion \citep[][]{feng16a} and species-dependent D-fractionation \citep[][{\it Paper I}]{feng19a}.

However, we cannot generalize our conclusions because of the small sample size.
To ground our pilot study results, we initiated a multiwavelength line-imaging survey project (MIAO) in 2017. Aiming to characterize the chemical processes (presented here) and gas motions (S. S. Feng et al. 2020, in preparation) in primordial high-mass clumps, we design this project to observe a sample of 24  extremely cold, dense clumps ({Table~\ref{sourceall}}) by using the IRAM 30\,m, the Nobeyama 45\,m telescope (NRO 45\,m), and the Atacama Large Millimeter/submillimeter Array (ALMA).
To have a robust analysis, we select the sources in the sample  based on the following criteria.

\begin{enumerate}
\item {\it Dense}. 
All of the regions in our sample are selected from a high-mass starless clump (HMSC) candidate catalog \citep{yuan17}, which is provided by  analyzing the millimeter and submillimeter continuum from the APEX/ATLASGAL \citep{schuller09}, {\it Spitzer}/GLIMPSE-MIPSGAL  \citep{benjamin03,churchwell10}, and  {\it Herschel}/Hi-GAL  \citep{molinari10} surveys throughout the entire inner Galactic plane.
For comparative study, each imaged region  covered a pair of 70\,$\mu$m dark and bright clumps.
Both  clumps in each region are high-mass, with $\rm M>870\,M_\odot(radius/parsec)^{1.33}$ \citep{kauffmann10} and mass surface density $\rm >0.2\,g\,cm^{-2}$, fulfilling the empirical threshold  of 0.05\,$\rm g\,cm^{-2}$ given by \citet{urquhart14} and \citet{he15} for HMSF.  Specifically, each 70\,$\mu$m dark clump is an HMSC candidate, with  high dust extinction and low luminosity ($\rm L_{bol}$/$\rm M_c<1\,L_\odot/M_\odot$, \citealp{molinari16}), and it is associated with neither a methanol maser nor an \HII\,region, indicating that they are young.

\item {\it Cold}. 
Using the spectral energy distribution (SED) method \citep[elaborated in ][ and Sect.~\ref{h2tdust}]{yuan17,lin17},  the dust temperature of each 70\,$\mu$m dark clump in our sample is low ($\rm <20$\,K).

\item {\it Relatively near}. 
The sources in our sample are selected within a kinematic distance of $d\rm<5\,kpc$. At an angular resolution from 30\arcsec (IRAM 30\,m observations) down to 2\arcsec (ALMA observations), a quantitative characterization of the star-forming activities
at a subparsec linear resolution  will help us interpret the star-forming activities of more distant regions.

\item {\it Well-constrained environmental properties of the parental cloud}.
Each 70\,$\mu$m dark clump is located at the morphological end of a filamentary cloud. Such objects have been proposed as prime targets to study the initial conditions of HMSF  because gravity-driven accretion (gravitational acceleration) is likely  enhanced around the morphological ends of the filaments and in the edges of sheet-like structures (edge effects) on a timescale shorter than the global collapse timescale \citep[e.g., ][]{burkert04,pon12,lig16}. At least one 70\,$\mu$m bright counterpart is  within a 1.5\arcmin ($\rm<$2\,pc) distance to the 70\,$\mu$m dark clump in the plane-of-sky. Pairs of 70\,$\mu$m dark and bright clumps show the same systemic velocity $\rm V_{sys}$ as previous point observations by using  single-dish telescopes \citep[][]{purcell12,wienen12,shirley13,dempsey13,csengeri14}, indicating that they are in the same parental cloud.  Being away from the filament ends, the 70\,$\mu$m bright clump may be dynamically more  evolved than the 70\,$\mu$m dark clump.

\end{enumerate}

Our comparative kinematic and chemical study toward the 70\,$\mu$m dark and bright clump pairs in each region includes the comparison of molecular line profiles, molecular spatial distributions, and relative abundances between different species. Such a study will minimize environmental differences  (e.g., interstellar UV heating, cosmic-ray ionization rate (CRIR), elemental abundances, magnetic fields), which goes a step ahead of previous studies that targeted  samples of spatially separated sources in different clouds.

The present work focuses on the parsec-scale chemical features, and the data were obtained from the following single-dish observations\footnote{Our ongoing ALMA observations will be presented in a future study ({\it Paper III}), focusing on the connection between the parsec-  and subparsec scale gas motions (Feng et al. 2020 in preparation).}.

\subsection{IRAM 30\,m observations}
The line-imaging survey of the entire sample of 24 regions was carried out using  the IRAM 30\,m telescope at 1.3\,mm, 3.4\,mm, and 4.0\,mm. Observations were performed in  on-the-fly (OTF) mode  from August 2017 to May 2018, and the map centers of the four sources considered in the present work are listed in Table~\ref{tab:source} (see the complete list of the entire sample in Table~\ref{sourceall}). 

 The broad bandpass of EMIR  simultaneously covers 16\,GHz bandwidth. By superpositioning two spectral tunings, our observations cover the frequency ranges 70.718--78.501, 82.058--94.183, and 217.122--224.842 GHz in total. These frequency ranges cover several dense gas tracers, shock tracers, and deuterated lines (see the targeted lines in Table~\ref{tab:lines}, {which will be analyzed in future studies}). Using the FTS200 backend, we achieve a frequency resolution of 195\,kHz (corresponding to $\rm 0.659\,km\,s^{-1}$ at 88.632\,GHz). The angular resolution of the IRAM 30\,m telescope is 29.3\arcsec at 88.632\,GHz. The weather conditions during the observations were good (radiometer opacity $\rm \tau$ at 255GHz $\rm <0.6$), and we used Saturn and Mars for pointing and focus. Using the corresponding forward efficiency ($F_{eff}$) and a main-beam efficiency ($B_{eff}$) at individual frequencies\footnote{http://www.iram.es/IRAMES/mainWiki/Iram30\,mEfficiencies}, we converted the data from antenna temperature ($T_A^*$) to main-beam temperature ($T_{mb}$ = $F_{eff} /B_{eff}\times T_A^*$). We used the GILDAS software package for data reduction and line identification. The typical root mean square (rms) noise levels in  $T_{mb}$ in the line-free channels are listed in Table~\ref{tab:source}.

 \subsection{NRO 45\,m observations}
 Using the  FOREST receiver \citep{minamidani16}  mounted on the NRO 45\,m telescope, our entire sample was observed with the NRO 45\,m telescope from 2018 January to 2018 February,  simultaneously targeting the ground transition lines of $\rm C^{18}O$, $\rm C^{17}O$, and $\rm ^{13}CO$.
 Employing the OTF scan mode \citep{sawada08}, each region was imaged with the same map size and center as those in the IRAM 30\,m observations (Table~\ref{sourceall}). 
 
 Using the SAM45 digital spectrometer \citep{kamazaki12}, we achieved a frequency resolution of 61.04\,kHz (corresponding to 0.120\,$\rm km\,s^{-1}$ at 109.783\,GHz). The effective angular resolution, i.e.,  Full Width at Half Maximum (FWHM) beam of the NRO 45\,m is 16$\farcs$4 at 109.783\,GHz. 
 
 The telescope pointing was established by observing the 43\,GHz SiO maser of OH397 or VX-SGR every 60\,minutes, achieving an accuracy of $\rm \sim 5\arcsec$ (FWHM beam as 42\arcsec at 43\,GHz). Using the corresponding main-beam efficiency $\eta_{mb}$ ($\rm 44\%\pm3\%$ at 110\,GHz), we converted the data from antenna temperature ($T_A^*$) to main-beam temperature ($T_{mb} =T_A^*/\eta_{mb}$).
 We used the NOSTAR software package \citep{sawada08} for data reduction. The rms noise levels in  $ T_{mb}$ in the line-free channels are listed in Table~\ref{tab:source}.

\begin{table*}
\caption{Sources in this work and their observation parameters
}\label{tab:source}
\scalebox{0.9}{
\begin{tabular}{ccccccccccc}
\hline\hline

Source$^a$  &Abbrev.  & R.A.$^b$      &DEC.$^b$  &$d$$^c$  &$\rm R_{\rm GC}$$^d$   &$\rm V_{sys}$          &$\rm rms_{4.0\,mm}$$^e$          &$\rm rms_{3.4\,mm}$$^f$  &$\rm rms_{1.3\,mm}$$^g$  &$\rm rms_{2.7\,mm}$$^h$\\
         &  &[J2000]  &[J2000]   &(kpc)   &$\rm (kpc)$          &$\rm (km\,s^{-1})$                     &(K)                         &(K)                         &(K)  &(K) \\
         \hline
{G\,015.2169-0.4267}   &{G\,15.22-0.43}              &$\rm 18^h19^m51^s.2$       &$\rm -15^\circ54^{'}50^{"}.8$        &1.9   &6.1  &22.7$^i$  &0.05 &0.04  &0.33   &0.26\\   
{G\,011.3811+0.8103}   &{G\,11.38+0.81}              &$\rm 18^h07^m36^s.4$        &$\rm -18^\circ41^{'}21^{"}.1$        &2.8  &5.2  &26.8$^j$  &0.02  &0.02 &0.17    &0.24\\
{G\,014.4876-0.1274}   &{G\,14.49-0.13}              &$\rm 18^h17^m19^s.0$       &$\rm -16^\circ24^{'}53^{"}.6$        &3.2  &4.9   &39.7$^k$    &0.03 &0.03  &0.31  &0.46\\
{G\,034.7391-0.1197}   &{G\,34.74-0.12}              &$\rm 18^h55^m09^s.7$        &$\rm +01^\circ33^{'}13^{"}.3$           &4.7  &5.2  &79.0$^l$  &0.02 &0.02  &0.18    &0.22\\

\hline\hline
\multicolumn{10}{l}{\color{black} { Note.} $a$. ATLASGAL name. A complete list of sources in our sample is given as Appendix Table~\ref{sourceall}. }\\
\multicolumn{10}{l}{\color{black} ~~~~~~~~~~~$b$. OTF mapping center. }\\
\multicolumn{10}{l}{\color{black} ~~~~~~~~~~~$c$. Kinematic distance, from \citet{yuan17}, with an uncertainty of $\rm \pm 0.5$\,kpc. }\\
\multicolumn{10}{l}{\color{black} ~~~~~~~~~~~$d$. Galactocentric distance, calculated by using \citet{wenger18}.}\\
\multicolumn{10}{l}{\color{black} ~~~~~~~~~~~$e.$ Measured by IRAM 30\,m in main-beam temperature  $\rm T_{mb}$ (K)  directly from observations without smoothing, }\\
\multicolumn{10}{l}{\color{black} ~~~~~~~~~~~ with an angular resolution of $\sim$36\arcsec~ and velocity resolution of $\sim$0.72\,$\rm km\,s^{-1}$   for 4.0\,mm lines.}\\
\multicolumn{10}{l}{\color{black} ~~~~~~~~~~~$f.$ Measured by IRAM 30\,m in main-beam temperature  $\rm T_{mb}$ (K)  directly from observations without smoothing, }\\
\multicolumn{10}{l}{\color{black} ~~~~~~~~~~~  with an angular resolution of $\sim$29\arcsec~ and velocity resolution of $\sim$0.56\,$\rm km\,s^{-1}$   for 3.4\,mm lines.}\\
\multicolumn{10}{l}{\color{black} ~~~~~~~~~~~$g.$ Measured by IRAM 30\,m in main-beam temperature  $\rm T_{mb}$ (K) directly from observations without smoothing, }\\
\multicolumn{10}{l}{\color{black} ~~~~~~~~~~~with an angular resolution of $\sim$11\arcsec~ and velocity resolution of $\sim$0.22\,$\rm km\,s^{-1}$   for 1.3\,mm lines.}\\
\multicolumn{10}{l}{\color{black} ~~~~~~~~~~~$h.$ Measured by NRO 45\,m in main-beam temperature  $\rm T_{mb}$ (K) directly from observations without smoothing, }\\
\multicolumn{10}{l}{\color{black} ~~~~~~~~~~~with an angular resolution of $\sim$16\arcsec~ and velocity resolution of $\sim$0.12\,$\rm km\,s^{-1}$   for 2.7\,mm lines.}\\
\multicolumn{10}{l}{\color{black} ~~~~~~~~~~~$i$. \citet{dempsey13}.}\\
\multicolumn{10}{l}{\color{black} ~~~~~~~~~~~$j$. \citet{csengeri14}. }\\
\multicolumn{10}{l}{\color{black} ~~~~~~~~~~~$k$. \citet{wienen12}.}\\
\multicolumn{10}{l}{\color{black} ~~~~~~~~~~~$l$. \citet{shirley13}. }
\end{tabular}
}
\end{table*}

\subsection{Archival data}\label{arch}
Moreover, we used the following archival data.

Continuum data were obtained from the {\it Herschel}/Hi-GAL  survey at 160\,$\mu$m (PACS) and 250, 350, 500\,$\mu$m (SPIRE; \citealp{molinari10}),   as well as from the combination of {Planck} \citep{planck14} and James Clerk Maxwell telescope {(JCMT) -SCUBA2} data at 850\,$\mu$m (G\,11.38+0.81 and G\,14.49-0.13, obsID: M11BEC30) or APEX-LABOCA data at 870\,$\mu$m \citep[G\,15.22-0.43 and G\,34.74-0.12,][]{schuller09}. 

We also used $\rm NH_3\,(\it J, \it K \rm)$=(1,1) and (2,2) lines  from the Radio Ammonia Mid-plane Survey (RAMPS; \citealp{hogge18}), observed with the Green Bank Telescope (GBT). 
The data achieve an angular resolution of $\rm \sim 34\farcs7$ and a velocity resolution of $\rm 0.018\,km\,s^{-1}$. Using a main-beam efficiency of 0.91,  the rms  in $T_{mb}$ for each source is $\rm \sim0.5\,K$.

%%%%%%%%%%%%%%%%%%%%%%%%%%%%%
\section{Results and analysis}\label{result}
\subsection{Spatial distribution of the continuum  and molecular line emission}\label{distribution}
Analyzing our entire sample of 24 regions, we found that over $\rm 50\%$  show parsec-scale CO depletion.
A complete statistical overview of the chemical and physical properties of the entire sample will be given {in a follow-up paper}.
Grouping these sources according to their kinematic-distances  ($d$) progressively further away from the Sun,  we picked out a pilot sample of four regions (G\,11.38+0.81,  G\,14.49-0.13,  G\,15.22-0.43,  and G\,34.74-0.12) that show the most obvious spatial anticorrelation between CO and deuterated species  from each kinematic distance group.

All four regions in the pilot sample contain a clump, for which the 70\,$\mu$m extinction and 870\,$\mu$m emission are spatially correlated  (P1 in Figure~\ref{fig:soumap} {\it left}). This indicates that these 70\,$\mu$m dark clumps are at an early stage in their evolution.

In each region, extracting the beam-averaged spectrum toward the  870 or 70\,$\mu$m continuum peaks, we found that, at a linear resolution of $\rm >0.1$\,pc,
neighboring clumps in the same cloud show  similar line profiles (Figures~\ref{spec1}), i.e., the differences in the centroid velocities and FWHM linewidths toward neighboring clumps are less than $\rm 2\,km\,s^{-1}$ (Table~\ref{tab:linedeu}).

To compare the chemical differentiations of molecules  in the same clouds, it is important to have the spatial distribution maps of all of the molecular lines  covered by our multiwavelength line-imaging survey project (listed in Table~\ref{tab:lines}). Considering their broad linewidths (FWHM $\sim$2--6\,$\rm km\,s^{-1}$),  we 
imaged their integrated intensities over the same velocity range  towards each source, covering all of the line wings down to the continuum level (given in Table~\ref{tab:linedeu}). 
In particular, lines with critical densities $\rm >10^5\,cm^{-3}$ in the temperature range of 10--20\,K are treated as high-density tracers, including  the 1--0 lines from HCN isotopologues ($\rm H^{13}CN$, $\rm HC^{15}N$, DCN), HNC isotopologues ($\rm HN^{13}C$, $\rm H^{15}NC$, DNC),  $\rm HCO^+$ isotopologues ($\rm H^{13}CO^+$, $\rm HC^{18}O^+$, $\rm DCO^+$), and $\rm N_2H^+$ isotopologues ($\rm N_2D^+$), which are covered by our observations (Table~\ref{tab:lines}). 
Morphologically, they are spatially coincident with the 870\,$\mu$m continuum emission. 
In contrast,  the integrated intensities of the 1--0 and 2--1 lines from  $\rm C^{17}O$, $\rm ^{13}CO$, and $\rm C^{18}O$ show anticorrelated spatial distributions with the dense gas tracers, as already found in low-mass star-forming regions \citep[e.g., ][]{caselli02a}.

We focus here on the spatial distributions of $\rm C^{18}O$\,(2--1) and $\rm DCO^+$\,(1--0) toward each source,
based on the consideration of the  observational uncertainties, chemical differentiation, and  data sensitivity. 
(1) Comparing the data obtained from the same observations  will exclude the uncertainty of pointing and calibration caused by using two different telescopes, so we do not show the 1--0 lines of CO isotopologues obtained from the NRO 45\,m here.
(2) Theoretically, 
 $\rm N_2H^+$ and $\rm HCO^+$ are formed exclusively in the gas phase \citep[e.g., ][]{parise02, aikawa05, aikawa12, graninger14}. Emission intensity peaks of $\rm N_2D^+$\,(1--0) and $\rm DCO^+$\,(1--0)  are strong indicators of the densest and coldest environment of each source. This has been proved by extensive observations, including our pilot study \citep[e.g., ][]{fontani14, feng19a}.
(3) Investigating the line profiles of the dense gas tracers and CO isotopologue lines (Figure~\ref{spec1}), we found that  the  signal-to-noise ratios (S/Ns) of $\rm C^{17}O$\,(2--1) and $\rm N_2D^+$\,(1--0) are too low ($\rm S/N<5$)  toward some regions, and that $\rm ^{13}CO$\,(2--1) seems to be optically thick at certain locations.

For each region, the dust continuum emission at 70 and 870\,$\mu$m is shown in the left panel of Figure~\ref{fig:soumap}, and a two-color image of  $\rm C^{18}O$\,(2--1) (red) and $\rm DCO^+$\,(1--0) (cyan) is shown in the right panel of Figure~\ref{fig:soumap}. 
Comparing these two panels, we visually separate each source into two or three zones:

{\it The $DCO^+$-dominant zone} appears cyan in the two-color image, where $\rm C^{18}O$ emission is weaker than elsewhere. The 870\,$\mu$m continuum peak in this region is labeled  P1, where the  dust emission at 70\,$\mu$m is $<3\sigma$ rms.

{\it The CO-dominant zone}  appears reddish in the two-color image, where the $\rm DCO^+$\,(1--0) emission shows $\rm S/N<3$. This region is 70\,$\mu$m bright, and we label the continuum peak at either 870\,$\mu$m  (if it exists; e.g., G\,14.49-0.13 and G\,11.38+0.81) or 70\,$\mu$m  (e.g., G\,34.74-0.12 and  G\,15.22-0.43)   P3. 

{\it The transition zone} exhibits equally  weak (e.g., G\,34.74-0.12, G\,14.49-0.13) or strong (G\,11.38+0.81) $\rm DCO^+$ and CO emissions. The 870\,$\mu$m continuum peak in this region is labeled  P2, and the  dust emission at 70\,$\mu$m here is brighter than that toward P1. Since G\,15.22-0.43 shows only two continuum peaks at 870\,$\mu$m in the imaged region, we do not separate the CO-dominant and the transition zones on its map.

  \begin{figure*}
  \begin{tabular}{lccc}
G\,15.22-0.43
&  \multirow{40}{*} {\begin{sideways}Dec. offset (\arcsec)\end{sideways}}
& \includegraphics[align=c,height=6.1cm] {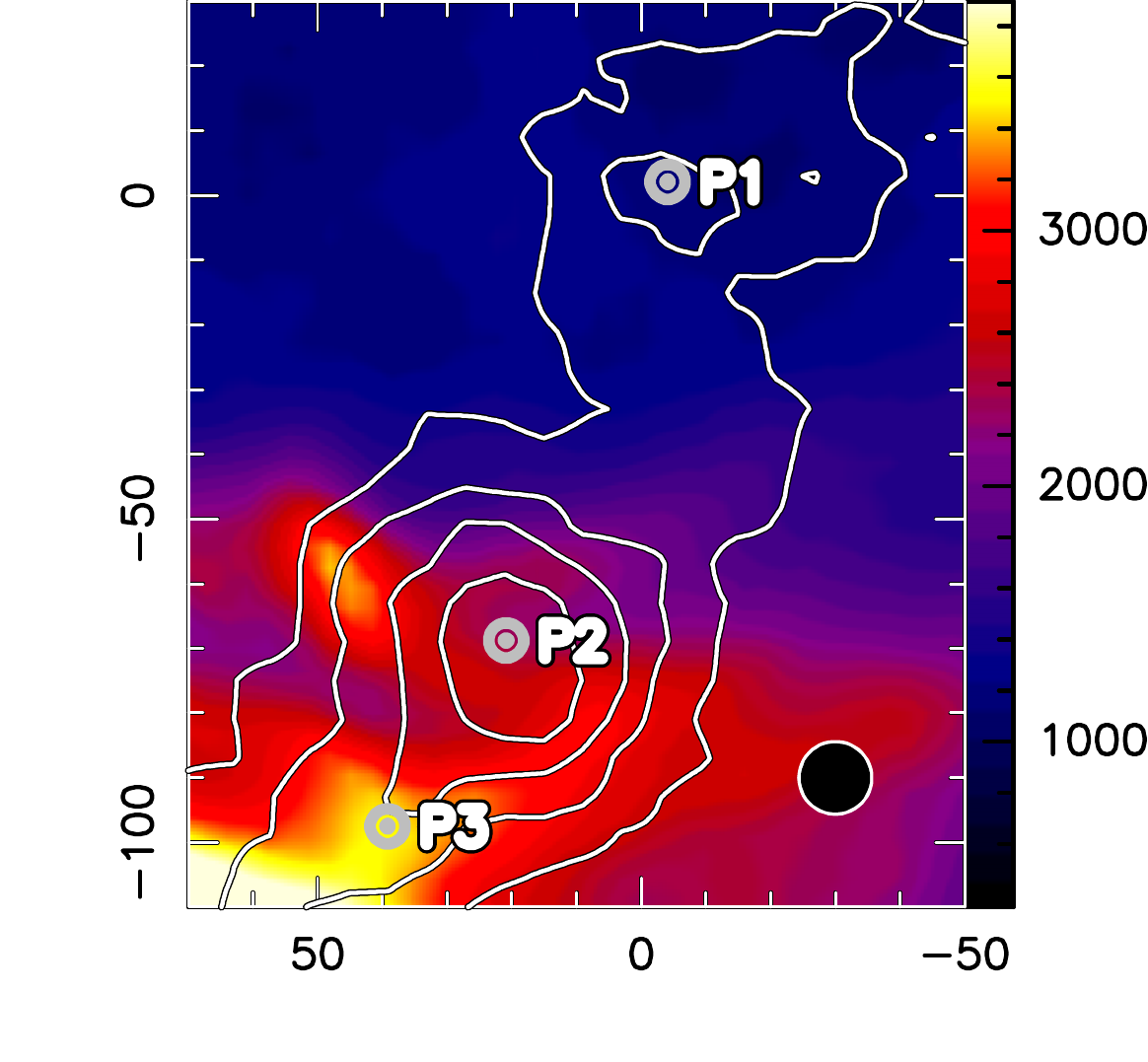}
& \includegraphics[align=c,height=6.1cm] {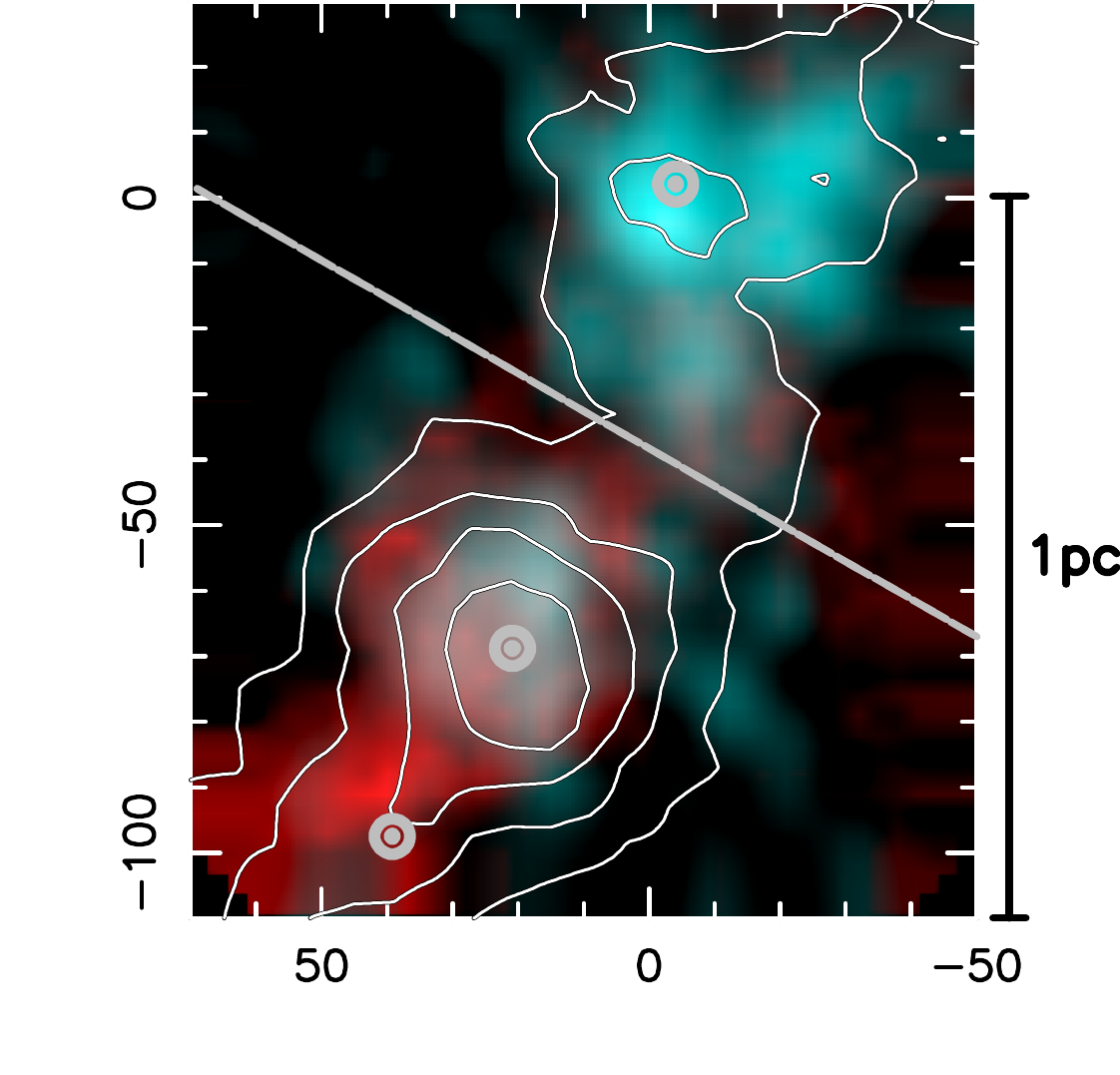}\\[-0.2cm]
G\,11.38+0.81 &
 & \includegraphics[align=c,height=8.cm] {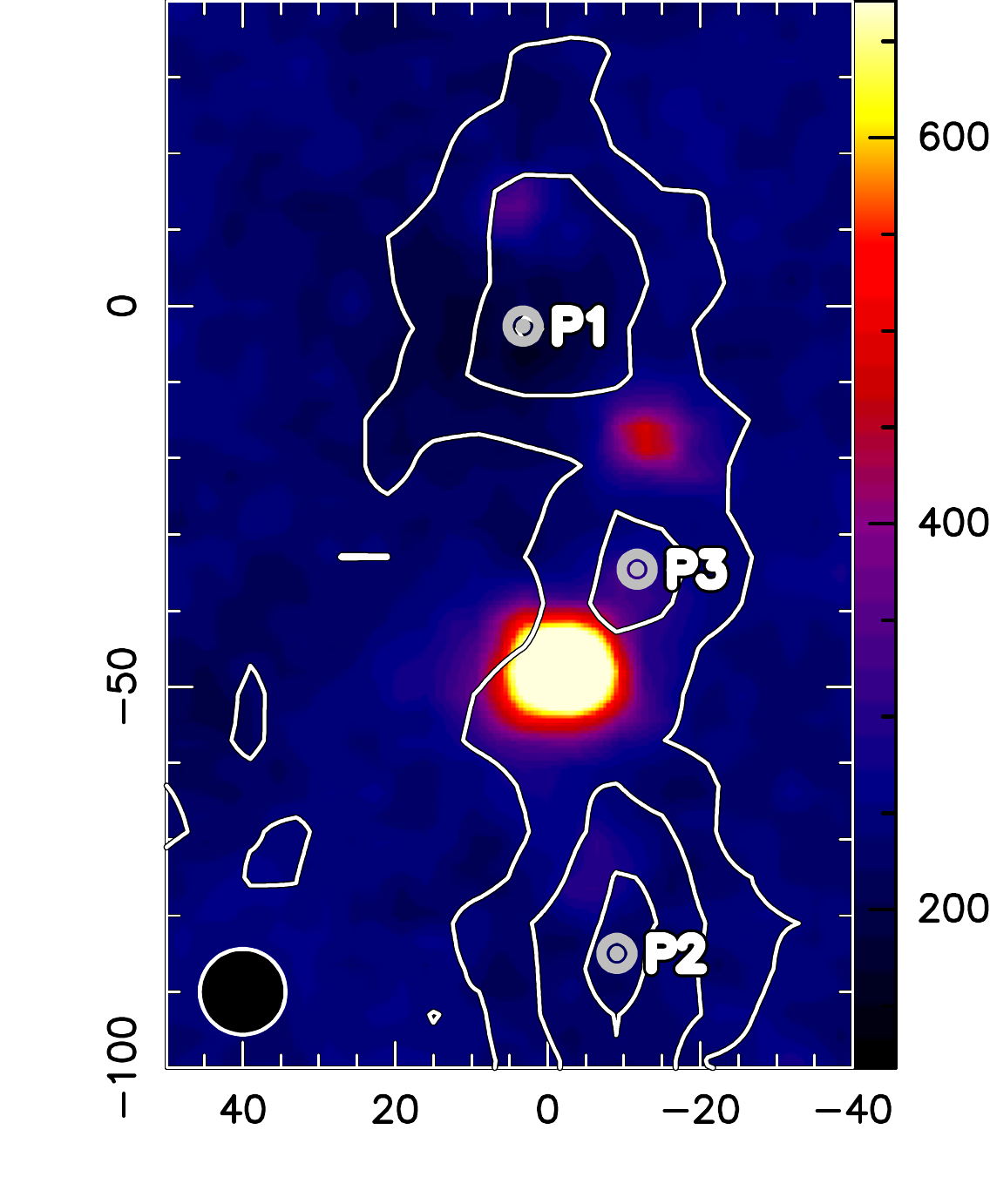}
 &\includegraphics[align=c,height=8.cm] {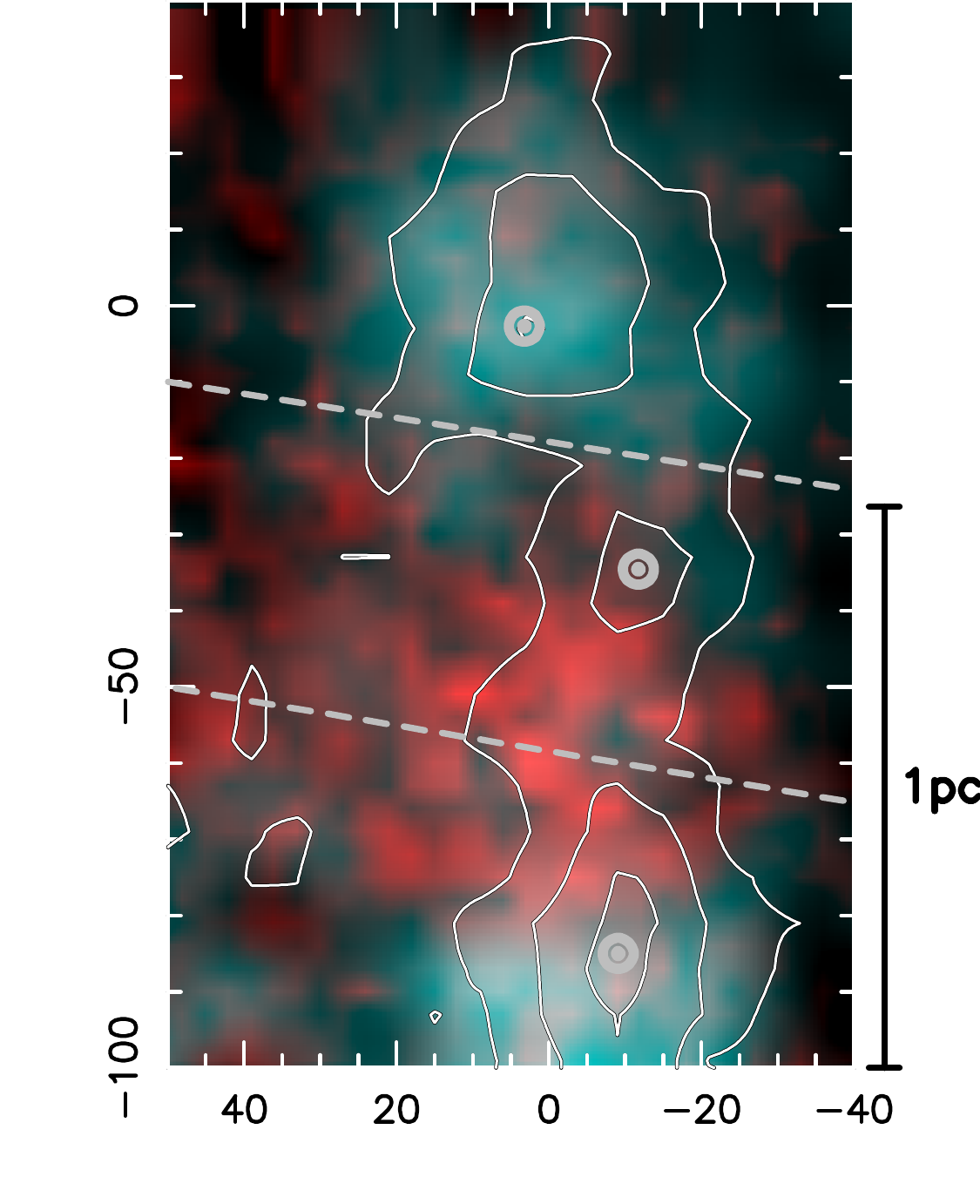}\\[-0.5cm]
G\,14.49-0.13 &
& \includegraphics[align=c,height=3.7cm] {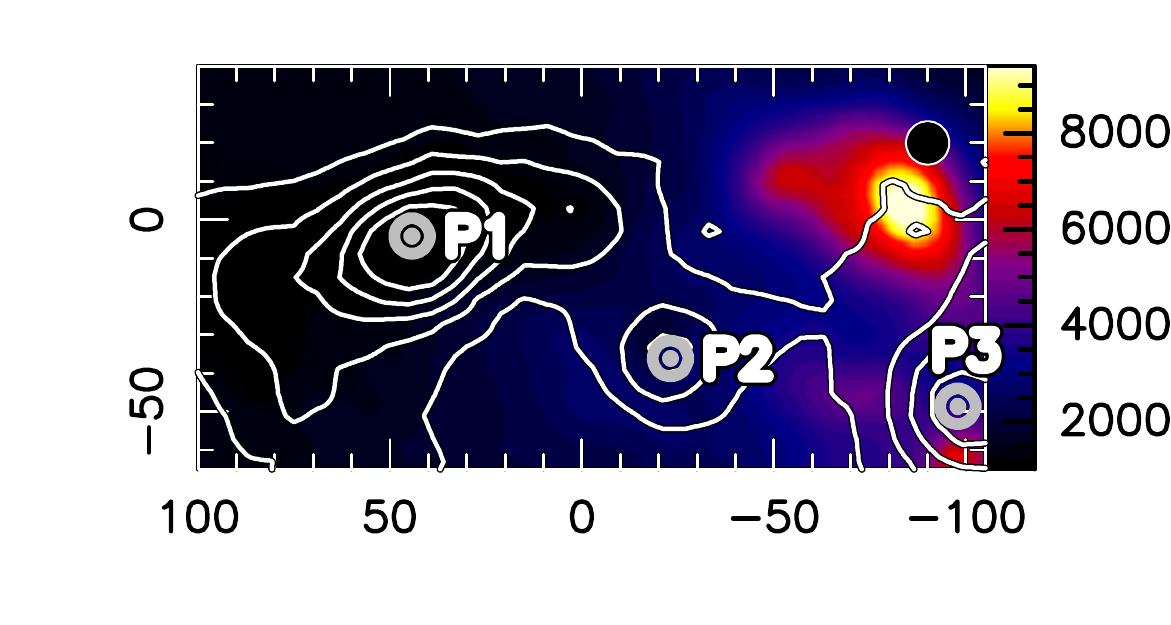}
& \includegraphics[align=c,height=3.7cm] {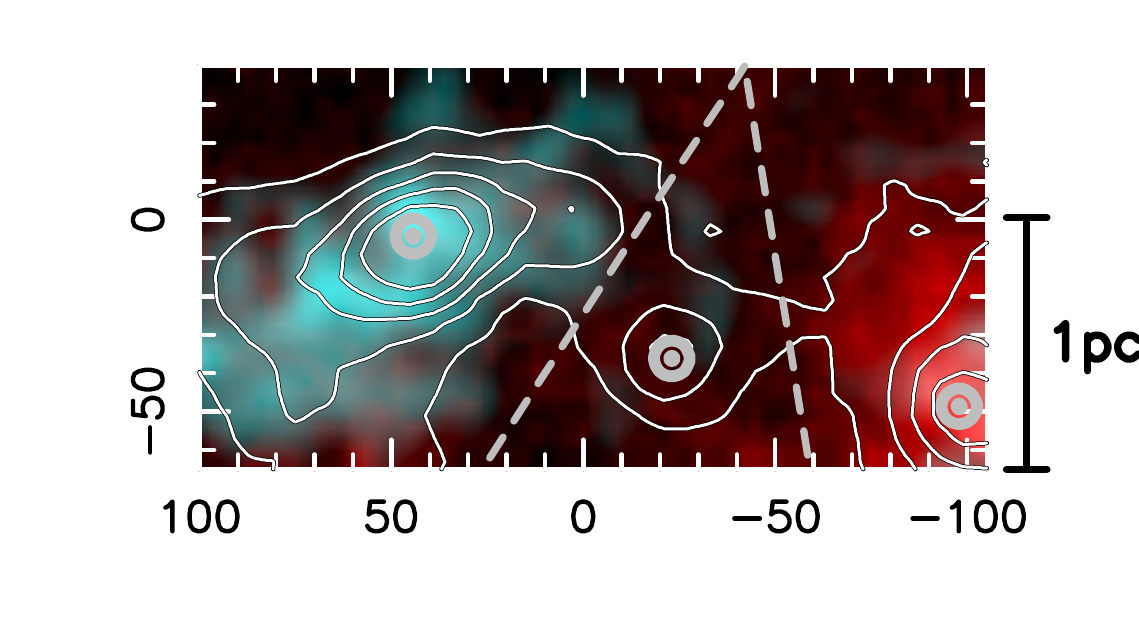}\\[-0.2cm]
G\,34.74-0.12 &
   & \includegraphics[align=c,height=5.5cm] {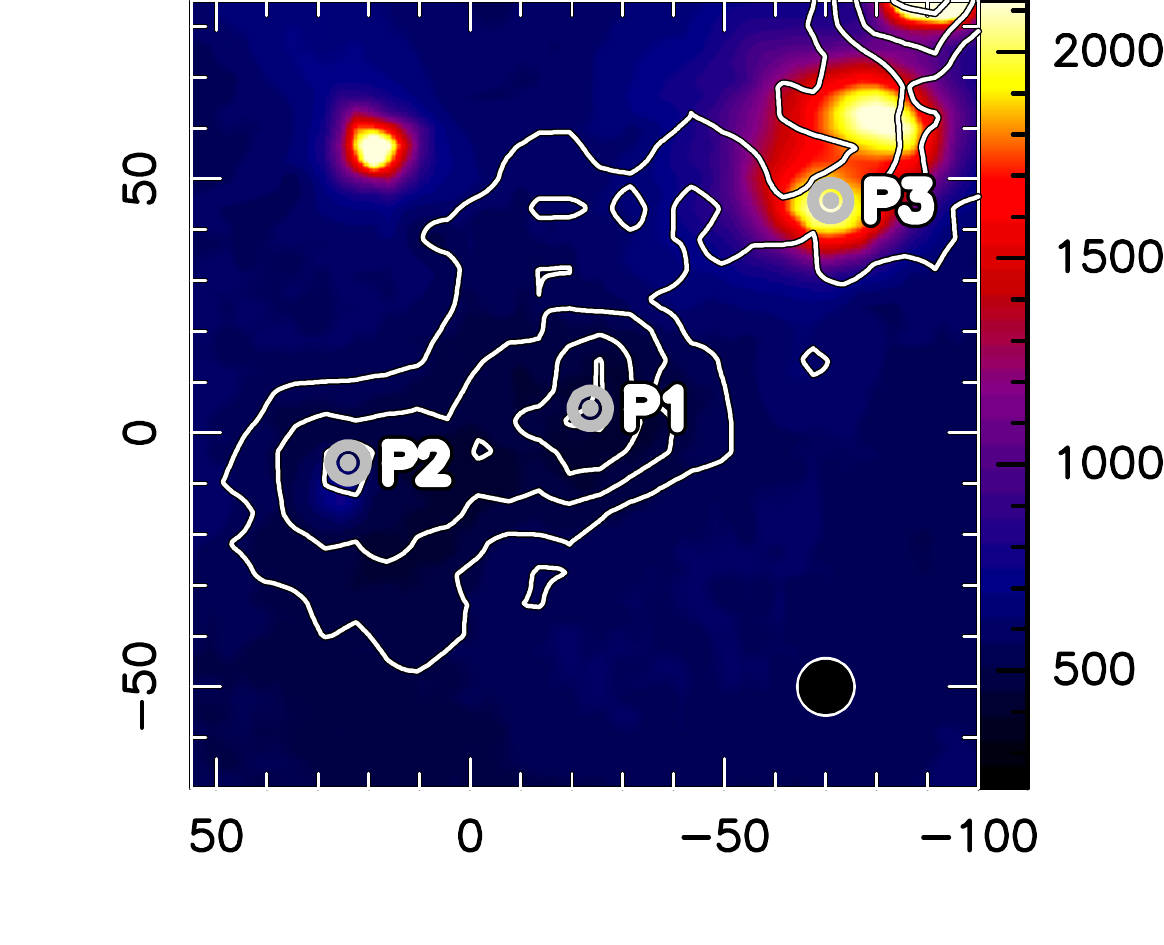}
 &\includegraphics[align=c,height=5.5cm] {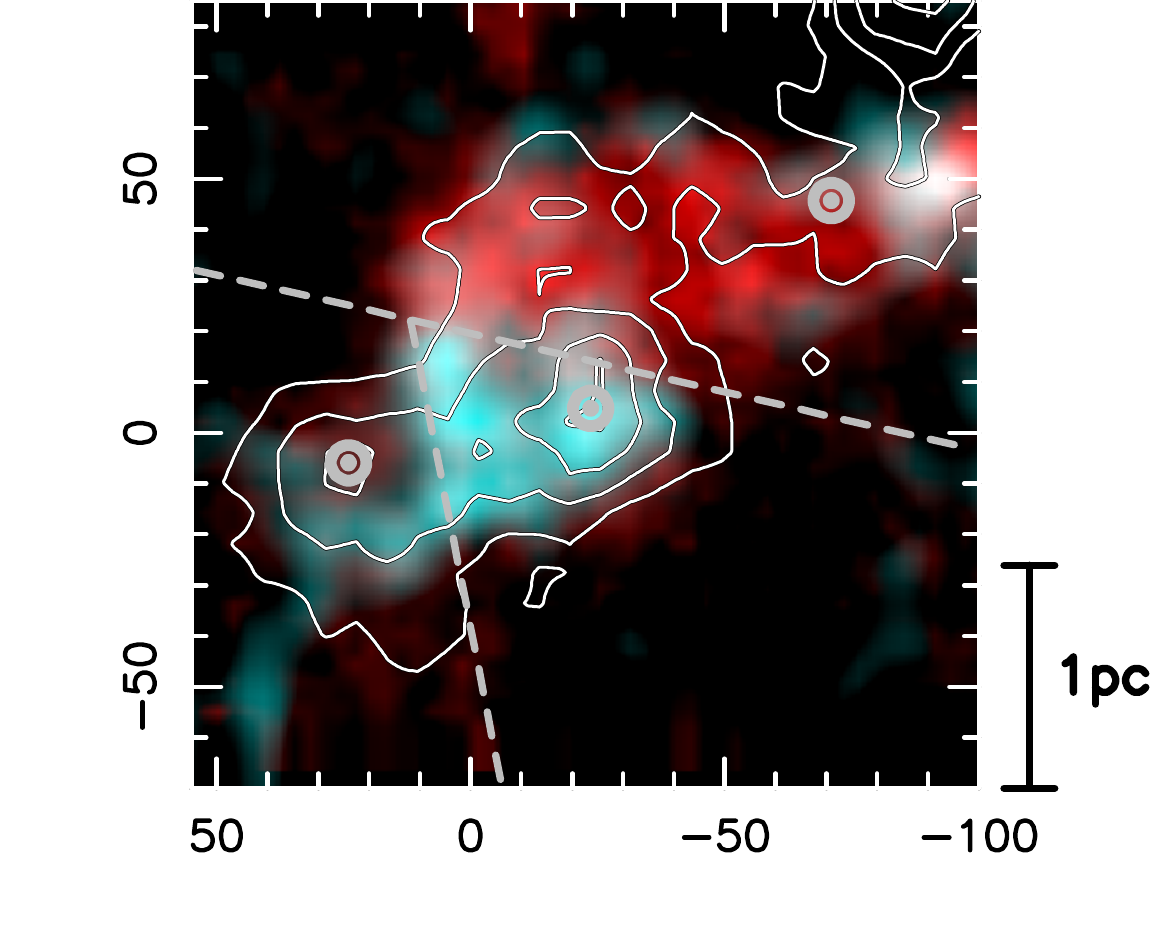}\\[-0.2cm]
&&\multicolumn{2}{c}{R.A. offset (\arcsec)}
\end{tabular}
\caption{Compilation of the continuum and line data for  G\,15.22-0.43, G\,11.38+0.81, G\,14.49-0.13, and G\,34.74-0.12.
 {\it Left column}: color maps of the dust emission observed by {\it Herschel} at 70\,$\mu$m (colorscale in unit of $\rm MJy\,sr^{-1}$).
{\it Right column}: two-color maps show the intensity of $\rm C^{18}O$\,(2--1, in red and with an angular resolution of 11\farcs8) and $\rm DCO^+$\,(1--0, in cyan and with an angular resolution of 36\farcs0) integrated in the same velocity range (given in Table~\ref{tab:linedeu}). 
 The white contours in each panel show continuum emission observed by {APEX} at 870\,$\mu$m \citep{schuller09}, starting from 3\,$\sigma$ rms and increasing by  3\,$\sigma$ steps. 
 The 1$\sigma$ of 870\,$\mu$m emission for 
G\,15.22-0.43, G\,11.38+0.81, G\,14.49-0.13, and G\,34.74-0.12 is 15.2, 10.6, 23.7, and 20.6\,$\rm MJy\,sr^{-1}$,  respectively, at an angular resolution of 18\farcs2 (shown in black in the corners of the left panels).
  The dashed lines and the labeled positions P1, P2, and P3  in each panel are described in Section~\ref{distribution}.
}\label{fig:soumap}
\end{figure*}

\subsection{CO depletion factor and D-fraction of $\rm HCO^+$}\label{calculation}
Following \citet{feng19a}, we derive the map of the CO depletion factor and the D-fraction map of $\rm HCO^+$  for each source in four steps.

\subsubsection {\it Dust temperature map and $\rm H_2$ column density map}\label{h2tdust}
Using our well-developed image combination and iterative SED fitting  method (see details in \citealt{lin16,lin17} {and our pilot study ({\it Paper I}; \citealt{feng19a}}), we established a reliable blackbody model and obtained the dust opacity index $\beta$ map for each source at a coarse angular resolution of 37\arcsec. In order to recover the missing flux of the parsec-scale structure, we used the continuum data from PACS 160\,$\mu$m, SPIRE 250, 350, 500\,$\mu$m and combined the Planck data with JCMT at 850\,$\mu$m or APEX at 870\,$\mu$m. 

Then, assuming that the $\beta$ map has no local variation from 37\arcsec~ to 18\arcsec~ resolution and  that  the gas-to-dust mass ratio $log(\gamma)=\rm 0.087 {\it R_{\rm GC}} (kpc)+1.44$  \citep{draine11,giannetti17b} changes with galactocentric distance $R_{\rm GC}$, we  fit the SED of each pixel by using the continuum data from PACS 160\,$\mu$m, SPIRE 250 and the combined PLANCK-JCMT 850\,$\mu$m or PLANCK-APEX 870\,$\mu$m.
 Therein, achieving  an angular resolution of 18\arcsec or 20\arcsec, we simultaneously obtain the maps of
 dust temperature $T_{\rm dust}$ and $\rm H_2$ column density $N_{\rm H_2}$ (Figure~\ref{fig:paramap}).

 The $\rm H_2$ column density  toward each clump is in the range of $\rm 10^{22}\text{--}10^{23}\,cm^{-2}$ (Table~\ref{tab:physical}), which is at least 1\,mag higher than that ($\rm \sim10^{21}\,cm^{-2}$) toward the outskirts of the natal cloud (the location where the continuum emission at 870\,$\mu$m is $\rm <5\sigma$ rms). Therefore, we believe that the background and foreground contamination have a negligible effect on the $\rm H_2$ column density estimates of our targeted regions.

\subsubsection {\it Gas temperature}\label{gastemp} 
Our line-imaging survey includes three thermometers:  para ($p$)-$\rm NH_3$ lines,  $p$-$\rm H_2CO$ lines, and CO isotopologue lines.

The N-bearing species are resilient to depletion \citep[e.g., ][]{caselli99,bergin02,caselli02b,jorgensen04}. 
The inversion lines from different rotational ladders ($J$=1,2...) are coupled only collisionally and have similar frequencies. Furthermore, the combination of its energy level structures and the numerical value of Einstein coefficients $A_{ij}$ makes the majority of the $\rm NH_3$ population stay in the metastable states. In the temperature range of 10--100\,K, inversion lines of $\rm NH_3$ have modestly high critical densities of $\rm \lesssim10^4\,cm^{-3}$  \citep[e.g.,][]{ho83,walmsley83,crapsi07,rosolowsky08,juvela11}. These unique qualities make $\rm NH_3$ a great interstellar thermometer for the gases of modestly high  densities \citep[e.g., inversion lines, see][]{lid02,lid03} and high \citep[{e.g.,  rotational transition lines, see}][]{caselli17}  densities.
 Two regions in our pilot sample (G\,14.49-0.13 and G\,34.74-0.12) are covered by the {\it RAMPS} program, so we  use the $p$-$\rm NH_3$ lines ($J$,$K$)=(1,1) and (2,2) to provide the gas temperature maps toward them. To derive the gas kinetic temperature $T_{kin}$($p$-$\rm NH_3$) maps at an angular resolution of 34\arcsec, we apply two Monte Carlo fitting tools; one is \texttt{HfS} developed by \citet{estalella17},  and another is a much faster temperature-fitting algorithm \citep{wangs20}\footnote{This tool is to measure the gas kinetic temperature by only using the line intensity ratios between $\rm NH_3$ hyperfine groups, which was first proposed by \citet{lid13}. The python package for this method \citep{wangs20} is publicly available at \url{https://github.com/plotxyz/nh3_trot.git}.  }. We found consistent results for $T_{kin}$($p$-$\rm NH_3$) from both tools, spanning the range of 11--21\,K in our pilot sample.

Comparing this gas temperature with the  dust temperature $T_{\rm dust}$ (Sect.~\ref{h2tdust})  toward P1, P2, and P3 of each region (Table~\ref{tab:physical}), we found that they are consistent at individual positions,  though the angular resolutions for their measurements are different.
 Therefore, we believe that dust and gas are thermally coupled 
  in G\,14.49-0.13 and G\,34.74-0.12 \citep{goldsmith01}. 
 $\rm NH_3$ images toward  G\,15.22-0.43 and G\,11.38+0.81 are not available. Nevertheless, they show similar dust properties (dynamic ranges of $T_{\rm dust}$ and  $N_{\rm H_2}$) as the other two, so we expect that the dust and gas towards  these regions are thermally coupled as well, i.e., the $T_{kin}$($p$-$\rm NH_3$) maps of  G\,15.22-0.43 and G\,11.38+0.81 are consistent with their  $T_{\rm dust}$ maps.

Using the IRAM 30\,m and NRO 45\,m, we observed the 1--0 and 2--1  lines  of all three CO isotopologues ($\rm C^{18}O$, $\rm C^{17}O$, and $\rm ^{13}CO$) at an angular resolution of 16\arcsec and 12\arcsec, respectively. Smoothing them to the same angular resolution as that  of the dust continuum observations (18\arcsec or 20\arcsec)  allows us to compare the gas and dust temperature at the same spatial scale toward individual regions. 
To test whether these low-$J$ lines can be treated as gas thermometers in each region, we estimate  the  H and $\rm H_2$ number density $n_{\rm H}$, the molecular column density $N_{\rm mol}$, and the gas kinetic temperature $T_{kin}$\,(CO) toward P1, P2, and P3  by employing the large velocity gradient (LVG) escape probability approximation.
Using the non-local thermal equilibrium (non-LTE) statistical equilibrium radiative transfer code RADEX \citep{vandertak07} along with a related solver (Fujun Du's myRadex)\footnote{See https://github.com/fjdu/myRadex.}, we 
apply the MultiNest Algorithm \citep{feroz07,feroz08,feroz13}, and derive the probability density function (PDF) of these variables  (Table~\ref{tab:physical}). From the best-fit results, $T_{kin}$\,(CO)  toward individual locations in G\,15.22-0.43 and G\,14.49-0.13 is generally higher than $T_{\rm dust}$, but with larger uncertainties (20\%--50\%). A possible reason might be the deeply embedded protostellar objects and young outflows, which have been resolved with our ALMA observations at higher angular resolution (1\farcs2, \citealt{sanhueza19,lis19}, S. Feng et al. 2020, in preparation). 
In contrast, the $T_{kin}$\,(CO)  toward individual locations in G\,11.38+0.81 and G\,34.74-0.12 are  in the range of 7--10\,K, which is lower than $T_{\rm dust}$ at the same locations. The 2--1 lines, with critical densities (Table~\ref{tab:lines}) not significantly less than $n_{\rm H}$ ($\rm \sim3\times10^4\,cm^{-3}$), may be subthermally excited. Therefore, we do not consider low-$J$ CO isotopologue lines as reliable gas temperature tracers in this work.
Nevertheless, the LVG estimates indicate that the $\rm C^{18}O$\,(2--1) and  $\rm C^{17}O$\,(2--1) lines are optically thin ($\tau\rm <1$) toward the pixels where they are detected with $\rm S/N>5$. 
Moreover, we also derive the  $\rm C^{18}O$ column densities by using $T_{\rm dust}$ and assuming that the 2--1 line is optically thin and in LTE. Compared to those,
 $\rm C^{18}O$ column densities derived from RADEX are  higher by a factor of 2--3  toward a few locations, such as  P1 and P2 in G\,11.38+0.81 and  G\,14.49-0.13 (see Table~\ref{tab:gaspara}). Nevertheless, the  differences in the above  estimates lie within the uncertainties.

Moreover, our IRAM 30\,m observations covered four lines of  $p$-$\rm H_2CO$ ($\rm 1_{0,1}-0_{0,0}$, $\rm 3_{0,3}-2_{0,2}$, $\rm 3_{2,2}-2_{2,1}$, $\rm 3_{2,1}-2_{2,0}$) that have been previously used as a gas thermometer \citep[e.g., ][]{mangum93,johnston03,leurini04,leurini07,ao13,ginsburg16,giannetti17,tang18,feng19a}. The $\rm 1_{0,1}-0_{0,0}$ ($\rm E_u/k_B\sim3\,K$) and $\rm 3_{0,3}-2_{0,2}$  ($\rm E_u/k_B\sim21\,K$) lines are detected with $\rm S/N>4$ toward all zones, while the $\rm 3_{2,2}-2_{2,1}$ and  $\rm 3_{2,1}-2_{2,0}$ lines ($\rm E_u/k_B\sim68\,K$) detected with low S/N ($\rm <3$) can be used to constrain the upper limit of the gas temperature. Smoothing them to the same angular resolution (35\farcs6), we used RADEX and found that the gas kinetic temperatures derived from these lines, $T_{kin}$($p$-$\rm H_2CO$), are in the range of 12--37\,K, higher than $T_{kin}$($p$-$\rm NH_3$) at the same spatial scale and $T_{kin}$\,(CO) at a smaller spatial scale. Possible reasons are as follows.
(1) The $\rm H_2CO$ is likely formed in the gas phase as reaction products of hydrocarbons \citep{yamamoto17}. The hydrocarbons are typically found in the outer regions of molecular clouds, where the gas temperature is  higher than in the dense regions of the cold clumps (see the $T_{\rm dust}$ maps in Figure~\ref{fig:soumap}).   
(2) In the dense and cold clumps,  $\rm H_2CO$ is probably frozen onto dust grains in the same way as CO and maybe transformed into $\rm CH_3OH$ in a relatively fast process. 
(3) The $\rm H_2CO$ lines have higher critical densities than the $\rm NH_3$ lines and CO isotopologue lines even at  the same temperature (Table~\ref{tab:lines}), so they may trace different gas.

\begin{table*}
\caption{Dust and gas properties of our target
}\label{tab:physical}
\scalebox{1}{
\begin{tabular}{p{2cm}p{2cm}|p{1cm}p{2.2cm}p{2.2cm}p{2.2cm}p{2.2cm}}
\hline\hline

Properties & &Source$^a$             &{G\,15.22-0.43}                          &{G\,11.38+0.81}                                                                 &{G\,14.49-0.13}    &{G\,34.74-0.12} 
   \\% &G\,12.6738 &G\,30.8523\\ 
\hline
\multirow{3}{*} {[R.A., Dec.]$^a$}  &\multirow{3}{*} {(J2000, J2000)}  
&P1        &[18$\rm ^h$19$\rm ^m$52$\rm ^s$.637, -15$\rm ^\circ$55\arcmin59\arcsec.95]
	   &[18$\rm ^h$07$\rm ^m$35$\rm ^s$.771, -18$\rm ^\circ$42\arcmin46\arcsec.37]    	        
	   &[18$\rm ^h$17$\rm ^m$16$\rm ^s$.750, -16$\rm ^\circ$25\arcmin21\arcsec.33]
	   &[18$\rm ^h$55$\rm ^m$12$\rm ^s$.803, +01$\rm ^\circ$33\arcmin01\arcsec.75]  \\
&&P2  	   &[18$\rm ^h$19$\rm ^m$50$\rm ^s$.907, -15$\rm ^\circ$54\arcmin49\arcsec.07]    
           &[18$\rm ^h$07$\rm ^m$36$\rm ^s$.638, -18$\rm ^\circ$41\arcmin24\arcsec.05]                
           &[18$\rm ^h$17$\rm ^m$22$\rm ^s$.106, -16$\rm ^\circ$24\arcmin58\arcsec.51]
           &[18$\rm ^h$55$\rm ^m$09$\rm ^s$.624, +01$\rm ^\circ$33\arcmin12\arcsec.44]  \\
&&P3   	   &[18$\rm ^h$19$\rm ^m$53$\rm ^s$.907, -15$\rm ^\circ$56\arcmin28\arcsec.66]
           &[18$\rm ^h$07$\rm ^m$35$\rm ^s$.584, -18$\rm ^\circ$41\arcmin55\arcsec.96]                       
           &[18$\rm ^h$17$\rm ^m$12$\rm ^s$.229, -16$\rm ^\circ$25\arcmin42\arcsec.85]
           &[18$\rm ^h$55$\rm ^m$06$\rm ^s$.466, +01$\rm ^\circ$33\arcmin53\arcsec.34]   \\
\hline
\multirow{3}{*} {$N_{\rm H_2}$ $^b$}  &\multirow{3}{*} {($\rm \times10^{22}\,cm^{-2}$)}  
&P1     &{$\rm 3.6\pm 0.1$}
	   &{$\rm 6.7\pm 0.1$}     	        
	   &{$\rm 9.3\pm0.8$}
	   &{$\rm 3.9\pm0.2$} \\
&&P2  &{$\rm 5.5\pm0.1$}      
           &{$\rm 5.9\pm 0.1$}                  
           &{$\rm 4.6\pm 0.2$}
           &{$\rm 3.7\pm 0.1$}  \\
&&P3   &{$\rm 2.1\pm 0.1$}  
            &{$\rm 4.7\pm 0.1$}                       
            &{$\rm 5.8\pm 0.1$}
            &{$\rm 1.6\pm 0.1$}   \\
 \hline
\multirow{3}{*} {$T_{\rm dust}$$^b$}  &\multirow{3}{*} {(K)} 
&P1  &{$\rm 17.2\pm 0.1$}    
        &{$\rm  11.5\pm 0.1$}          
         &{$\rm 15.6\pm 0.3$}
         &{$\rm  14.6\pm0.1$}   \\
&&P2    &{$\rm  17.9\pm 0.1$}
             &{$\rm 12.6\pm 0.1$}                   
             &{$\rm 18.0\pm 0.2$}
             &{$\rm 14.9\pm 0.1 $} \\
&&P3   &{$\rm 21.9\pm 0.1$}    
            &{$\rm 12.2\pm 0.1 $}                
            &{$\rm 18.9\pm 0.1 $}
            &{$\rm 18.8\pm 0.1$} \\
  \hline                                                    
\multirow{3}{*} {$T_{kin}$($p$-$\rm NH_3$)$^e$}  &\multirow{3}{*} {(K)}  
&P1 &{--$^c$}  
        &{--$^c$}           
        &{$\rm 16.1\pm 0.9 $}
        &{$\rm  14.9\pm 1.7$}   \\
&&P2  &{--$^c$}     
           &{--$^c$}                  
           &{$\rm 17.9\pm 5.6 $}
           &{$\rm  13.3\pm1.0$}  \\
&&P3   &{--$^c$}   
           &{--$^c$}                     
           &{$\rm 17.1\pm 6.5 $}
           &{$\rm 15.8\pm 3.5$} \\
 \hline                                                                                                                                                           
\multirow{3}{*} {$T_{kin}$($p$-$\rm H_2CO$)$^f$}  &\multirow{3}{*} {(K)} 
&P1  &{$\rm 25.2\pm1.5$}      
         &{$\rm 27.5\pm14.0$}   
         &{$\rm 22.7\pm1.5$}    
         &{$\rm 28.0\pm1.9$}\\
&&P2  &{$\rm 25.5\pm1.5$}      
           &{$\rm 37.6\pm3.9$}   
           &{$\rm 19.6\pm12.4$}   
           &{$\rm 20.2\pm1.0$} \\
 &&P3   &{$\rm 24.8\pm1.4$}      
             &{$\rm 25.4\pm1.5$}   
             &{$\rm 12.0\pm0.5$}   
             &{$\rm 31.2\pm2.4$}\\
                \hline          

\multirow{3}{*} {$n_{H,\rm lvg, {\it p}-H_2CO}$$^d$}  &\multirow{3}{*} {($\rm \times10^4\,cm^{-3}$)} 
&P1  &{$\rm 5.4\pm3.5$}      
        &{$\rm 13.4\pm7.6$}   
        &{$\rm 11.4\pm2.2$}    
        &{$\rm 5.5\pm1.2$}\\
&&P2  &{$\rm 4.7\pm3.3$}      
           &{$\rm 6.6\pm1.4$}   
           &{$\rm 9.3\pm2.1$}   
           &{$\rm 11.6\pm2.1$} \\
 &&P3   &{$\rm 7.8\pm5.1$}      
             &{$\rm 6.3\pm4.2$}   
             &{$\rm 12.5\pm0.3$}   
             &{$\rm 7.0\pm1.4$}\\
   \hline      
     \multirow{3}{*} {$T_{kin}$(CO)$^g$}  &\multirow{3}{*} {(K)} 
&P1 &$\rm 10.9\pm2.6$
       &$\rm 7.1\pm1.5$
       &$\rm 12.0\pm5.9$
       & $\rm 10.2\pm1.5$\\
&&P2  &$\rm 25.1\pm4.8$
       &$\rm 7.4\pm1.2$
       &$\rm 11.1\pm9.4$
       & $\rm 9.5\pm0.9$\\

 &&P3   &$\rm 36.2\pm7.6$
       &$\rm 8.6\pm0.6$
       &$\rm 22.9\pm12.9$
       &$\rm 9.0\pm3.0$ \\

   \hline     
   \multirow{3}{*} {$n_{H,\rm lvg, CO}$$^d$}  &\multirow{3}{*} {($\rm \times10^4\,cm^{-3}$)} 
&P1  & {$\rm 3.7\pm2.6$}     
        & {$\rm 3.0\pm2.5$} 
        & {$\rm 3.3\pm2.8$}   
        &{$\rm 3.1\pm2.6$} \\
&&P2  &{$\rm 3.7\pm2.8$}      
           & {$\rm 3.0\pm2.5$} 
           &{$\rm 3.1\pm2.7$}    
           & {$\rm 3.1\pm2.7$}\\
 &&P3   &{$\rm 3.3\pm2.8$}    
             &{$\rm 3.3\pm2.8$}   
             & {$\rm 3.1\pm2.7$}  
             &{$\rm 3.1\pm2.6$}\\

%%%%%%%%%%%%%%%%%%%%%%%%%%%%%

\hline \hline
\multicolumn{7}{l}{\color{black} { Note.} $a$. P1, P2, and P3 denote the $\rm DCO^+$-dominant,   transition, and CO-dominant  zones, respectively.  The given coordinates of P1 and P2 }\\
\multicolumn{7}{l}{\color{black} ~~~~~~~~~~~~~correspond to  the  870\,$\mu$m emission peaks, and those of P3  correspond to the  870\,$\mu$m  (G\,14.49-0.13 and G\,11.38+0.81) }\\
\multicolumn{7}{l}{\color{black}  ~~~~~~~~~~~~~ or 70\,$\mu$m (G\,34.74-0.12 and G\,15.22-0.43) emission peaks.}\\
\multicolumn{7}{l}{\color{black} ~~~~~~~~~ $b$. From the SED fit, achieving an angular resolution of 18\arcsec or 20\arcsec.}\\
\multicolumn{7}{l}{\color{black} ~~~~~~~~~ $c$. Here ``--" indicates a the location where we do not have  $\rm NH_3$ observations.}\\
\multicolumn{7}{l}{\color{black} ~~~~~~~~~ $d$. The $\rm H_2$ and H number density is derived by an LVG fit of the $p$-$\rm H_2CO$ lines. }\\
\multicolumn{7}{l}{\color{black} ~~~~~~~~~ $e$. $T_{kin}$ derived from $p$-$\rm NH_3$ (2,2)/(1,1) lines at an angular resolution of 34\farcs7. }\\
\multicolumn{7}{l}{\color{black} ~~~~~~~~~ $f$. $T_{kin}$  derived from an LVG fit of four $p$-$\rm H_2CO$ lines at an angular resolution of 35\farcs6. }\\
\multicolumn{7}{l}{\color{black} ~~~~~~~~~ $g$. $T_{kin}$  derived from an LVG fit of  $\rm C^{18}O$ (2-1)/(1-0), $\rm C^{17}O$ (2-1)/(1-0), and $\rm ^{13}CO$ (2-1)/(1-0) lines at an angular resolution of 16\farcs4. }\\

\end{tabular}
}
\end{table*}

\subsubsection  {\it Depletion factor  map of $\rm C^{18}O$}\label{dc18o}
{ 
Assuming that $\rm C^{18}O$\,(2--1) is optically thin and under LTE toward each pixel (Sect~\ref{gastemp}), we derived the column density of $\rm C^{18}O$ toward P1, P2, and P3 by using the dust temperature $T_{\rm dust}$ and the gas kinetic temperature $T_{kin}$($p$-$\rm NH_3$) at an angular resolution of $\rm \sim35\arcsec$.
For testing the effect of the gas temperature uncertainty on the measurement of the column density uncertainty, we also use $T_{kin}$($p$-$\rm H_2CO$) at the same angular resolution. 
We  found that the estimates of $\rm C^{18}O$ column density by using different temperature sets at individual pixels are consistent within the uncertainty  (Table~\ref{tab:gaspara}).
Furthermore, when the gas temperature is in the range of 11--40\,K, an uncertainty of 10\,K (at most) brings in 15\% uncertainty on the accuracy of  the $\rm C^{18}O$ column density estimates.} In the interest  of higher angular resolution and less uncertainty, we use the $T_{\rm dust}$ map to derive the observed $\rm C^{18}O$ column density (denoted as $N_{\rm C^{18}O}^o$) map.

Statistically, in a star-forming environment  without CO depletion, its relative abundance $\chi^E\rm (^{12}CO)$ with respect to $\rm H_2$  is expected  to change with galactocentric distance $R_{\rm GC}$. Moreover, the $\mathcal{R}_{\rm ^{16}O/^{18}O}$ isotopic ratio changes with galactocentric distance as well, so the  $\rm C^{18}O$ column density is expected  (denoted as $N_{\rm C^{18}O}^E$) to be correlated with the observed $\rm H_2$ column density as \citep{frerking82,wilson94,giannetti14}:  
\begin{equation}
\begin{split}
N_{\rm C^{18}O}^E&=\frac{N_{\rm ^{12}CO}^E}{\mathcal{R}_{\rm ^{16}O/^{18}O}^E}=\frac{\chi^E(\rm ^{12}CO)}{\mathcal{R}_{\rm ^{16}O/^{18}O}^E} N_{\rm H_2}\\
&=  \frac{9.50\times10^{-5} {\it e}^{(1.11-0.13{\it R}_{\rm GC}({\rm kpc}))}}{58.80{\it R}_{\rm GC}({\rm kpc})+37.10}  N_{\rm H_2}.
\end{split}
\label{eq:nc18oex}
\end{equation}
Then, the $\rm C^{18}O$ depletion factor  (Figure~\ref{fig:paramap}) can be derived as
\begin{equation}
f_D({\rm C^{18}O})=\frac{N_{\rm C^{18}O}^E}{N_{\rm C^{18}O}^o}. \label{eq:dc18o}
\end{equation}

The sources in our pilot sample are at the galactocentric distance $R_{\rm GC}=\rm 5.4\pm0.5$\,kpc. 
Apart from G\,11.38+0.81, where $f_D\rm(C^{18}O)$ is higher than the rest of the sources by a factor of 2--3, the maximum of the $f_D\rm(C^{18}O)$ appears toward the P1 (or P2) of each region, reaching $\rm 4.2\pm0.5$ at a linear scale of 0.18--0.46\,pc (an angular resolution of 20\arcsec). 
Moreover, smoothing the dust and $\rm C^{18}O$ line emission from 20\arcsec to 35\arcsec does not change the  $f_D\rm(C^{18}O)$ estimates.  Without analyzing the entire sample of 24 regions, we are not able to test whether or not the absolute value of $f_D\rm(C^{18}O)$ shows a correlation with the source $R_{\rm GC}$.} 
Nevertheless, when comparing the relative  $\rm C^{18}O$ depletion factor toward locations in individual sources, we note that  the largest $f_D\rm(C^{18}O)$ values towards P1 or P2 are higher than those toward P3 (the CO-dominant zone) by a factor of 1.4--3, with small uncertainties (Table~\ref{tab:parafDD}).

\subsubsection  {\it D-fraction map of $\rm HCO^+$}\label{dfrac}
\begin{figure*}
    \begin{tabular}{lcc}
    \vspace{0.3cm}
G\,15.22-0.43 &  \multirow{40}{*} {\begin{sideways}Dec. offset (\arcsec)\end{sideways}}
&\includegraphics[align=c,width=16cm] {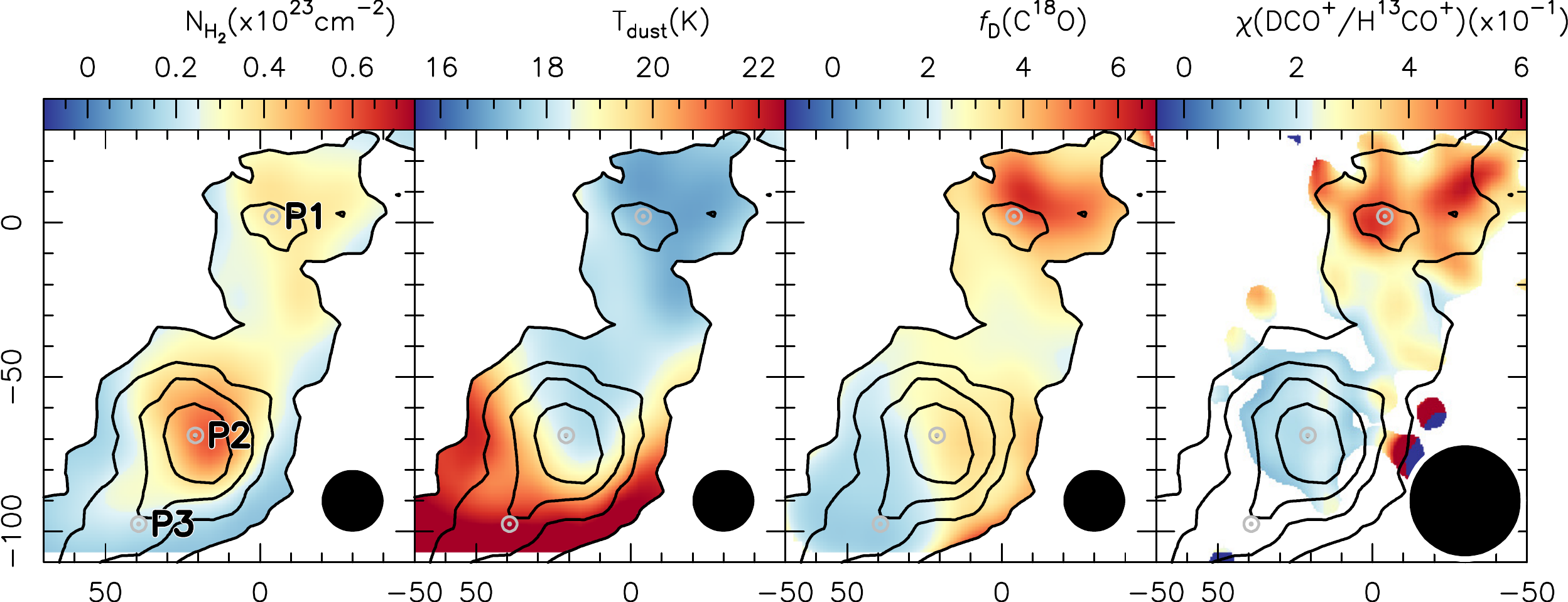}\\
\vspace{0.3cm}
G\,11.38+0.81  &
&\includegraphics[align=c,width=16cm] {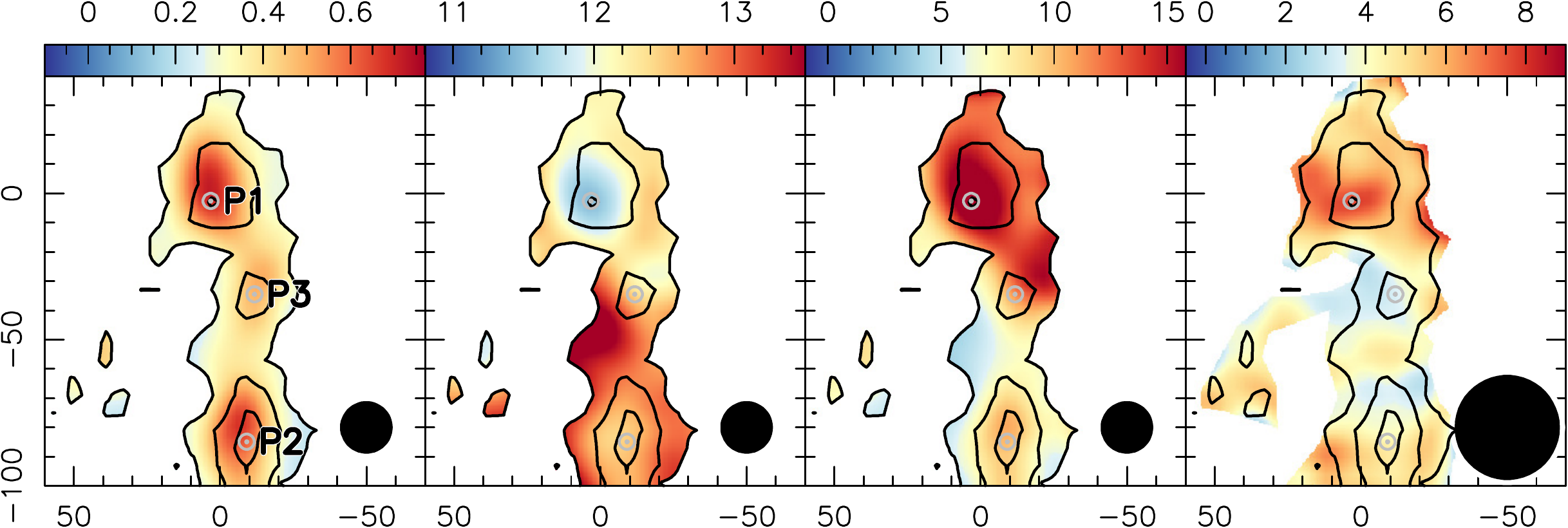}\\
\vspace{0.3cm}
G\,14.49-0.13 &
&\includegraphics[align=c,width=16cm] {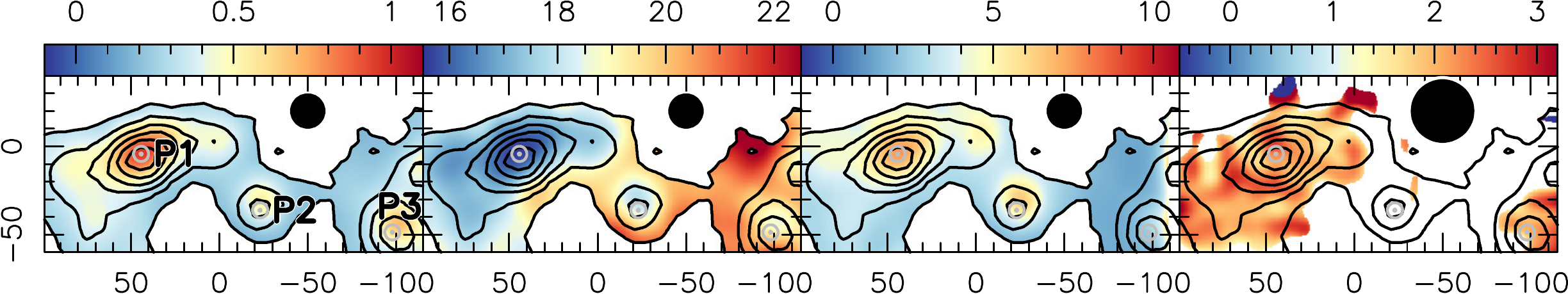}\\
\vspace{0.3cm}
G\,34.74-0.12 &
&\includegraphics[align=c,width=16cm] {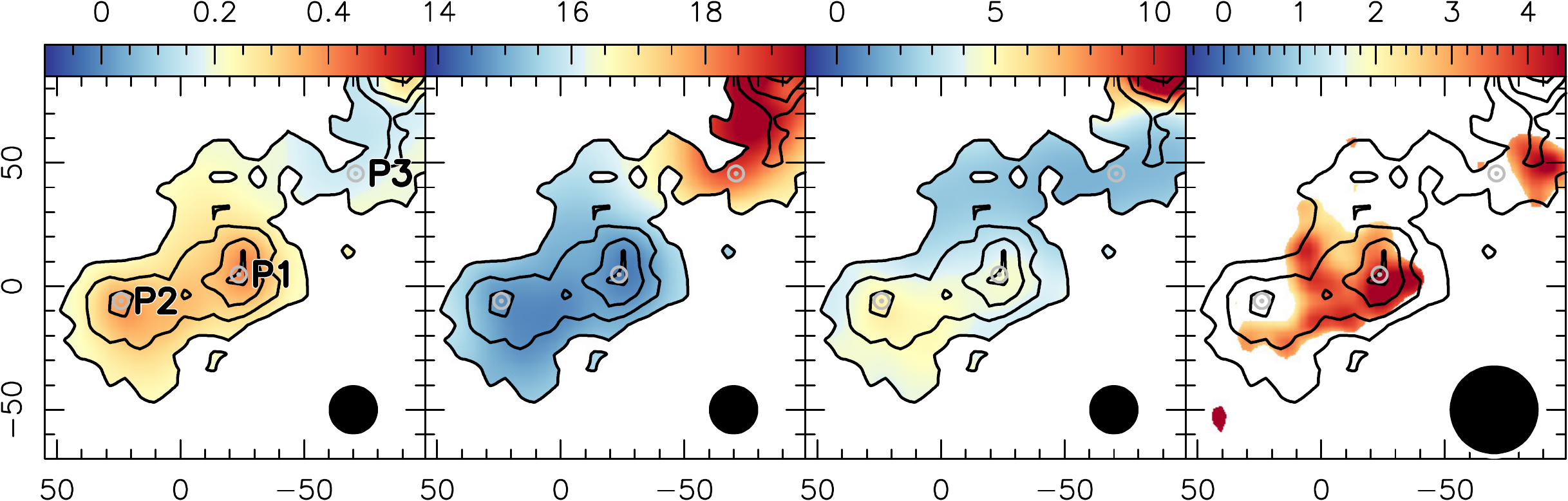}\\
&&\multicolumn{1}{c}{R.A. offset (\arcsec)}
\end{tabular}

\caption{Color maps of $\rm H_2$ column density ({\it the 1st column}) and dust temperature  from SED fitting ({\it the 2nd column}), $\rm C^{18}O$ depletion factor ({\it the 3rd column}), and relative abundance ratio between $\rm DCO^+$ and $\rm H^{13}CO^+$ ({\it the 4th column}) toward regions G\,15.22-0.43,    G\,11.38+0.81, G\,14.49-0.13, and G\,34.74-0.12.
 The black contours show continuum emission observed by {APEX} at 870\,$\mu$m \citep{schuller09}, with the contour levels   as in Figure~\ref{fig:soumap}.
The blanking threshold for the 1st to 3rd  panels is $\rm <3\sigma$ continuum emission at 870\,$\mu$m;  for the 4th panel, it is the pixels where $\rm DCO^+$\,(1-0) shows $\rm <3\sigma$ emission.
The angular resolution for each map is given in the top or bottom right corner by the black circles.
}\label{fig:paramap}
\end{figure*}

Assuming that the $\mathcal{R}_{\rm ^{12}C/^{13}C}$ isotopic ratio changes with $R_{\rm GC}$ \citep{giannetti14}, having no variation within each source, and that the 1--0 lines from $\rm H^{13}CO^+$ and $\rm DCO^+$ are optically thin \citep[e.g., ][]{sanhueza12,feng16b}, the D-fraction map of $\rm HCO^+$ toward each source can be derived from the relative abundance ratio map of $\rm DCO^+$ with respect to $\rm H^{13}CO^+$ as (Figure~\ref{fig:paramap}),  
\begin{equation}
D_{\rm HCO^+}=\frac{\chi\rm (DCO^+/H^{13}CO^+)}{\mathcal{R}_{\rm ^{12}C/^{13}C}}=\frac{\chi\rm (DCO^+/H^{13}CO^+)}{6.1{\it R}_{\rm GC}({\rm kpc}) + 14.3}.
\end{equation}

Using the three temperature sets $T_{\rm dust}$,  $T_{kin}$($p$-$\rm NH_3$), and $T_{kin}$($p$-$\rm H_2CO$) at an angular resolution of $\rm \sim35\arcsec$, we found that $D_{\rm HCO^+}$ toward the same location does not show much difference ($<10\%$, see Table~\ref{tab:gaspara}). Instead, each $D_{\rm HCO^+}$ map shows a trend, dropping from P1, the $\rm DCO^+$-dominant region ($\rm 1\%-2\%$), to P3 (where $\rm C^{18}O$  shows the  maximum abundance in G\,15.22-0.43  and G\,11.38+0.81) and P2 (where both $\rm C^{18}O$ and $\rm DCO^+$ are deficient in G\,34.74-0.12 and G\,14.49-0.13) by a factor of more than 2 (Table~\ref{tab:parafDD}).

\subsubsection{Error budget}\label{error}
Above, we discussed the  validity  of our assumptions to treat the $\rm H^{13}CO^+$\,(1--0), $\rm DCO^+$\,(1--0), and $\rm C^{18}O$\,(2--1) lines as optically thin and in LTE condition (Sect~\ref{gastemp}--\ref{dfrac}).
However, we are not able to test the validity of the other assumptions in our measurements, such as 
the unity beam-filling factor for the low-$J$ lines with extended emission, the constant conversion factor of gas-to-dust mass ratio $\gamma$, the expected gas-phase abundance of CO with respect to $\rm H_2$ (without depletion), the same depletion factor for CO and $\rm C^{18}O$, and the fractionation of $\mathcal{R}_{\rm ^{12}C/^{13}C}$  and $\mathcal{R}_{\rm ^{16}O/^{18}O}$. 
Nevertheless,  in the context of only the relative trend in the $f_D\rm (C^{18}O)$ map  and $D_{\rm HCO^+}$ maps toward the same cloud, these uncertainties are canceled out (see Table~\ref{tab:parafDD}).

\begin{table*}[htbp]
\caption{Relative ratios of $\rm C^{18}O$ depletion factor and  $\rm HCO^+$  D-fraction between locations
}\label{tab:parafDD}
\scalebox{1}{

\begin{tabular}{p{2cm}|p{3cm}p{3cm}|p{3cm}p{3cm}}
\hline\hline

                   &\multicolumn{2}{c|}{$f_D(\rm C^{18}O)$$^{b,c}$}                  &\multicolumn{2}{c}{$D$$\rm _{HCO^+}$$^{b,d,e}$}                                                                   
   \\
   \hline
Zones$^a$      &P1 /P3$^f$  &P2  /P3$^f$  &P1 /P3$^f$ &P2  /P3$^f$ \\
 \hline                                                         
                                                                     
\multirow{1}{*} {G\,15.22-0.43} 
    &$\rm 2.7\pm 0.0 $ 		&$\rm 1.8\pm 0.0 $ 
    &$\rm >2.9$ 		&$\rm >1.0$ \\

\hline                                                                          
\multirow{1}{*} {G\,11.38+0.81} 
    &$\rm 1.4\pm 0.1 $ 		&$\rm 0.8\pm 0.0 $  
    &$\rm >2.2$ 		&$\rm >1.3$ \\

\hline                                                                          
\multirow{1}{*} {G\,14.49-0.13} 
     &$\rm 1.9\pm 0.1 $ 		&$\rm 1.3\pm 0.1 $  
     &$\rm 1.2\pm0.1$ 		&$\rm <0.3$ \\
\hline                                                                          
\multirow{1}{*} {G\,34.74-0.12} 
     &$\rm 1.8\pm 0.0 $ 		&$\rm 2.1\pm 0.0 $  
    &$\rm >1.4$ 		&-- \\

\hline \hline
\multicolumn{5}{l}{\color{black} { Note.} $a$. P1, P2, P3 denotes the $\rm DCO^+$-dominant,   transition, and the CO-dominant  zones, respectively. }\\
\multicolumn{5}{l}{\color{black} ~~~~~~~~~ $b$. Value derived by using $T_{\rm dust}$, at an angular resolution of 34\farcs7. }\\
\multicolumn{5}{l}{\color{black} ~~~~~~~~~ $c$. The $\rm C^{18}O$ depletion is derived by assuming the expected abundance with respect to $\rm H_2$  as Equs.~\ref{eq:nc18oex}--\ref{eq:dc18o}}\\
\multicolumn{5}{l}{~~~~~~~~~~~~% $\rm  9.50\times10^{-5} {\it exp}(1.11-0.13{\it D}_{GC,kpc})/ (58.80{\it D}_{GC,kpc}+37.10)$,      
and assuming a gas-to-dust mass ratio $log(\gamma)=\rm 0.087 {\it R_{\rm GC}} (kpc)+1.44$  \citep{draine11,giannetti17}.}\\
\multicolumn{5}{l}{\color{black} ~~~~~~~~~$d$. The D-fraction is derived from the  $\rm DCO^+$\,(1--0) and $\rm H^{13}CO^+$\,(1--0) lines by assuming that they are optically thin,}\\
\multicolumn{5}{l}{~~~~~~~~~~~~   have the same beam filling toward each pixel, and have a fraction of $\mathcal{R}_{\rm ^{12}C/^{13}C}\sim  6.1{\it R}_{\rm GC}({\rm kpc})+ 14.3 $ \citep{giannetti14}.}\\
\multicolumn{5}{l}{\color{black} ~~~~~~~~~ $e$. A lower or upper limit is given when the detected $\rm DCO^+$\,(1--0) shows $\rm <3\sigma$ emission toward P3 or P2.}\\
\multicolumn{5}{l}{\color{black} ~~~~~~~~~ $f$. Values are given as the relative ratio between two locations. }\\

\end{tabular}

}
\end{table*}

\section{Discussion}\label{discussion}
In the sources of our pilot sample, the $f_D\rm (C^{18}O)$ is high ($\rm >3$) towards the $\rm DCO^+$-dominant zone with high  optical extinction \text{Av} ($\rm >20$\,mag, see \citealp{guever09}). The $N_{\rm H_2}$ toward this zone is denser than that toward the CO-dominant zone ($\text{Av}\sim$10--15\,mag) by a factor of 2. Similar to the findings in previous studies \citep[e.g., ][]{pagani05}, the difference in self-shielding of CO may not be responsible for the trend in the $f_D\rm (C^{18}O)$ towards the same natal cloud.  Therefore, it is worth investigating whether the variation in gas number density and/or the source temperature leads to  such an $f_D\rm (C^{18}O)$ trend, and whether the chemical relation between the CO depletion and D-fraction of $\rm HCO^+$ can give a constraint on the chemical age of our sources.

\subsection{Comparison with previous works}
Our sources are selected at different kinematic distances, i.e., progressively further away from the Sun by 1\, kpc.
Comparing the absolute value of $f_D\rm (C^{18}O)$  toward the pilot sample at the same angular resolution (16\arcsec or smoothing to 35\arcsec, corresponding to 0.2--1\,pc), 
{we find that the maxima  of $f_D\rm (C^{18}O)$  toward the pilot sample regions are similar.} In general, they appear as  4--6 at the locations with $T_{\rm dust}$ in the range of 14--18\,K (Figure~\ref{fig:paramap}). The  exception  is G\,11.38+0.81, where the maximum  of $f_D\rm (C^{18}O)$  is higher  (up to 15) than the rest of the sources towards the region with colder $T_{\rm dust}$ (12\,K).

Compared to previous studies, the absolute values of $f_D\rm (C^{18}O)$ in our regions are generally consistent with those toward  low-mass clouds \citep[e.g., ][]{bacmann03,ceccarelli07,christie12} and high-mass clumps  \citep[e.g., ][]{hernandez11,rygl13, liut13,sabatini19}.
Moreover, the high value toward G\,11.38+0.81 is consistent with those found  at a comparable linear resolution from large sample studies of high-mass clumps in \citet[][{in a range of 50--80, being two to three times larger because of the use of a different dust opacity}]{fontani12} and \citet[][up to 20 toward the cold and young clumps]{giannetti14}, where $\rm \gamma$ were adopted as 100. 
A similar case of higher $f_D\rm (C^{18}O)$ is also found towards G\,35.39-0.33 at a linear resolution of 0.2\,pc, where $f_D\rm (C^{18}O)$ is up to 4 in a region with $n_{\rm H}\sim\rm 10^3\,cm^{-3}$ \citep{hernandez11}, and up to 12  in  regions with $n_{\rm H}\sim\rm 10^4\,cm^{-3}$ \citep{jimenez14b}.

We also note that  $f_D\rm (C^{18}O)$ measured in our regions at a linear scale of $\rm >0.1$\,pc is smaller than that measured at 0.01\,pc scale. This is consistent with   $f_D\rm (C^{18}O)$  found towards our pilot source, G\,28.34+0.06 ($R_{\rm GC}\rm \sim$4.6\,kpc), $\rm \sim5$ at a linear resolution of 0.8\,pc \citep{feng19a}\footnote{The $f_D\rm (C^{18}O)$ is measured  up to 10 by adopting $\rm \gamma\sim150$ in \citep{feng19a}, and corrected  up to 5 by adopting $\rm \gamma\sim70$ at $R_{\rm GC}\rm \sim$4.6\,kpc.}, while it is $\rm 10^2-10^3$ at a linear resolution of 0.01\,pc \citep{zhang09,urquhart18}. The gas number density at different scales, as well as beam dilution for relatively compact $\rm C^{18}O$ emission, could be  reasons for the different  magnitudes in measuring $f_D\rm (C^{18}O)$.

Although we give the absolute values of $f_D\rm (C^{18}O)$ in Figure~\ref{fig:paramap}, in the following, we focus on the relative trends observed from the  70\,$\mu$m dark region  (P1) to the 70\,$\mu$m bright region  (P3) for  the uncertainty of the gas and dust conversion constants used in our analysis (see Sect~\ref{error}).
The depletion factor $f_D\rm (C^{18}O)$ decreases toward individual sources from P1 to P3 by a factor of 2--4, behaving the same as those found from the less evolved to the more evolved high- and low-mass clumps \citep[e.g., ][]{christie12,fontani12,giannetti14}.

The projected distance from the depletion maximum (P1) to the minimum (P3) in our regions is in the range of 0.5--2\,pc. This is comparable to, or at most twice, the width of each filament (0.5--1parsec), obtained from the size of the  870\,$\mu$m continuum contour with $\rm S/N<5$. This feature is also found in a nearby high-mass region, G\,351.77-0.51 ($R_{\rm GC}\rm<1$\,kpc), where \citet{sabatini19} suggested that a depletion radius (0.02--0.15\,pc) is comparable to the filament width (0.1\,pc).

\subsection{Possible spatial correlation between the  dust and gas properties}

\begin{figure*}
      \begin{tabular}{p{4.5cm}p{4.5cm}p{4.5cm}p{4.5cm}}
G\,15.22-0.43\\
\includegraphics[align=c,height=4.cm] {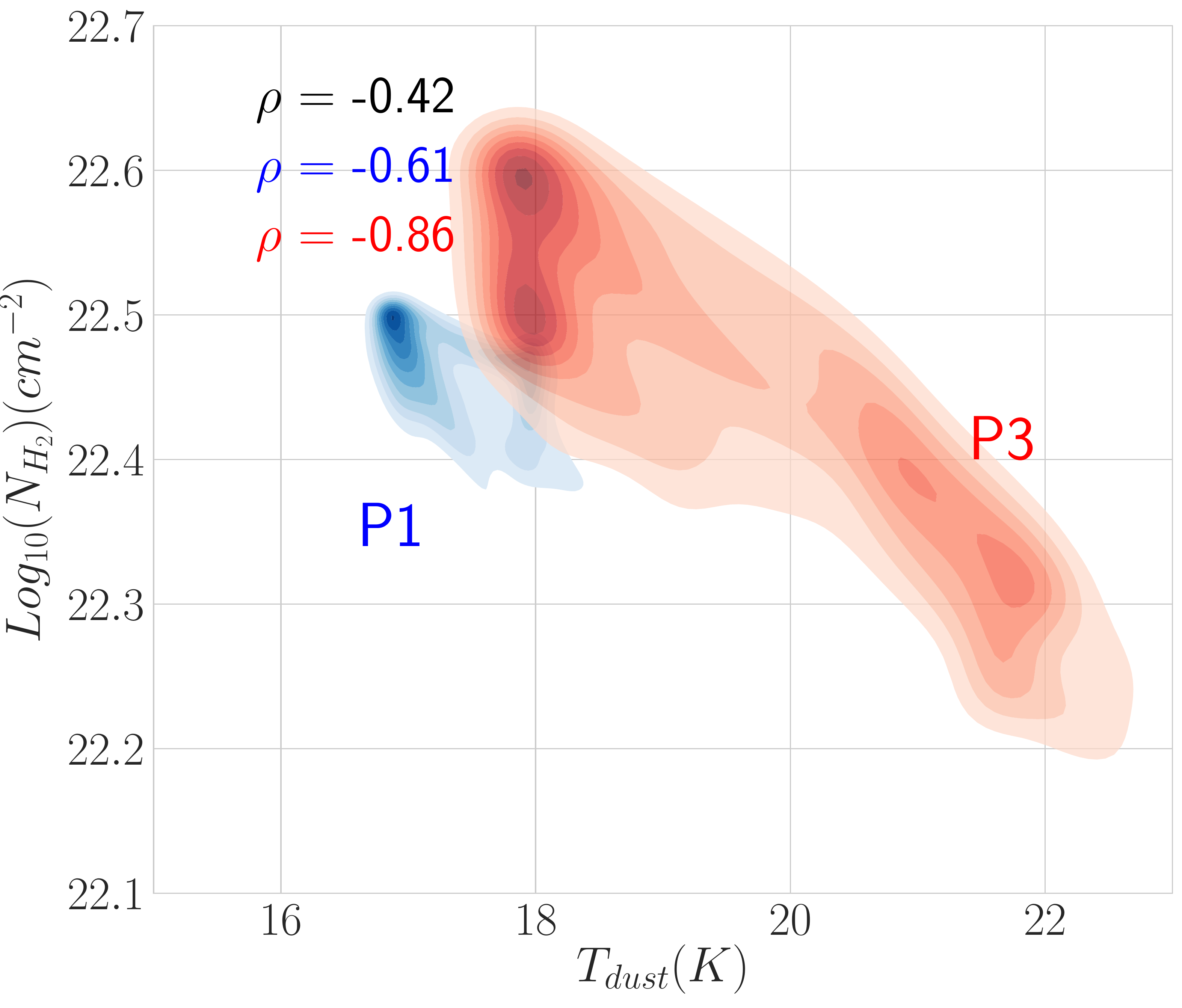} %{G15_2169-NH2-fDC18O-density-group.pdf}
&\includegraphics[align=c,height=4.cm] {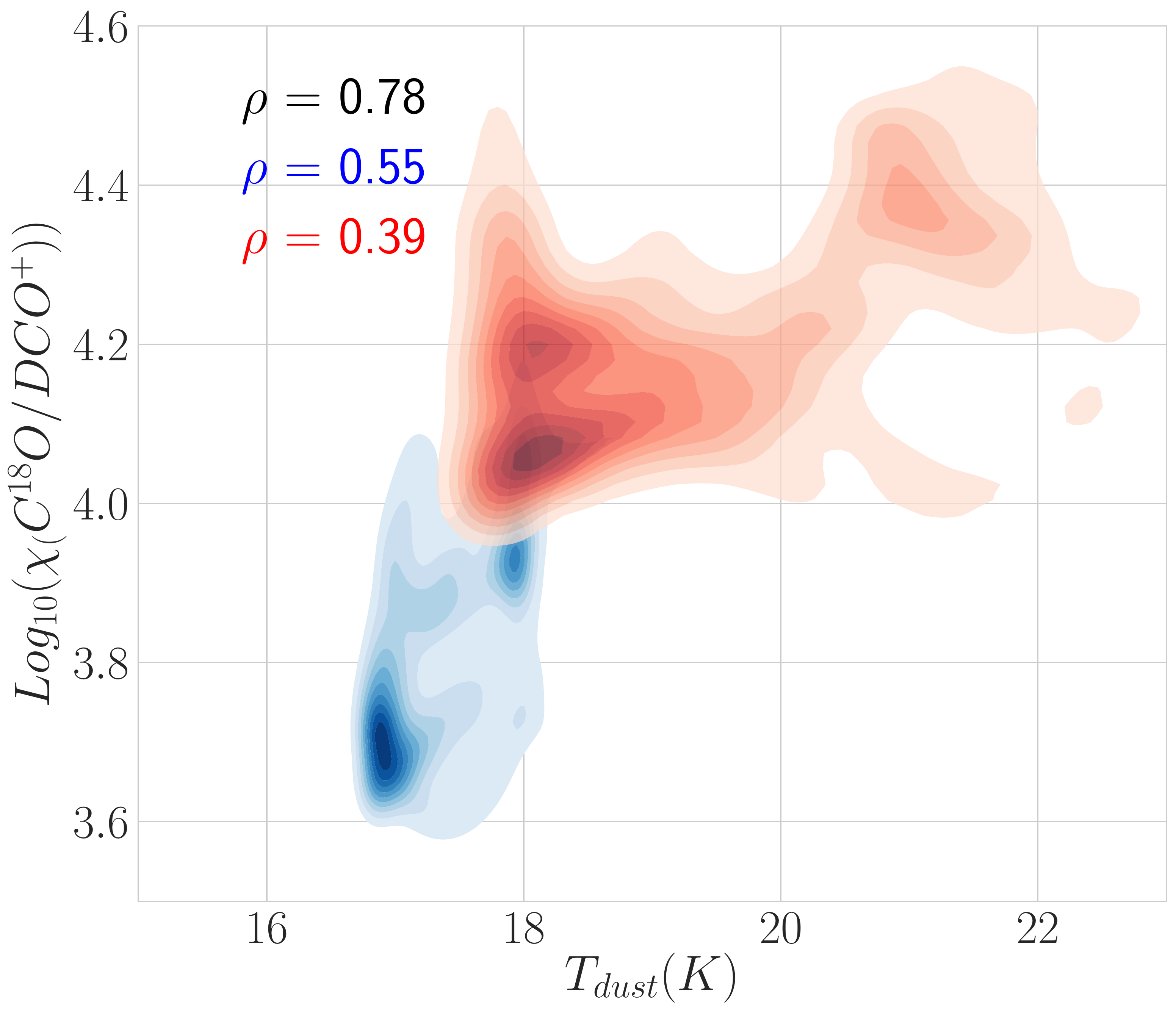}
&\includegraphics[align=c,height=4.cm] {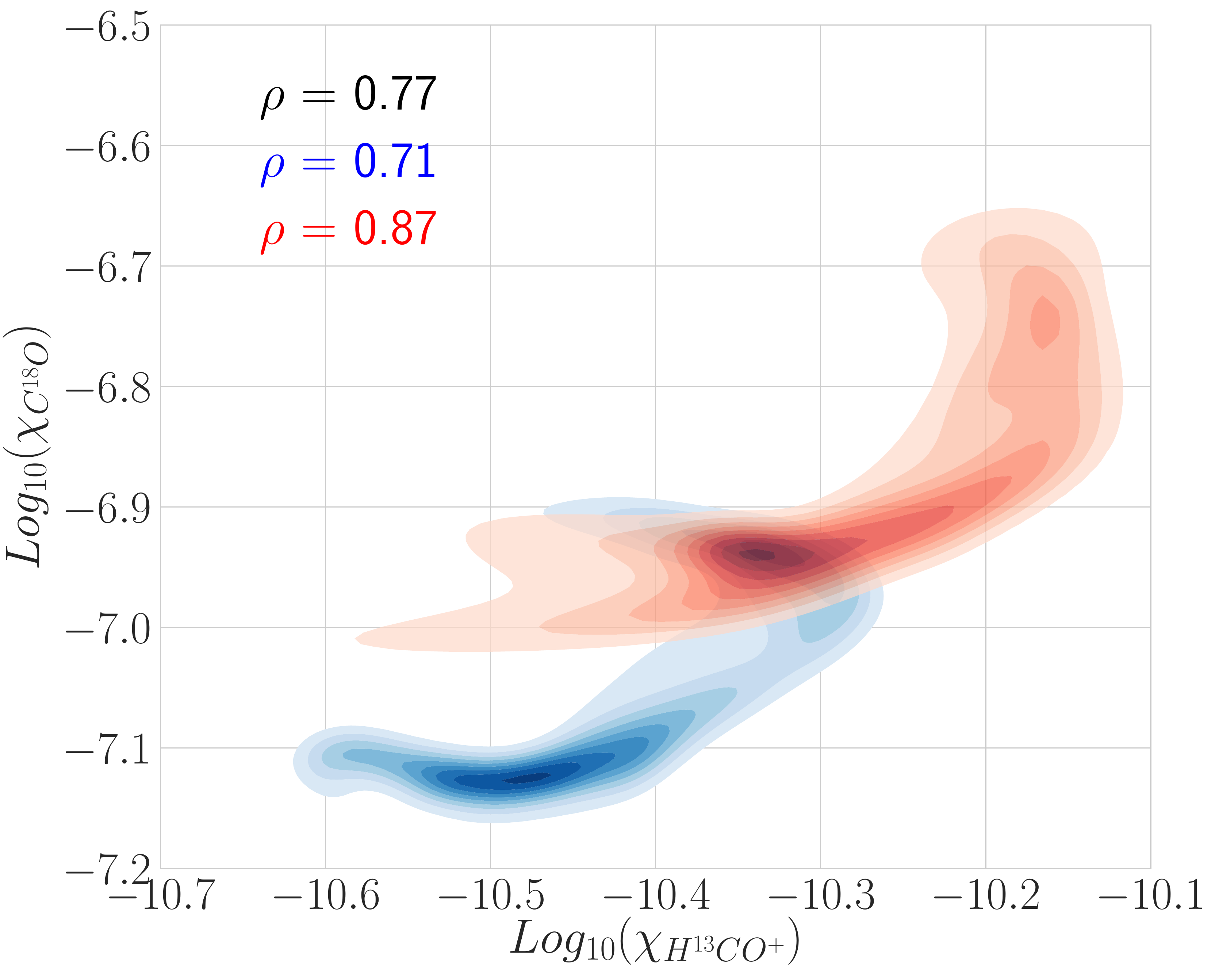}
&\includegraphics[align=c,height=4.cm] {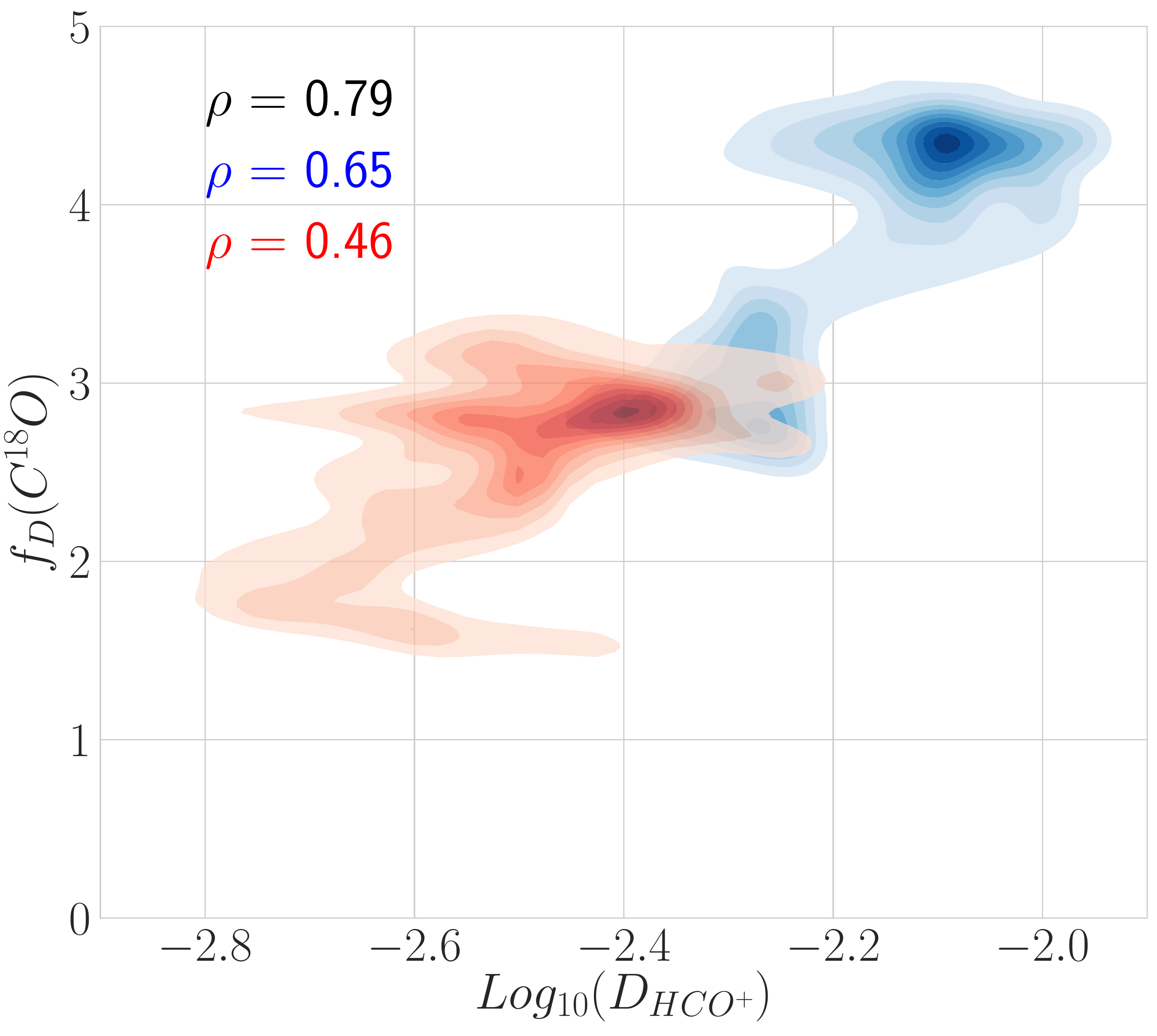}
\\[-0.2cm]
G\,11.38+0.81\\
\includegraphics[align=c,height=4.cm]  {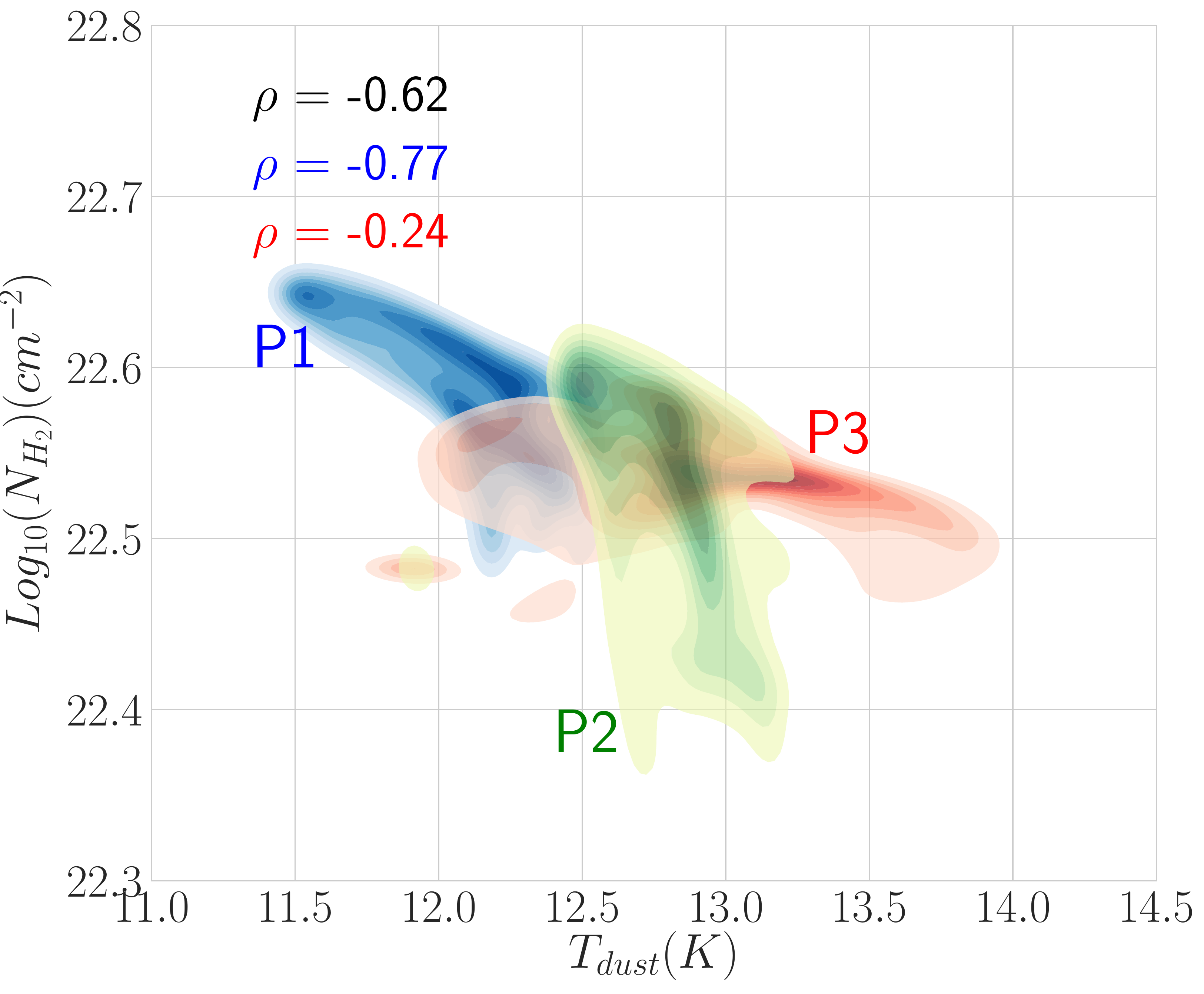} %{G11_3811-NH2-fDC18O-density-group.pdf}
&\includegraphics[align=c,height=4.cm] {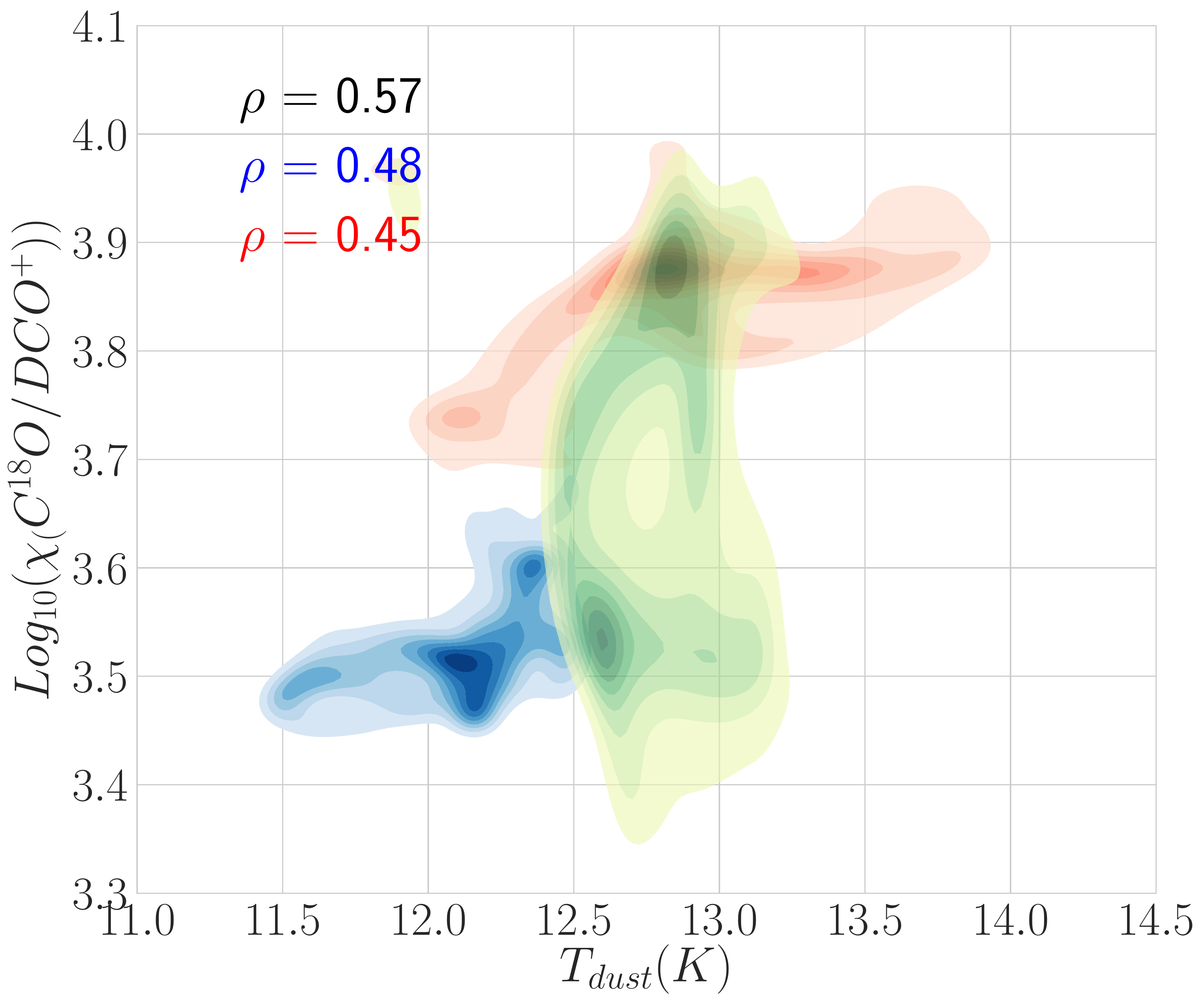}
&\includegraphics[align=c,height=4.cm] {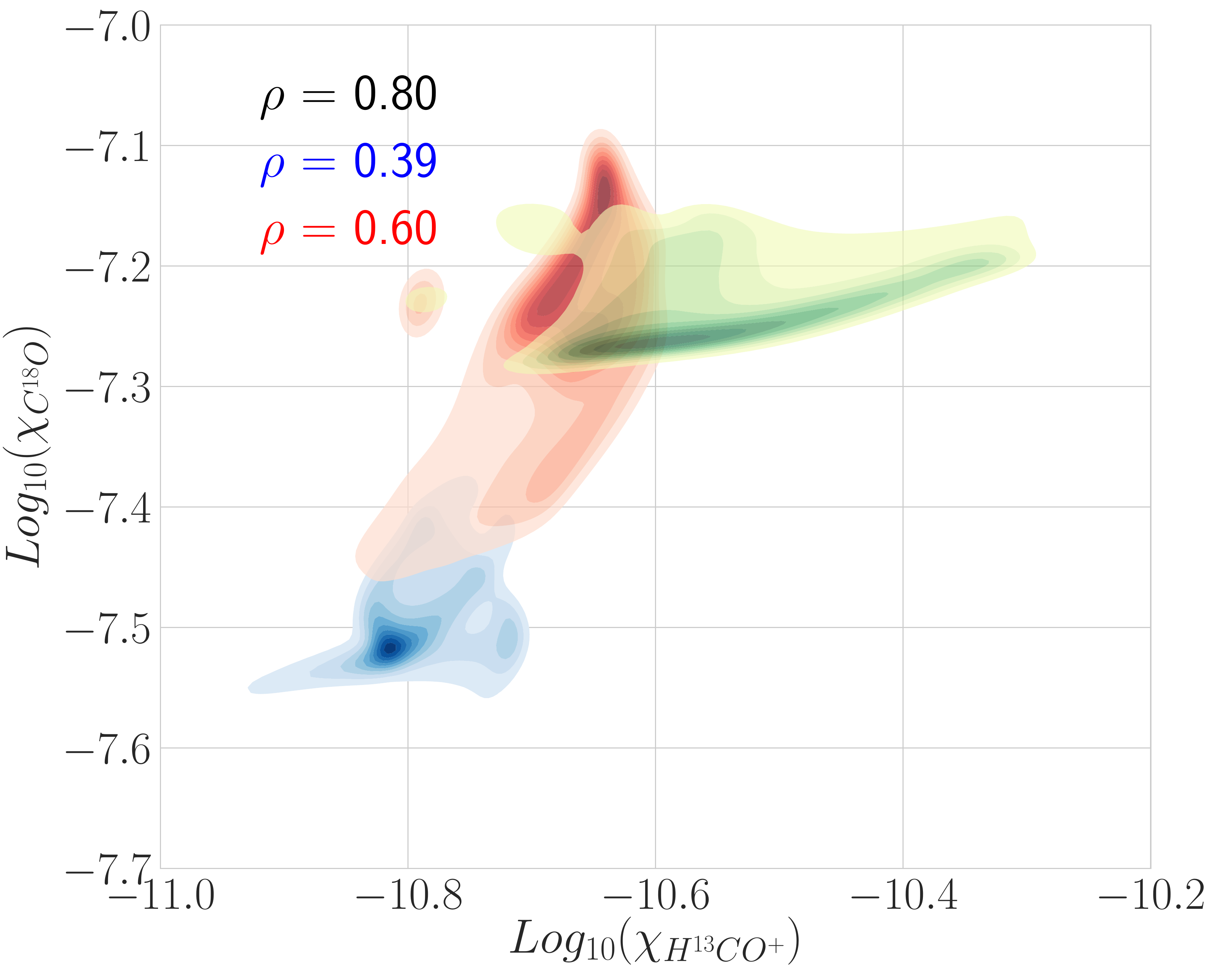}
&\includegraphics[align=c,height=4.cm] {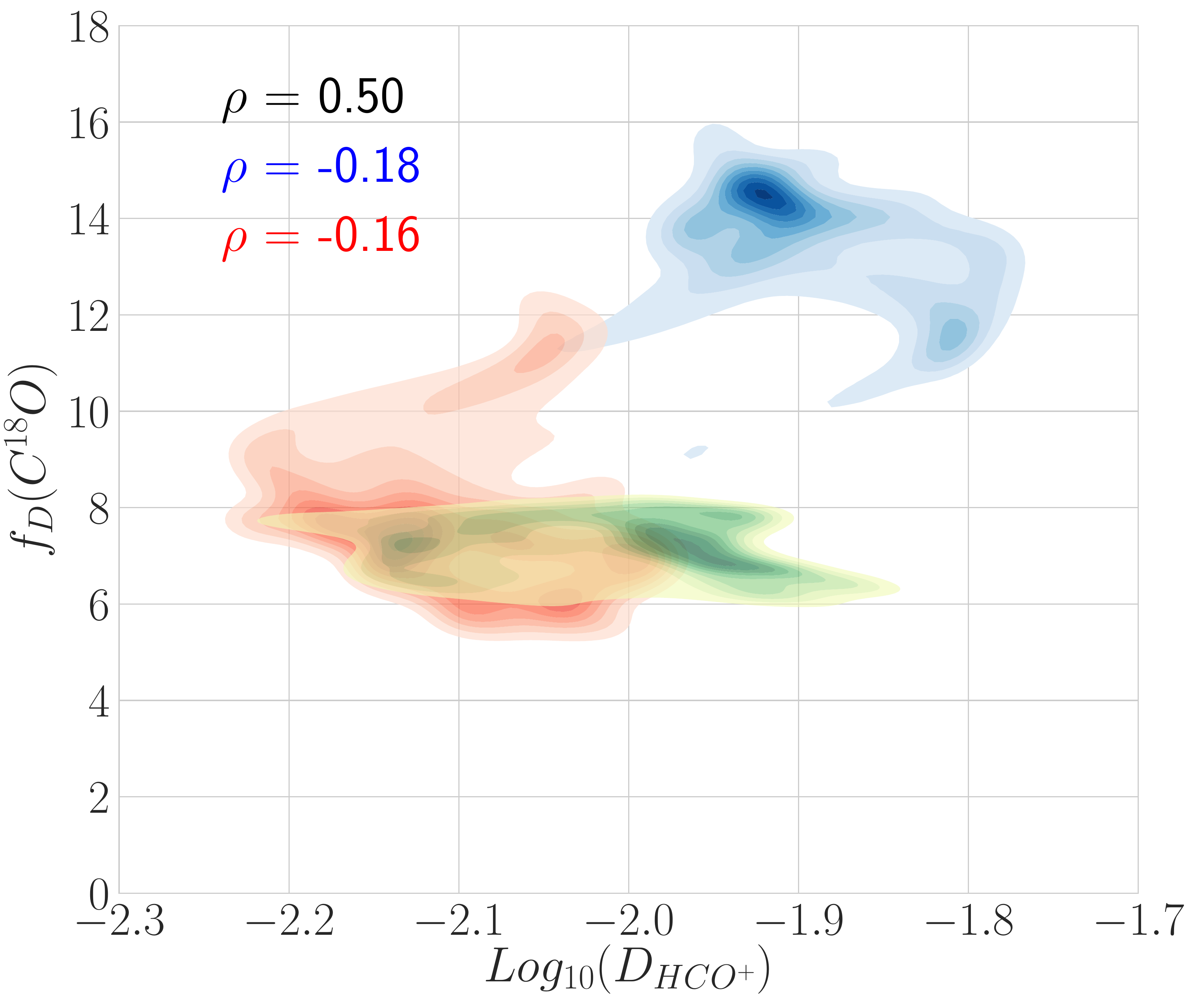}
\\[-0.2cm]
G\,14.49-0.13\\
\includegraphics[align=c,height=4.cm]  {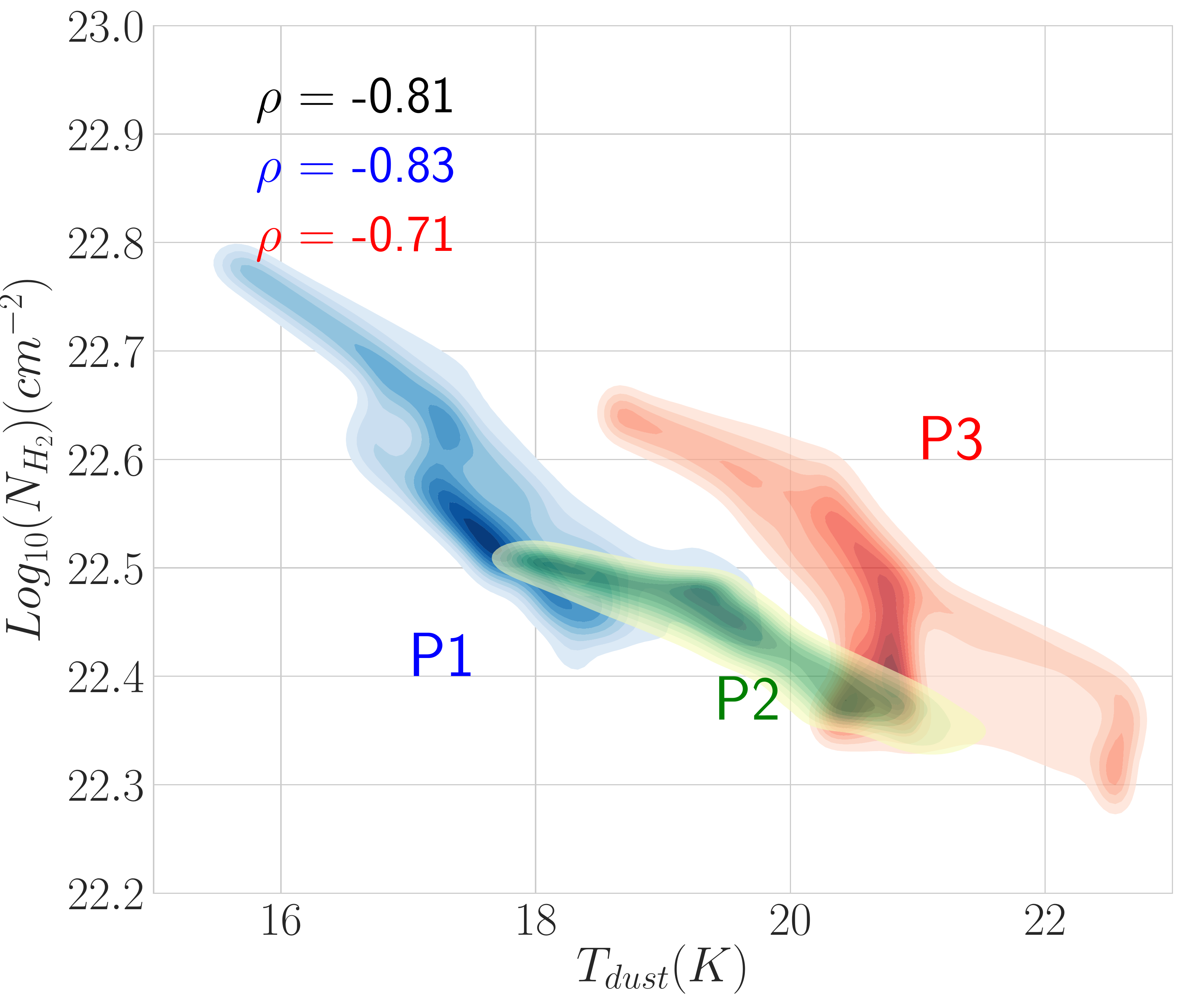} %{G14_4876-NH2-fDC18O-density-group.pdf}
&\includegraphics[align=c,height=4.cm] {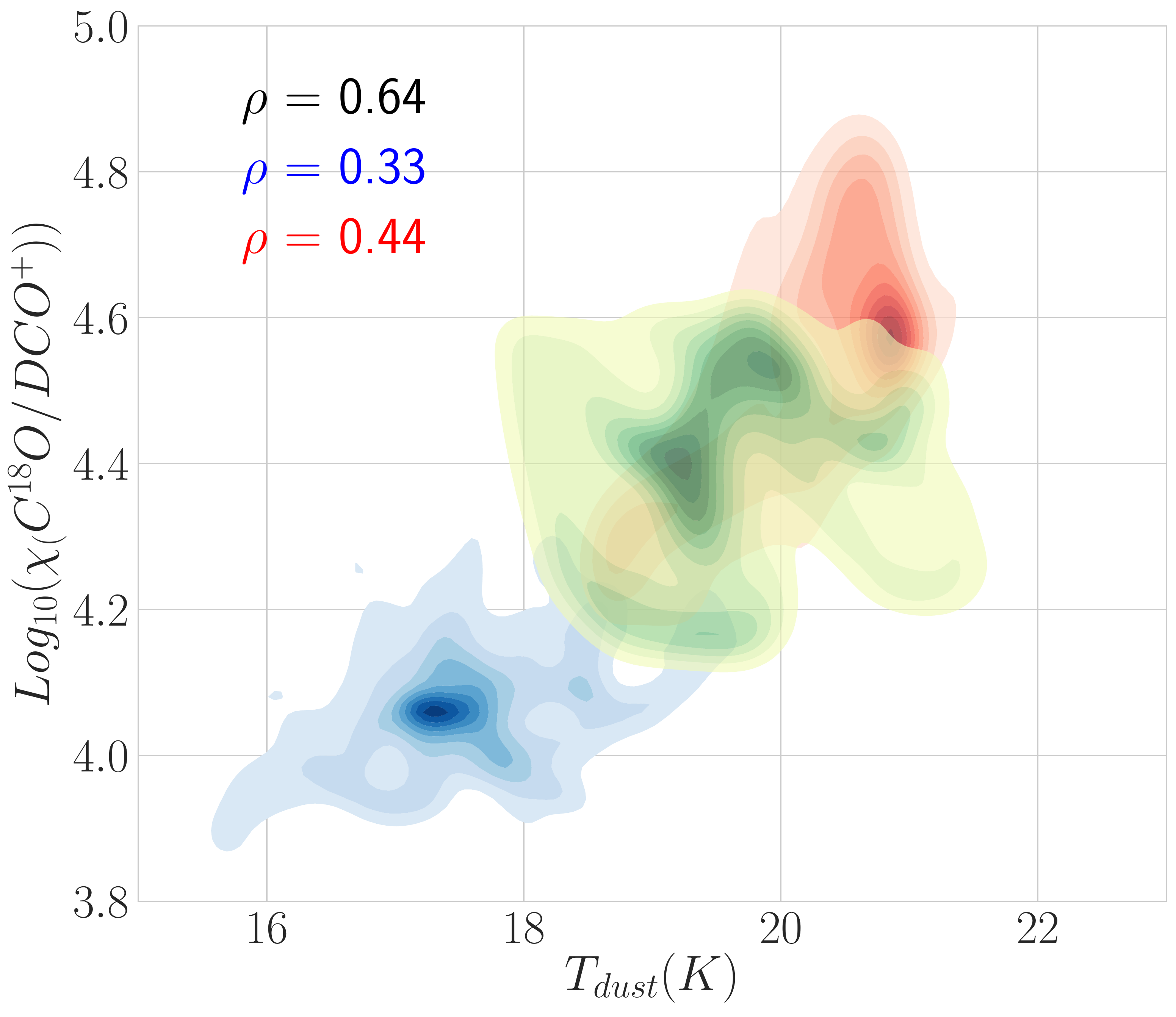}
&\includegraphics[align=c,height=4.cm] {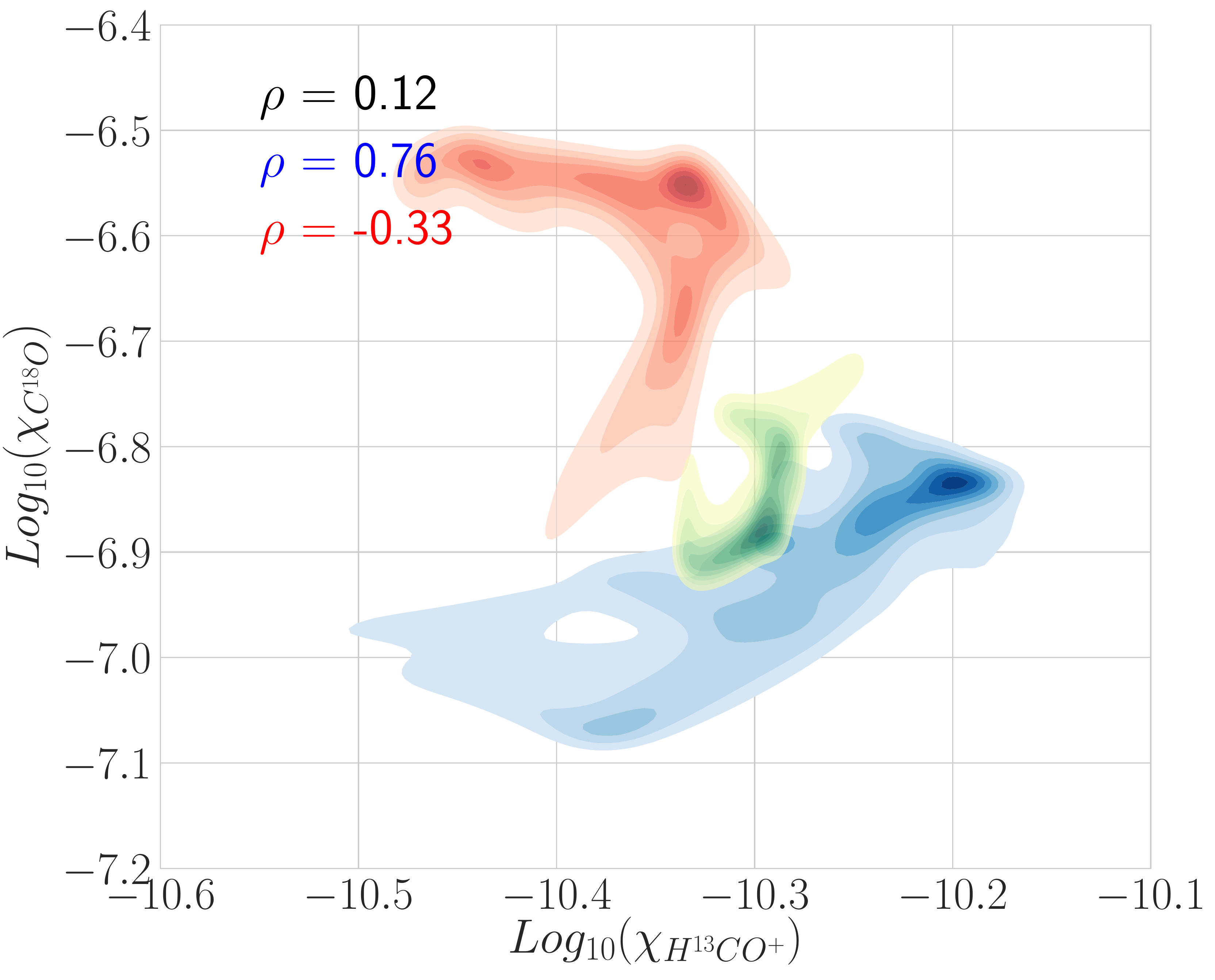}
&\includegraphics[align=c,height=4.cm] {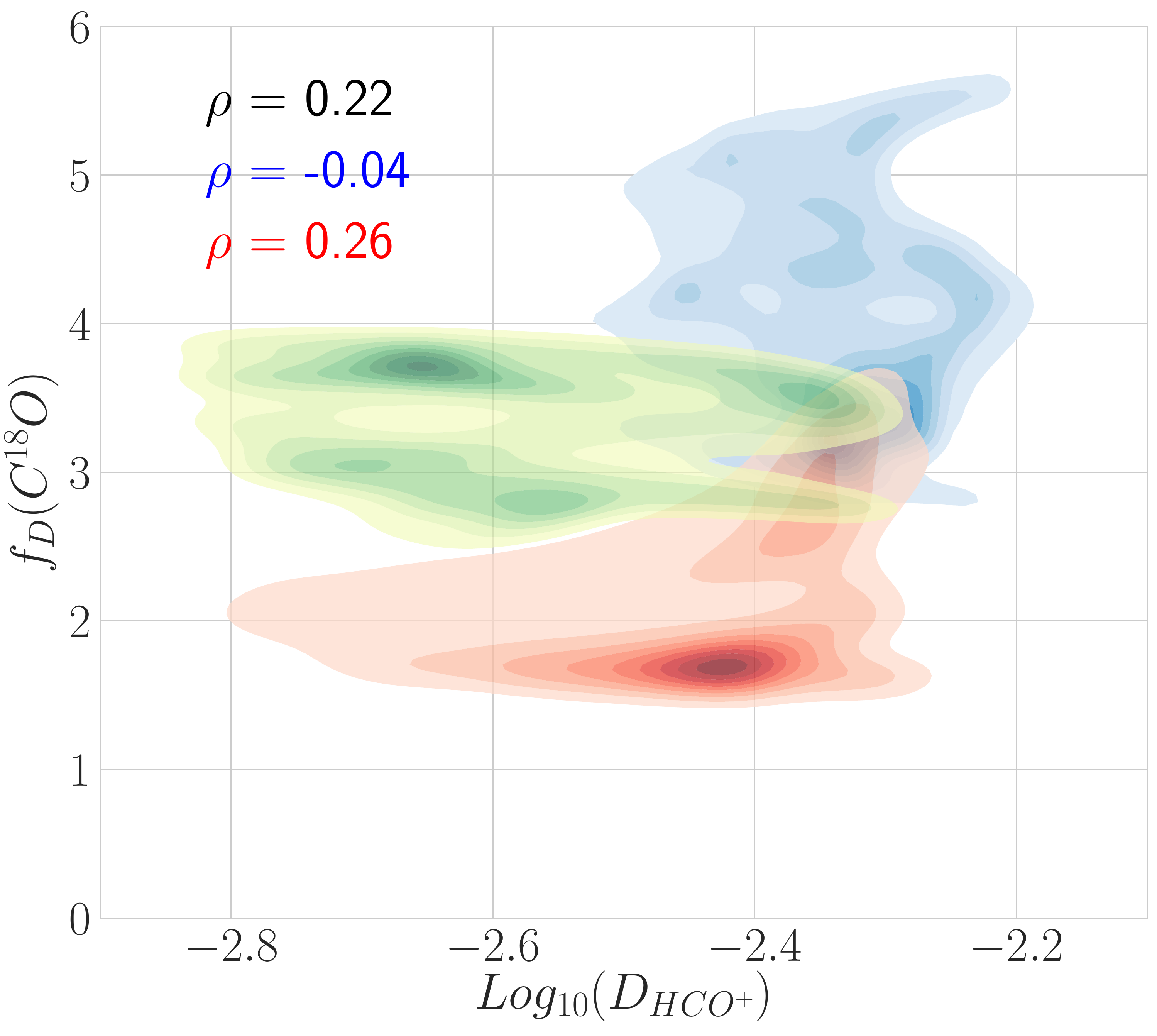}
\\[-0.2cm]
  G\,34.74-0.12\\
\includegraphics[align=c,height=4.cm]  {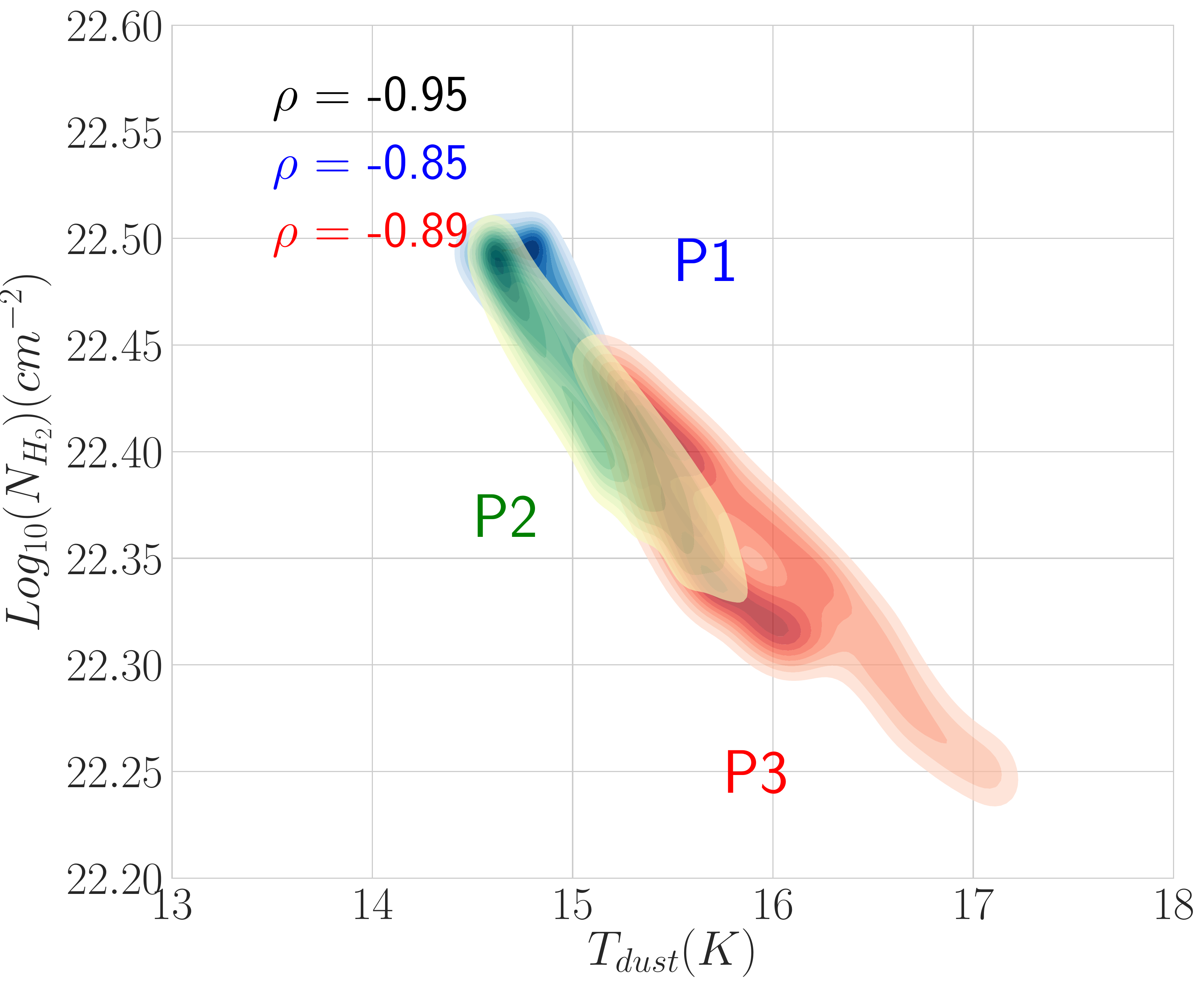} %{G34_7391-NH2-fDC18O-density-group.pdf}
&\includegraphics[align=c,height=4.cm] {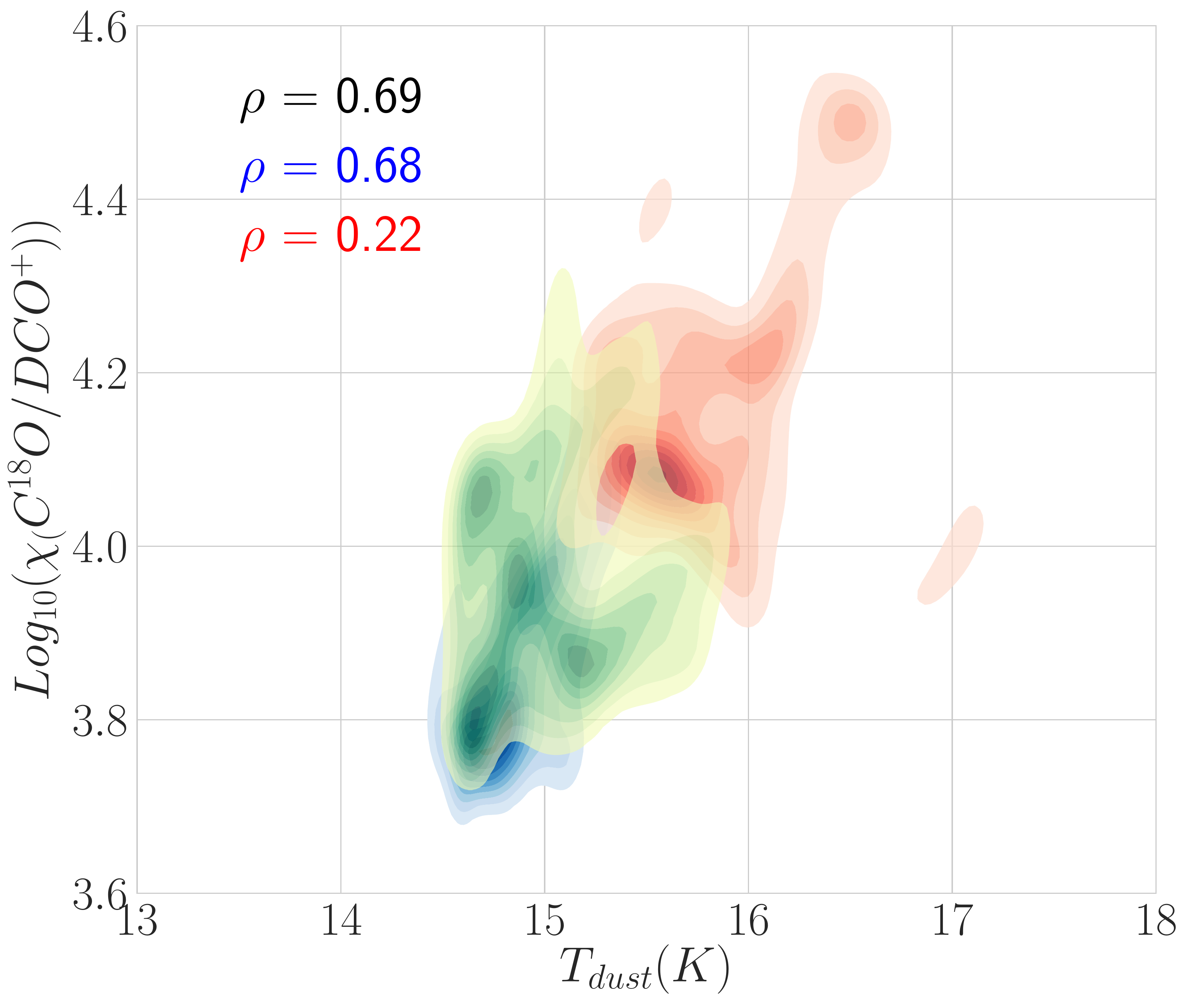}
&\includegraphics[align=c,height=4.cm] {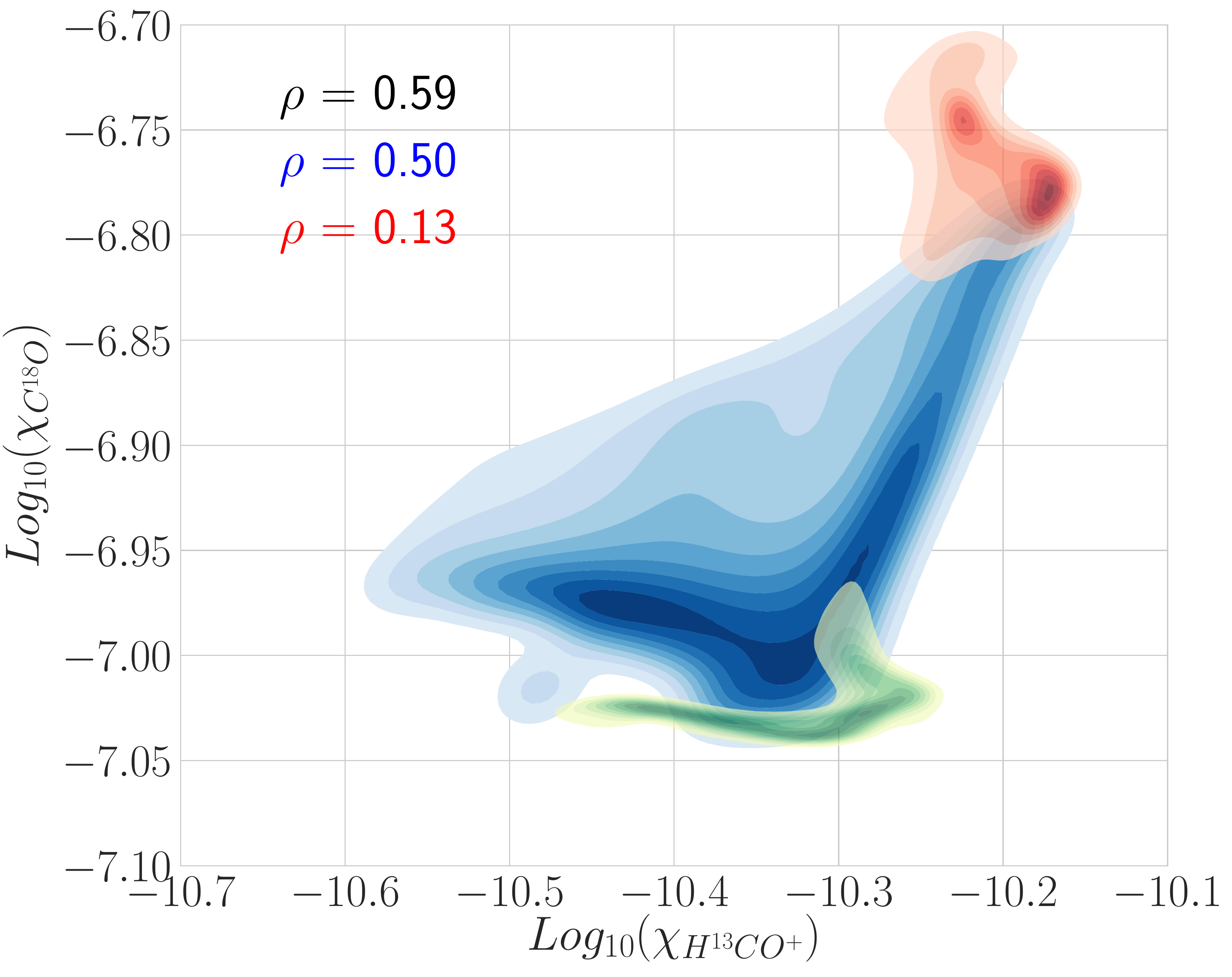}
&\includegraphics[align=c,height=4.cm] {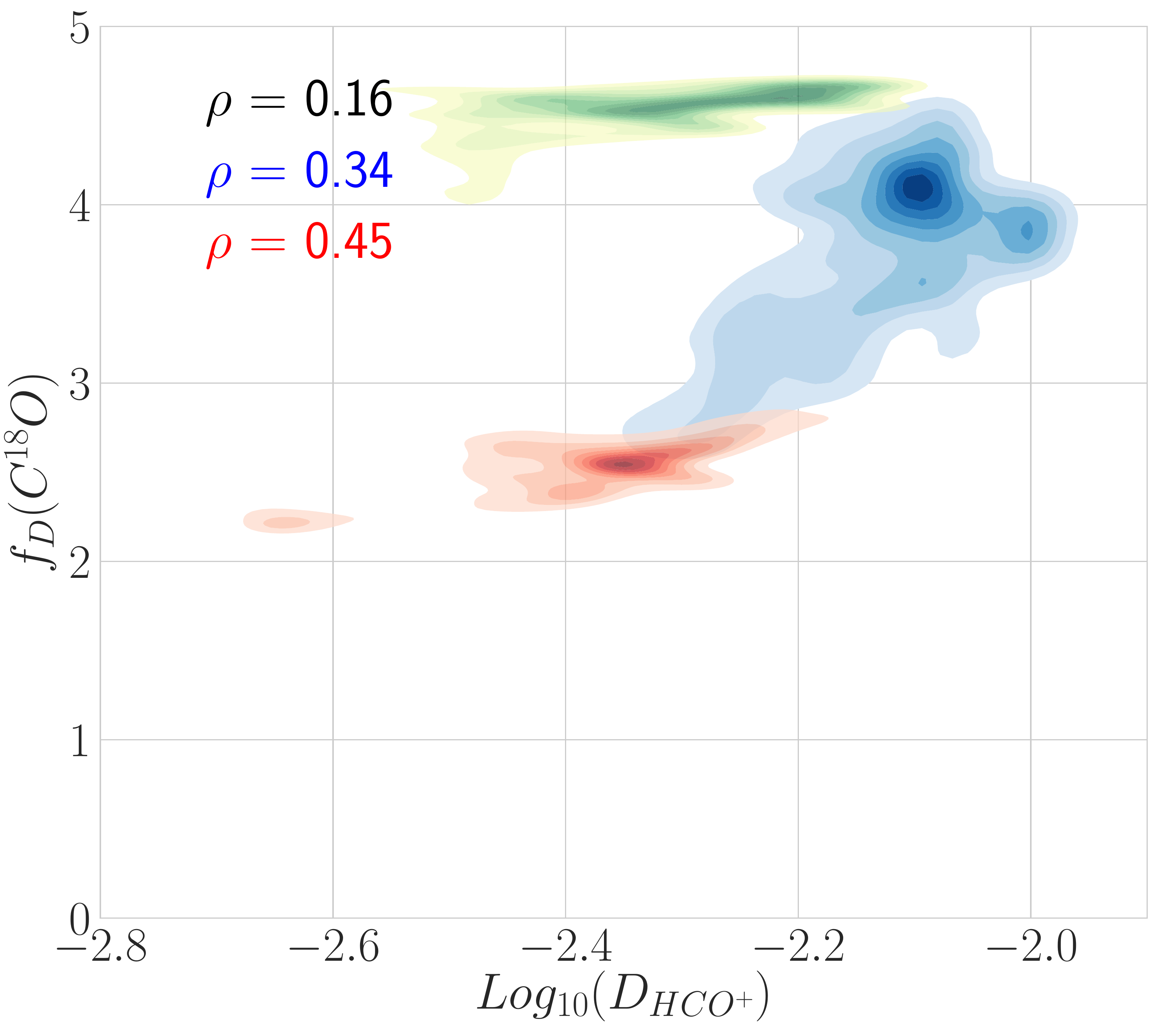}
\\[-0.2cm]

\end{tabular}
\caption{Possible correlation between variables. Values are extracted from pixels after smoothing the parameter maps to the same angular resolution (36\arcsec), and  plotted with a bivariate Gaussian kernel density estimate as contours. {\color{black}The CO- and $\rm DCO^+$-dominant zones are plotted  in red and blue, with the Spearman's rank correlation  coefficient $\rho$ given in red and blue, respectively.  The transition zone is plotted in yellowish-green. The  Spearman's rank correlation correlation coefficient $\rho$ of the entire region is given in black. } The pixels where continuum  at 870\,$\mu$m shows $\rm <5\sigma$ emission or  $\rm DCO^+$\,(1--0)  shows $\rm <3\sigma$ emissions are blanked.
}\label{fig:correlation1}
\end{figure*}

In our observations, the dust and gas appear to be thermally coupled ($T_{\rm dust}$ is close to $T_{kin}$($p$-$\rm NH_3$) towards individual pixels), and $T_{\rm dust}$ does not significantly change with angular resolution from 20\arcsec to 36\arcsec.  Five parameters derived from dust and gas emissions--$T_{\rm dust}$, the $\rm H_2$ column density ($N_{\rm H_2}$), the gaseous column densities of $\rm DCO^+$ ($N_{\rm DCO^+}$), $\rm C^{18}O$ ($N_{\rm C^{18}O}^{o}$), and $\rm H^{13}CO^+$  ($N_{\rm H^{13}CO^+}$)-- show variations as a function of location within each region. Smoothing these variable maps to the same angular resolution (36\arcsec), we extract their absolute values from each pixel and plot the bivariate Gaussian kernel density maps of several variable pairs (Figures~\ref{fig:correlation1} and ~\ref{fig:correlation2}). The red, blue, and yellowish-green areas represent the variables extracted from the CO-dominant, $\rm DCO^+$-dominant, and  transition zones, respectively. Moreover, to understand whether each pair of variables is correlated or not, we measure their Spearman's rank correlation\footnote{Spearman's rank correlation coefficient  $\rho$ is a nonparametric measure of statistical dependence between two variables \citep{cohen88}. 
This coefficient can assess how well a monotonic function (no matter whether linear or not) can describe the relationship between two variables. The coefficient  $\rho$ is in the range from -1 (decreasing monotonic relation) to 1 (increasing monotonic relation), with zero indicating no correlation.
}
coefficient $\rho$ \citep{cohen88} toward different zones, as well as toward the entire mapping region.
In the following discussion, we define the relationship between two variables as a ``strong correlation" when $\rm |\rho|\ge0.5$, a ``moderate correlation" when $\rm 0.5>|\rho|\ge0.3$, a ``weak correlation" when $\rm 0.3>|\rho|\ge0.1$, and  ``no correlation" when $\rm |\rho|<0.1$.

From Figures~\ref{fig:correlation1} and \ref{fig:correlation2}, we find the following.

\begin{itemize}

\item The $N_{\rm H_2}$ and  $T_{\rm dust}$ are strongly anticorrelated ($\rho<-0.5$). In general, the $\rm DCO^+$-dominant zone (P1) in each region is 3--6\,K colder and two to three times higher than the neighboring CO-dominant zone (P3).  Although a more robust fit is required to be applied to the full sample of regions, the linear proportion index between  the logarithm of $N_{\rm H_2}$ and the  $T_{\rm dust}$ of all four sources appears similar (will be discussed in Sect~\ref{model}), so this pair of variables seems to be dependent.

\item   The abundance of gaseous $\rm C^{18}O$  is strongly correlated with $T_{\rm dust}$ ($\rho>0.5$), and the $D\rm_{HCO^+}$ shows a strong or a moderate  anticorrelation with $T_{\rm dust}$ ($\rho<-0.4$; Figure~\ref{fig:correlation2}).
For all four regions, the colder gas toward P1 has  consistently lower values of the relative gaseous abundance ratio $\chi\rm (C^{18}O/DCO^+)$ than the warmer gas towards P3, showing a robust trend of increasing  $\chi\rm (C^{18}O/DCO^+)$  with the evolutionary stage of the star-forming clump. 
This is consistent with chemical model predictions (see Sect.~\ref{model}), where higher temperatures enhance the $\rm C^{18}O$ abundance  (lower depletion) and suppress the deuteration of other species \citep[e.g., ][]{roberts00,caselli08}.

\item The abundances of gaseous $\rm H^{13}CO^+$ and $\rm C^{18}O$ are strongly correlated  ($\rho>0.5$), except for G\,14.49-0.13. 
Apart from G\,14.49-0.13 , denser gas traced by a higher abundance  of $\rm H^{13}CO^+$  ($\rm 6\times10^{-11}$) towards each region shows a  relatively higher abundance of gas-phase  $\rm C^{18}O$ ($\rm 1.5\times10^{-7}$) than the rest by a factor of more than 3. 
 As for G\,14.49-0.13, gaseous  $\rm C^{18}O$ and $\rm H^{13}CO^+$ show a strong correlation only towards the $\rm DCO^+$-dominant zone, while the maximum  abundance of gaseous $\rm C^{18}O$ is not spatially coincident with the $\rm H^{13}CO^+$. On the one hand, this could be an apparent effect, due to the fact that the $\rm H^{13}CO^+$\,(1--0) line with high critical density is more  efficiently excited (i.e., showing stronger emission) in the dense regions where CO is frozen out.
On the other hand,  several protostellar cores with outflows were detected toward G\,14.49-0.13 at a linear resolution of 0.01\,pc \citep{lis19,sanhueza19}. 
Therefore, zones with an enhanced ionization fraction  in the vicinity of protostellar sources may show a larger abundance of  $\rm H^{13}CO^+$ \citep[see, e.g., ][]{ceccarelli14}.

\item The $f_D(\rm C^{18}O)$ and  $D\rm_{HCO^+}$ show a strong correlation  toward the entire region of G\,15.22-0.43 and G\,11.38+0.81 and a moderate correlation toward G\,14.49-0.13 and G\,34.74-0.12 when excluding the transition zone. 
The primary chemical process in the low-temperature ($\rm <$20\,K) $\rm DCO^+$-dominant zone, {after the onset of CO freeze-out, is the conversion of the remaining gaseous CO into $\rm DCO^+$ in reactions involving $\rm H_2D^+$ and $\rm D_2H^+$ \citep[e.g., ][]{watson76,gerlich02,caselli08,aikawa18}. }Therefore, with more CO depleted,  more $\rm {H_3}^+$ takes part in deuterium enrichment and increases the D-fraction of species, including $\rm HCO+$. 
This trend is also seen in, e.g., \citet[][]{caselli02b,tielens13,radaelli19}. 
In a warm protostellar environment, CO desorbs to the gas phase, producing the CO-dominant zone. Moreover, $\rm DCO^+$ is not efficiently formed; instead, {it is destroyed}  mainly through electron recombination.

\iffalse
In fact, three major chemical reactions are active \citep[e.g.,][]{caselli08}:
(1) $\rm HCO^+$ is mainly formed from gaseous CO through the reaction:
\begin{equation}
\rm {H_3}^+ + CO\rightarrow H_2+ HCO^+, \label{equ:co2hcop}
\end{equation}

(2) In the low-temperature ($\rm <$20\,K) environment where a large amount of CO is depleted, $\rm {H_3}^+$ is no longer destroyed efficiently by the gaseous CO (Equ.~\ref{equ:co2hcop}), so $\rm {H_3}^+$ takes part in deuterium enrichment via:
\begin{equation}
\rm {H_3}^+ +HD\leftrightarrow H_2D^++ H_2+232\,K,  \label{equ:h2dp}
\end{equation}
and further $\rm D_2H^+$,  except when a substantial fraction of $o$-$\rm H_2$ is present \citep[e.g., ][]{watson76,gerlich02,caselli08}.

(3) The enhancement of the $\rm {H_2D}^+$ and / or $\rm D_2H^+$  abundances drives the formation of deuterated species such as $\rm DCO^+$, by analogs of the reaction \citep[e.g., ][]{aikawa18}:
\begin{eqnarray}
\rm H_2D^++ CO\rightarrow H_2+ DCO^++250\,K.  \label{equ:dcop2}
\end{eqnarray}

\fi

\end{itemize}

\subsection{Chemical modeling}\label{model}

To understand the trends (at least qualitatively) seen in the observational data, {we first put together the observational data points of the four sources with the coordinates as ($T$, $\chi[{\rm CO}]$) and ($T$, $D[{\rm HCO^{+}}]$) 
(the colored bivariate Gaussian kernel density contours in Figure~\ref{fig:model}). {\color{black}Since the density contours from all sources  are overlapped  or well connected with the same slop in both plots, we assume that they have a similar nature.}
Then we aim to reproduce the correlations seen in the plots by running a set of models using a chemical code called \texttt{chempl}  \citep[][]{du20}. }
This chemical code is based on the ``three-phase'' description of interstellar chemistry, namely, species in the model can be in the gas phase, on the dust grain surface, and in the dust grain mantle.
The chemical network is based on the UMIST 2012 database \citep{mcelroy13}, ``deuterated'' by adding deuterium to the network \citep{roberts03,roberts04}, and augmented by adding grain surface reactions from \citet{hasegawa92} and recent experimental results.  In total, 35,457 reactions are included in the calculation.

{
Our models are ``pseudo-time-dependent'', in the sense that the physical conditions are kept constant; namely, parameters such as temperature and density do not change with time \citep[see, e.g., ][]{hassel10}.
The abundances of different species do evolve with time, starting from an assumed initial distribution; i.e., all of the elements are atomic, except for H and D which are assumed to be in H$_2$ and HD molecules.  
This types of initial conditions are traditionally used in astrochemical modeling (e.g. \citealt{hasegawa92,lee96,roberts04,garrod08,pagani11}).

We did not conduct a complete parameter search to find the ``best fit'', partly because the uncertainties associated with the data may make a best fit not very meaningful  {and partly because, due to the uncertainties of the many different parameters used by the model,}  a parameter search will be computationally very expensive.  Hence, we choose to model the data  heuristically  by adopting a set of physically reasonable parameters.

When adopting a constant density, with the temperature in the range of ${\sim}14$--$20$~K (assuming $T_\text{dust}=T_\text{gas}$), we were not able to reproduce the observed correlation between the temperature and CO abundance (Figure~\ref{fig:correlation2}).  {Namely, in such models, the gas-phase CO abundance \emph{decreases} with temperature.}  This may seem counterintuitive at first sight.  The underlying reason is that, as the dust grain temperature increases, reactions on the dust grain surface become more efficient to form species such as $\rm CO_2$ \citep[e.g., ][]{garrod13}, 
which cannot easily evaporate into the gas phase at such low temperatures.
Thus, in the current set of models, we let the density vary as a function of temperature.  Specifically, we let
\begin{equation}
  n_{\rm H}(\text{cm}^{-3}) = \left(23 - T_{\rm dust}(\text{K})\right) \times 10^4, \label{equ:NvsT}
\end{equation}
to semiquantitatively reflect the anticorrelation between density and temperature seen in the observational data (Figure~\ref{fig:correlation1}).  

The modeling results are shown in Figure~\ref{fig:model}, {in which the curves show the CO abundance and $D_{\rm HCO^+}$ as a function of temperature and gas number density at different times.  Though} quite simple, the models already provide some interesting insights. 
For example,
(1) As part of the heuristics, to match the observed CO abundance and its correlation with temperature, {we have used} enhanced elemental abundances of carbon and oxygen ($2{\times}10^{-4}$ and $5{\times}10^{-4}$ {instead of the frequently adopted $1.4\times  10^{-4}$ and $3.2{\times}10^{-4}$; \citealp{garrod08})}.

(2) To match the observed $D_{\rm HCO^+}$ trend, the D/H abundance ratio is set to $3{\times}10^{-6}$, a factor of ${\sim}5$ lower with respect to the usual value of $(1.5\text{--}2){\times}10^{-5}$.  
This is consistent with previous works on the Galactic elemental abundance gradient of deuterium, carbon, and oxygen \citep{smartt97,lubowich00,carigi05,lubowich10,esteban18}, considering that the sources in the current work have {galactocentric distances $R_{\rm GC}\sim$5 kpc.}

(3) 
An approximate ``fitting'' to the observed trends can be obtained from Figure~\ref{fig:model} for a chemical age of ${\sim}8\times10^4$~yr (solid orange curve).  Although there is no agreed-upon definition for the point of age zero in modeling, here we implicitly define it as the stage in which all of the elements are atomic except for H and D (in the form of H$_2$ and HD).  
For the fitting to $\chi(\rm CO)$ (panel (a) of Figure~\ref{fig:model}), the CO abundance is mainly determined by adsorption and desorption.  A longer age would lead to CO abundances lower than observed.  As noted before, the increase of CO abundance with temperature in that panel is \emph{not} caused by the increased evaporation rate but rather by the $T-n_{\rm H}$ relation (Equation (\ref{equ:NvsT})) implemented in our model.
For the fitting to $D_{\rm HCO^+}$ (panel (b) of Figure~\ref{fig:model}), a longer age would cause a $D_{\rm HCO^+}$ higher than observed.  The age cannot be shorter than ${\sim}10^5$~yr as well, otherwise the abundance of DCO$^+$ would not be high enough ($\gtrsim10^{-11}$) to be detectable.

{
Since we covered only a small fraction of the parameter space, there are caveats associated with the fitting and the derived nominal chemical age.  
First of all, putting together the observational data of CO abundance and D-fraction of $\rm HCO^+$, we have made a rather strong hypothesis that these sources are of similar nature because the data show the same trend in the plots of  $(T,\,\chi[\text{CO}])$ and $(T,\,D_{\text{HCO}+})$ (Figure~\ref{fig:model}). Moreover, our definition of the age zero-point is from the point of view of the formation of a molecular cloud.  However,  an appropriate assumption for the initial conditions depends on a ``proper" choice of chemical age tracer.  A better approach would be to look at the overall chemical inventory and to see whether or not one can identify a variety of ``early type" molecules in the cloud(s). 

Second, our model does not take into account the spin states of H$_2$ and other related species.  It is known that the abundances of deuterated species can be significantly affected by the ortho-para ratio ($o$/$p$ ratio) of H$_2$ \citep{sipilae10,pagani11,furuya15,sipila17}, because $o$-H$_2$ has a higher-energy ground state than $p$-H$_2$, and it can more efficiently destroy the deuterated isotopologues of H$_3^+$ (H$_2$D$^+$, D$_2$H$^+$, D$_3^+$), thus reducing the abundances of deuterated species derived from them.
It has been experimentally demonstrated by \citet{watanabe10} that
{\color{black}H$_2$ molecules freshly formed on amorphous solid water have a statistical $o$/$p$ ratio of 3, and that this $o$/$p$ ratio can change when H$_2$ molecules are retrapped by the water ice. Moreover, the $o$/$p$ ratio  of H$_2$ can  also be altered by gas-phase processes. The initial $o$/$p$ ratio of H$_2$ that was adopted by chemical models is subject to large uncertainties \citep[e.g., ][]{pagani11,bovino17}. One issue is that we do not know how long it takes for atomic hydrogen to become molecular, which affects the evolution of the $o$/$p$ ratio, especially under the circumstance when a molecular cloud may have gone through many dispersal-reassembly cycles \citep[with H$_2$ molecules may mostly be kept intact, while other species may be destroyed and reformed; e.g., ][]{chevance20}. 
If we take into account the spin states of  H$_2$ in our model, the deuteration process would be delayed, i.e. the chemical age would be longer, and this would render CO abundances lower than observed. }

Third, we used a ``canonical'' CRIR of $1.36{\times}10^{-17}\,\text{s}^{-1}$.  Cosmic-ray ionization is the main driving force of chemistry in shielded regions.  A moderately high CRIR can shorten the chemical evolution timescale, and, specifically,  help the conversion from $o$-H$_2$ to $p$-H$_2$.  It is known that the CRIR is higher in the inner region of the Galactic disk \citep{indriolo15,neufeld17}.  Since the four sources are at galactocentric distances of $\sim$5~kpc, their CRIR could be higher than the canonical value.  Scaling the canonical CRIR value up or down by a factor of 10  (dashed and dotted curves in Figure~\ref{fig:model}) does not improve the fitting to the  CO abundance and $D_{\rm HCO^+}$ trends. {\color{black}More comprehensive parameter studies, such as an MHD model coupled with chemical properties from the entire sample of sources, are needed to get more quantitative constraints.}

}

  \begin{figure*}
\includegraphics[width=18cm]{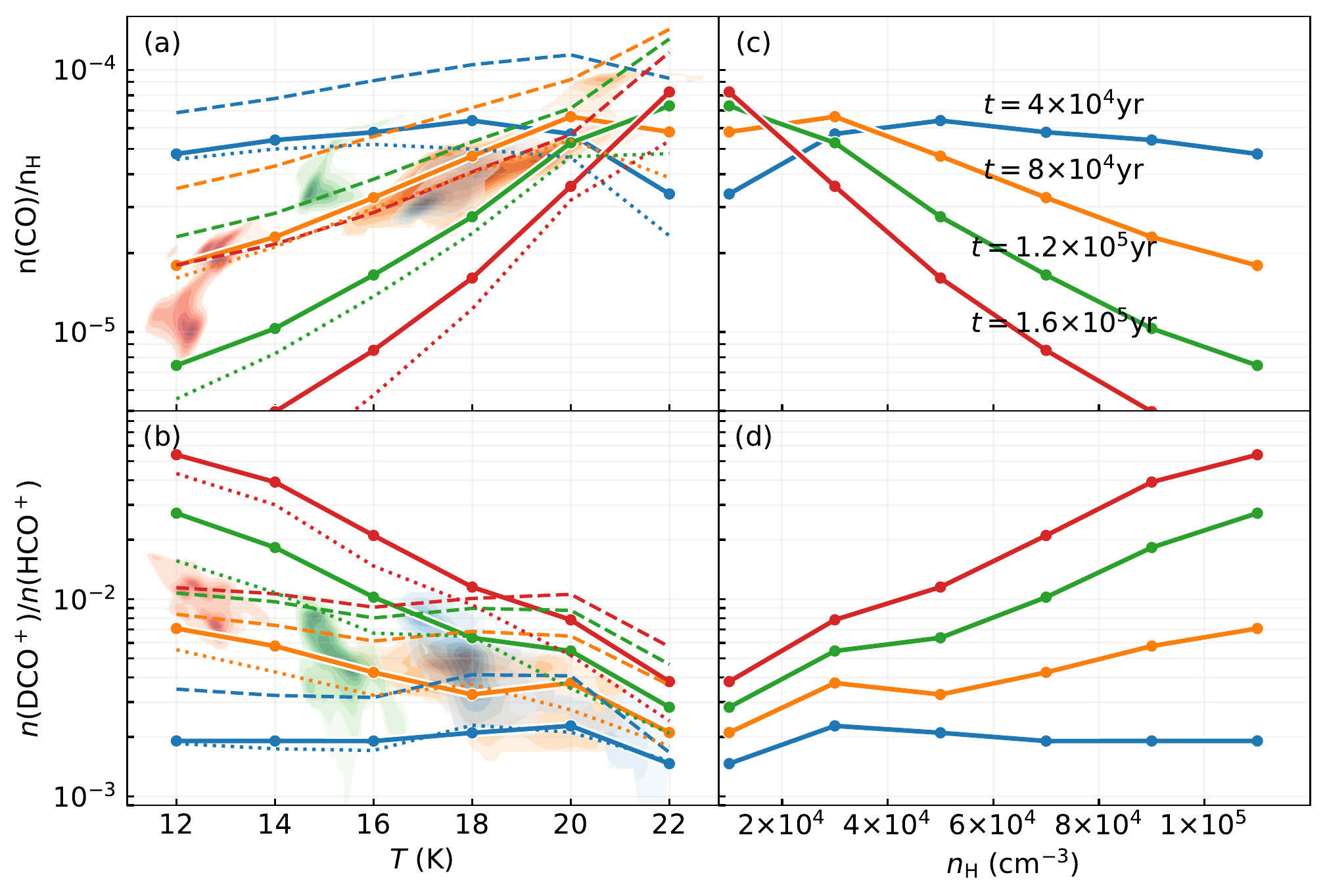}  %{n_T_vs_CO_DCO+_HCO+.pdf}
\caption{Modeled CO abundance (relative to hydrogen) and D-fraction of HCO$^+$ as a function of temperature and gas number density, evaluated at different time points (shown with different colored lines).  The left hand side and the right hand side panels appear to be mirror reflections of each other because temperature and density are linearly anticorrelated in the models. { The observational data of each source are plotted as a bivariate Gaussian kernel density map in the background (red: G\,11.38+0.81, orange: G\,14.49-0.13, blue: G\,15.22-0.43, and green:  G\,34.74-0.12, 
see also Figure~\ref{fig:correlation2}).  The solid orange curve is considered to be a ``fit'' to the four regions.  
The solid curves are calculated with a canonical CRIR of $1.36{\times}10^{-17}\,\text{s}^{-1}$, while the dashed and dotted curves correspond to 10 times higher or lower than this value.}
}\label{fig:model}
\end{figure*} 

\subsection{Dynamical and chemical timescales}
Three timescales can be used to characterize the evolutionary status of our sources.

In an ideal case of supercritical collapse, the free-fall timescale of a cloud is defined as
\begin{equation}
 t_\text{ff}=\left(\frac{3\pi}{32G\rho_\text{gas}}\right)^{1/2}\simeq 1.38\times10^6\;\left(\text{yr}\right)\left(\frac{n_\text{H}}{10^3\;\text{cm}^{-3}}\right)^{-1/2} \label{equ:tff}
 \end{equation}
 
In reality, physical mechanisms such as magnetic field and turbulence provide support against gravitational collapse. Therefore, the contraction speed of the clouds is in general  observed as a fraction ($\eta\sim 20\%\text{--}50\%$) of the free-fall speed \citep[e.g., ][and also found in our entire sample of sources, S. Feng et al. 2020, in preparation]{evans03,wyrowski12,wyrowski16}. The contraction timescale can be computed from the observation  as 
\begin{equation}
t_\text{contr}\sim  t_\text{ff}/\eta. 
\end{equation}

The timescale for CO molecules to freeze out onto dust grains is
\begin{equation}
\begin{split}
t_\text{ads,CO} &=\left(S\pi a^2\sqrt{\frac{8k_\text{B}T_\text{kin}}{\pi m_\text{CO}}}n_\text{grain}\right)^{-1} \\
&\simeq 1.2{\times}10^6\,\text{(yr)}\;
S^{-1}
\left(\frac{T_\text{kin}}{10\;\text{K}}\right)^{-1/2}\\
&\times \left(\frac{n_\text{grain}}{10^{-8}\;\text{cm}^{-3}}\right)^{-1}
\left(\frac{a}{0.1\;\mu\text{m}}\right)^{-2},
\end{split}\label{equ:tads}
\end{equation}
where $S$ is the sticking coefficient, $a$ is the dust grain radius, and the meaning of the other symbols should be self-evident \citep{caselli99,aikawa13}.

 \iffalse
The grain number density can be expressed as
\begin{equation}
 \begin{split}
   n_\text{grain} &= \frac{n_\text{H}m_\text{H}\mu}{\gamma \rho_\text{grain} 4\pi a^3/3} \\
   &= 2.8{\times}10^{-12}\; n_\text{H}\;
      \left(\frac{\mu}{1.4}\right)
      \left(\frac{\gamma}{100}\right)^{-1}
      \left(\frac{\rho_\text{grain}}{2\;\text{g}\;\text{cm}^{-3}}\right)^{-1}
      \left(\frac{a}{0.1\;\mu\text{m}}\right)^{-3},
 \end{split} 
\end{equation} 
\fi

Expressed in terms of gas density $n_{\rm H}$, the adsorption timescale can be written as
\begin{equation}
  \begin{split}
    t_\text{ads,CO} &=\left(\frac{3S}{4a}\sqrt{\frac{8k_\text{B}T_\text{kin}}{\pi m_\text{CO}}} \frac{n_\text{H}m_\text{H}\mu}{\gamma \rho_\text{grain}} \right)^{-1} \\
    &= 4.2{\times}10^5\;\text{(yr)}\;
       \left(\frac{T_\text{kin}}{10\;\text{K}}\right)^{-1/2}
       \left(\frac{n_\text{H}}{10^4\;\text{cm}^{-3}}\right)^{-1}
       \left(\frac{\mu}{1.4}\right)^{-1} \\
     &\quad \times \left(\frac{a}{0.1\;\mu\text{m}}\right)
       \left(\frac{\gamma}{100}\right)
       \left(\frac{\rho_\text{grain}}{2\;\text{g}\;\text{cm}^{-3}}\right).
  \end{split} 
\end{equation}

Hence, we have
\begin{equation}
\begin{split}
  \frac{t_\text{ff}}{t_\text{ads,CO}} &\simeq 0.75\times
       \left(\frac{T_\text{kin}}{10\;\text{K}}\right)^{1/2}
       \left(\frac{n_\text{H}}{10^4\;\text{cm}^{-3}}\right)^{1/2}
       \left(\frac{\mu}{1.4}\right) \\
     &\quad \times \left(\frac{a}{0.1\;\mu\text{m}}\right)^{-1}
       \left(\frac{\gamma}{100}\right)^{-1}
       \left(\frac{\rho_\text{grain}}{2\;\text{g}\;\text{cm}^{-3}}\right)^{-1}.
\end{split} 
\end{equation}

To match with both the observed CO abundance and $D_{\rm HCO^+}$ in our source environment, with $n_\text{H}\sim 10^4-10^5$ cm$^{-3}$ in our 70\,$\mu$m dark sources (Table~\ref{tab:physical})\footnote{Assuming that the clouds have a spherical structure, with a length along the line of sight {\it L} close to its projected width, i.e., 0.5--1\,pc in our sources, $n_{\rm H} \rm \sim N_{H_2}/{\it L}$. According to our SED fit, $N_{\rm H_2}$  in our sources is in the range of  $\rm 10^{22}\text{--}10^{23}\,cm^{-2}$, and $n_{\rm H}$ is in the range of $\rm 10^4\text{--}10^5\,cm^{-3}$. }, our preliminary chemical modeling prefers
a chemical age of the sources, in terms of  $t_{\rm ads,CO}$, of ${\sim}8\times10^4$\,yr (Section~\ref{model} and Figure~\ref{fig:model}).
 This chemical age appears to be comparable to or slightly shorter than the free-fall timescale  $t_{\rm ff}$ ((1--4)\,$\rm \times10^5$\,yr), and also shorter than the  $t_{\rm contr}$ ($\rm \sim10^6$\,yr). 
 {The above timescale estimates are based on the assumption that $n_\text{H}$ does not change with time. In reality, $n_\text{H}$ is increasing from a less dense initial condition, e.g., $\sim 10^3$ cm$^{-3}$,  to the current  status. Therefore, it takes longer time from the age zero-point till now ($t_{\rm ads,CO}$) than the above estimates, and will take more years for the cloud to contract ($t_{\rm contr}$) than the above prediction. }

{In these dense gas clumps, the CO depletes fast, and so  the clumps are expected to  dynamically evolve to young protostellar objects. It is possible that the observed sources in this work are at a crossing point of fully developed CO depletion and the onset of global collapse. 
These are expected to be rare objects, given their short lifetime.
 This may also be a reason why only a few cases of parsec-scale CO depletion were reported towards  high-mass star-forming regions so far, compared to the commonly reported subparsec-scale CO depletion towards much closer and more compact low-mass regions \citep[e.g., ][]{caselli99,tafalla02,pineda10}.
}

Of course, many details in these processes need to be further scrutinized. 
In the future, such a comparative analysis will be applied to the entire sample.
{More comprehensive chemical modeling taking into account the spin states of H$_2$ and other relevant species \citep{hugo09,sipilae10,kong15,bovino17} and coupling with dynamical evolution \citep[e.g., ][]{goodson16,bovino19} is needed  to understand the feedback of protostellar heating on the depletion efficiency and therefore to  profile the entire evolutionary process of these sources}.

\section{Conclusions}\label{conclusion}
With the aim of characterizing the kinematic and chemical properties of the initial conditions for HMSF, we  carried out a  line-imaging survey project (MIAO) toward a sample of 24 relatively near ($d\rm <5$\,kpc) IRDCs.  This project uses single-dish  (IRAM 30\,m and NRO 45\,m) and interferometric (ALMA) telescopes to image pairs of neighboring 70\,$\mu$m bright and dark clumps  at different spatial scales of individual regions, from parsec-scale filamentary clouds down to 0.01\,pc-scale dense cores. The comparative analysis is applied to each region, which improves the robustness by canceling out calibration uncertainties.

In the present work, we focus on  a detailed study of the parsec-scale CO depletion toward four regions (G\,11.38+0.81, G\,15.22-0.43, and G\,14.49-0.13, and G\,34.74-0.12) from IRAM 30\,m and NRO 45\,m observations. Showing the spatial correlation between the CO depletion factor and the source physical structure (gas and dust temperature and density), we discuss the interplay between CO depletion and D-fractionation of $\rm HCO^+$.  

Our conclusions are as follows. 

\begin{enumerate}
\item Our observations cover two transitions (1--0 and 2--1) from three CO isotopologues ($\rm ^{13}CO$, $\rm C^{18}O$, and $\rm C^{17}O$). They show anticorrelated spatial distributions with the dense gas tracers (1--0 lines of $\rm H^{13}CO^+$ and $\rm DCO^+$) in our sample sources, indicating that a high degree of CO depletion appears toward the cold, dense, 70\,$\mu$m dark clumps.

\item The SED fits to multiwavelength continuum data indicate strong spatial anticorrelation between $\rm H_2$ column density and the dust temperature of each source.

\item  The LVG analysis indicates that the {\color{black}kinetic temperature derived from $\rm NH_3$ is consistent with the dust temperature}, and that the $\rm C^{18}O$\,(2--1), $\rm H^{13}CO^+$\,(1--0), and $\rm DCO^+$\,(1--0) lines are reasonably assumed as optically thin and under LTE condition  in our source environment (with $\rm T<20$\,K and $n_{\rm H}\sim \rm 10^4$--$\rm 10^5\,cm^{-3}$).

\item The gas kinetic temperature measured with different thermometers ($p$-$\rm NH_3$ lines, $p$-$\rm H_2CO$ lines, and CO isotopologue lines) varies by a factor of up to 2. Although such a difference increases the uncertainty of the  molecular column density measurement toward a certain location, it does not result in a large uncertainty in $f_D(\rm C^{18}O)$ or $D\rm(HCO^+)$  in terms of the relative abundance ratio between molecules. 

\item Separating each region into a $\rm DCO^+$-dominant zone (P1), a CO-dominant zone (P3),  and a transition zone  (P2), we find that $f_D(\rm C^{18}O)$ and $D\rm(HCO^+)$ vary as a function of location, showing a robust decrease from P1 (with $f_D(\rm C^{18}O)$ as 5--20 and $D\rm(HCO^+)$ as $\rm 0.5\%\text{--}2\%$) to P3 by a factor of more than 3 within a spatial extension of 2\,pc.
The main reason for such a trend is the different evolutionary stages of the neighboring clumps in the same cloud, which show a distinctive difference in temperatures at a linear scale of 0.1--0.5\,pc.

\item To match the observed molecular abundances and trends,  our preliminary chemical modeling prefers chemical ages for our sources as of $\rm {\sim}8\times10^4$ yr, which is comparable to their free-fall timescales and smaller than their contraction timescales. This indicates that our sources are {at an early dynamical and chemical evolution}. With future modeling incorporating the effects of the spin states of $\rm H_2$ and dynamical evolution, we expect to get a more thorough understanding of the evolution of these sources.

\item {Limited by the sensitivity of previous observational instruments, CO depletion was  commonly reported at subparsec scale towards much closer and more compact low-mass star-forming  regions. Fast-growing high-quality spectral imaging projects will allow us to reduce observational bias; thus, parsec-scale CO depletion is expected to be commonly observed towards more distant high-mass star-forming regions. }

 \end{enumerate}

\begin{acknowledgements}
We would like to thank the IRAM 30\,m staff for
their helpful support during the performance of the 
IRAM 30\,m observations in service mode.

S.F. acknowledges the support of National Natural Science Foundation of China No. 11988101, the support of the CAS International Partnership Program No.114A11KYSB20160008, and the support of the EACOA fellowship from the East Asia Core Observatories Association (EACOA). EACOA consists of the National Astronomical Observatory of China, the National Astronomical Observatory of Japan, the Academia Sinica Institute of Astronomy and Astrophysics, and the Korea Astronomy and Space Science Institute.

P.C. acknowledges the financial support from the Max Planck Society.

H.B. and Y.W. acknowledge support from the European Research Council under the Horizon 2020 Framework Program via the ERC Consolidator Grant CSF-648505.

H.B. also acknowledges support from the Deutsche Forschungsgemeinschaft in the Collaborative Research Center (SFB 881) ``The Milky Way System" (subproject B1)

F.D. is supported by the One Hundred Person Project of the Chinese Academy of Sciences through grant 2017-089 and by  NSFC grant No. 11873094.

I.J.-S. has received partial support from the Spanish FEDER (project number ESP2017-86582-C4-1-R) and the State Research Agency (AEI; project number PID2019-105552RB-C41).

P.S. was partially supported by the Grant-in-Aid for Scientific Research (KAKENHI) No. 18H01259 of the 
Japan Society for the Promotion of Science (JSPS).

K.W. acknowledges support by the National Key Research and Development Program of China (2017YFA0402702, 2019YFA0405100), the National Science Foundation of China (11973013, 11721303), and the starting grant at the Kavli Institute for Astronomy and Astrophysics, Peking University (7101502016).

S.Z. acknowledges support by the NAOJ ALMA Scientific Research grant No. 2016-03B.

Part of this work was supported by the NAOJ Visiting Joint Research program (grant No. 1901-0403).

This work also benefited from the International Space Science Institute (ISSI/ISSI-BJ) in Bern and Beijing, thanks to the funding of the team ``Chemical abundances in the ISM: the litmus test of stellar IMF variations in galaxies across cosmic timeÓ (Principal Investigator D.R. and Z.-Y.Z.).

 \end{acknowledgements}

\software{GILDAS/CLASS \citep{pety05}, NOSTAR \citep{sawada08}, \texttt{HfS} \citep{estalella17}, RADEX \citep{vandertak07}, \texttt{chempl} \citep{du20}}

\bibliographystyle{apj}
\bibliography{70micron_CO-accepted.bbl}
%\bibliography{/Users/siyifeng/link2GD/HMSFR.bib}

%%%%%%%%%%%%%%%%%%%%%%%%%%%%%%%%%%%%%%%
%%%%%%%%%%%%%%%%%%%%%%%%%%%%%%%%%%%%%%%%
%%%%%%%%%%%%%%%%%%%%%%%%%%%%%%%%%%%%%%%% 

\newpage
\setcounter{section}{0}
\renewcommand{\thetable}{A\arabic{section}}
\setcounter{table}{0}
\renewcommand{\thetable}{A\arabic{table}}
\setcounter{figure}{0}
\renewcommand{\thefigure}{A\arabic{figure}}
\appendix

%\newpage
%%%%%%%%%%%%%%%%%%%Figure

%\section{Figures}
Figures~\ref{spec1}  shows the profile of the CO and $\rm HCO^+$ isotopologue lines we use to measure the molecular column densities toward the four sources. 

Figure~\ref{fig:correlation2}  shows the possible correlation between variables and gas density or dust temperature.

%\section{Tables}
Table~\ref{sourceall}  lists all of the sources in our sample for IRAM 30\,m, NRO 45\,m, and ALMA observations.

Table~\ref{tab:lines} lists the targeted lines covered by our IRAM 30\,m and  NRO 45\,m observations.

Table~\ref{tab:linedeu} lists the line profile fitting results using the GAUSS method in GILDAS package toward  the P1, P2, and P3 of each source.

Table~\ref{tab:gaspara} lists the gas parameters of our target in this work derived by using different temperature measurements.

\newpage

  \begin{figure}

\begin{tabular}{lll}
\begin{sideways}
\Large$\rm T_{mb}$\,(K)
\end{sideways}
&\includegraphics[width=8.5cm] {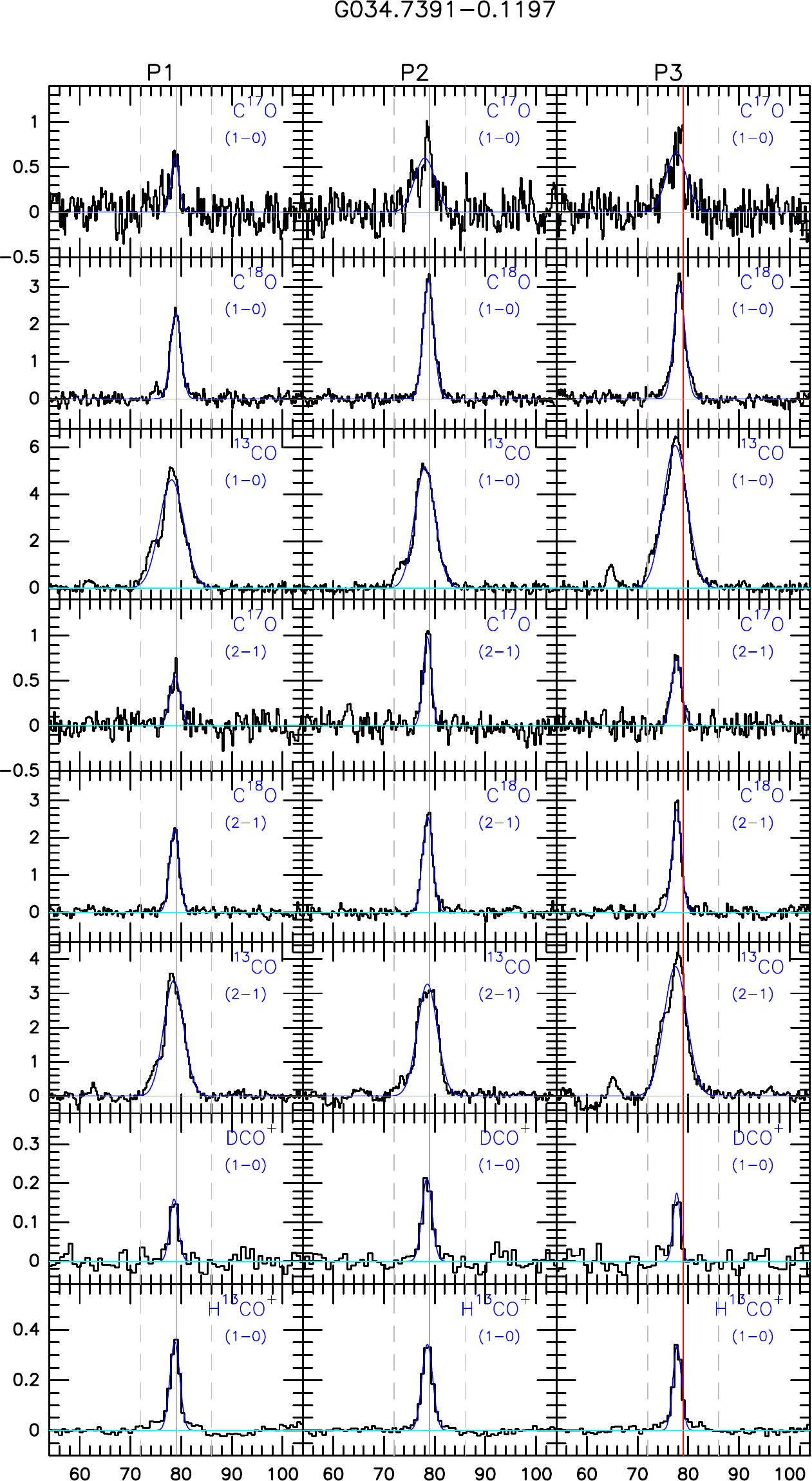}
&\includegraphics[width=8.5cm] {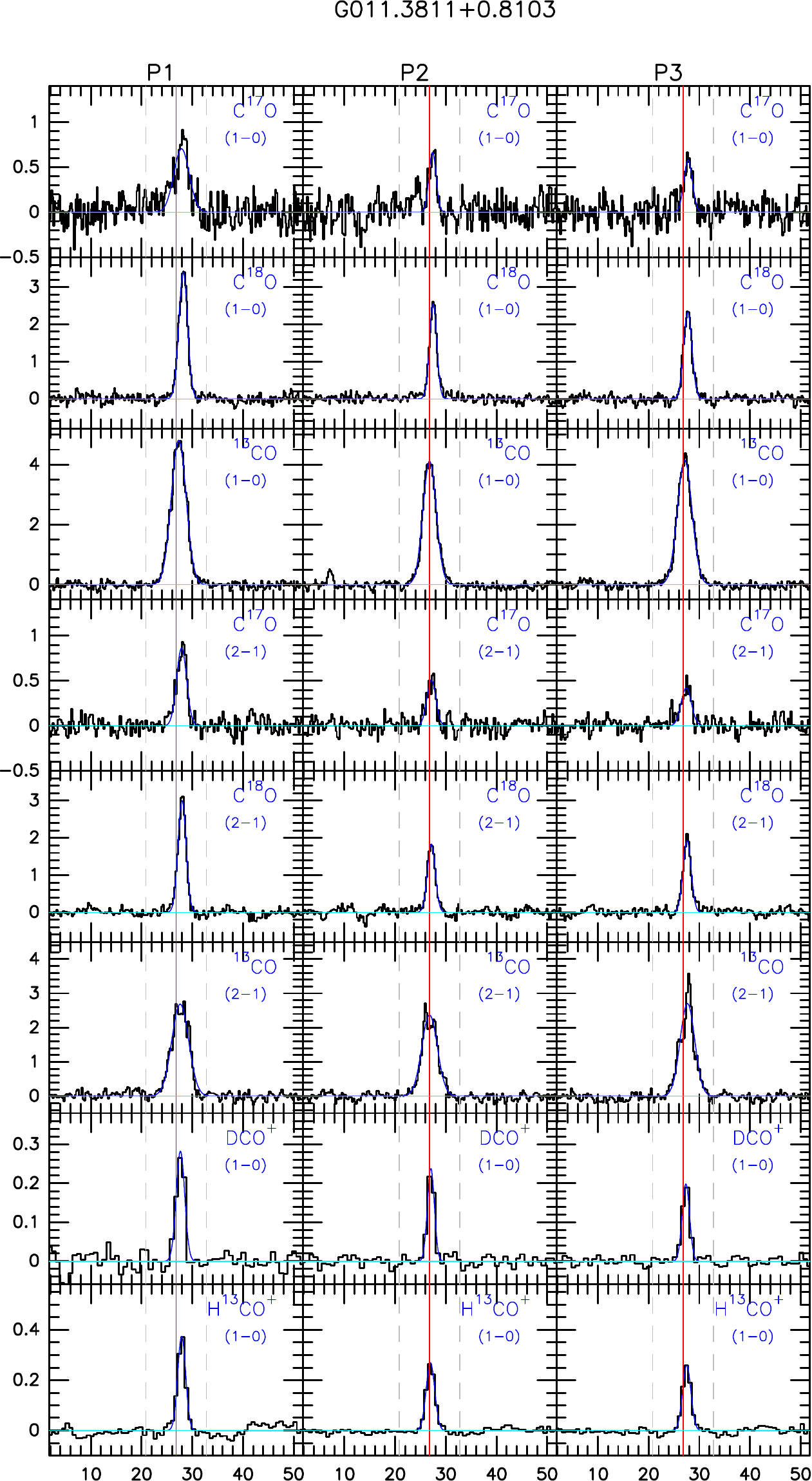}\\
%&\includegraphics[width=8.5cm] {G15_2169-beam_vel}
%&\includegraphics[width=8.5cm] {G14_4876-beam_vel}\\
&\multicolumn{2}{c}{\Large$\rm V_{lsr}\,(km\,s^{-1})$}
\end{tabular}
\caption{Profiles of the CO and $\rm HCO^+$ isotopologue lines observed using the IRAM 30\,m and NRO 45\,m, averaged from a beam-sized region with the center  toward P1, P2, and P3 of each source in the plane of the sky. All lines are extracted from images that we regridded to the same pixel size, but whose native angular and velocity resolution we kept as in the observations (see beam information in Table~\ref{tab:lines}). In each panel, two gray dashed vertical lines  indicate the velocity range for which we integrate the intensity; the red  vertical line indicates the $\rm V_{sys}$ of each source. The horizontal cyan line indicates the baseline ($\rm T_{mb}$=0\,K).}\label{spec1}
\end{figure} 

\newpage

\setcounter{figure}{0}
\begin{figure}
\begin{tabular}{lll}
\begin{sideways}
\Large$\rm T_{mb}$\,(K)
\end{sideways}
%&\includegraphics[width=8.5cm] {G34_7391-beam_vel}
%&\includegraphics[width=8.5cm] {G11_3811-beam_vel}\\
&\includegraphics[width=8.5cm] {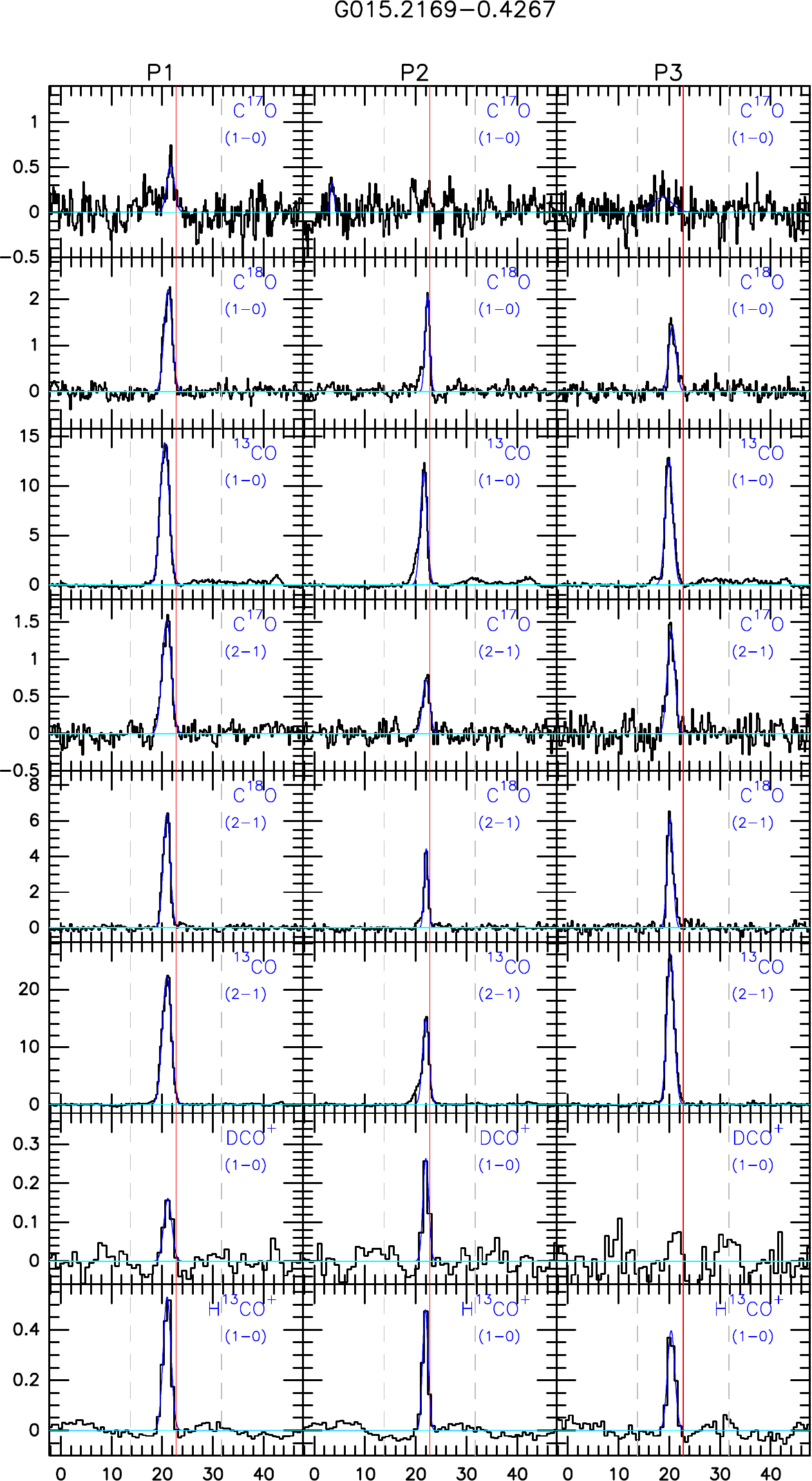}
&\includegraphics[width=8.36cm] {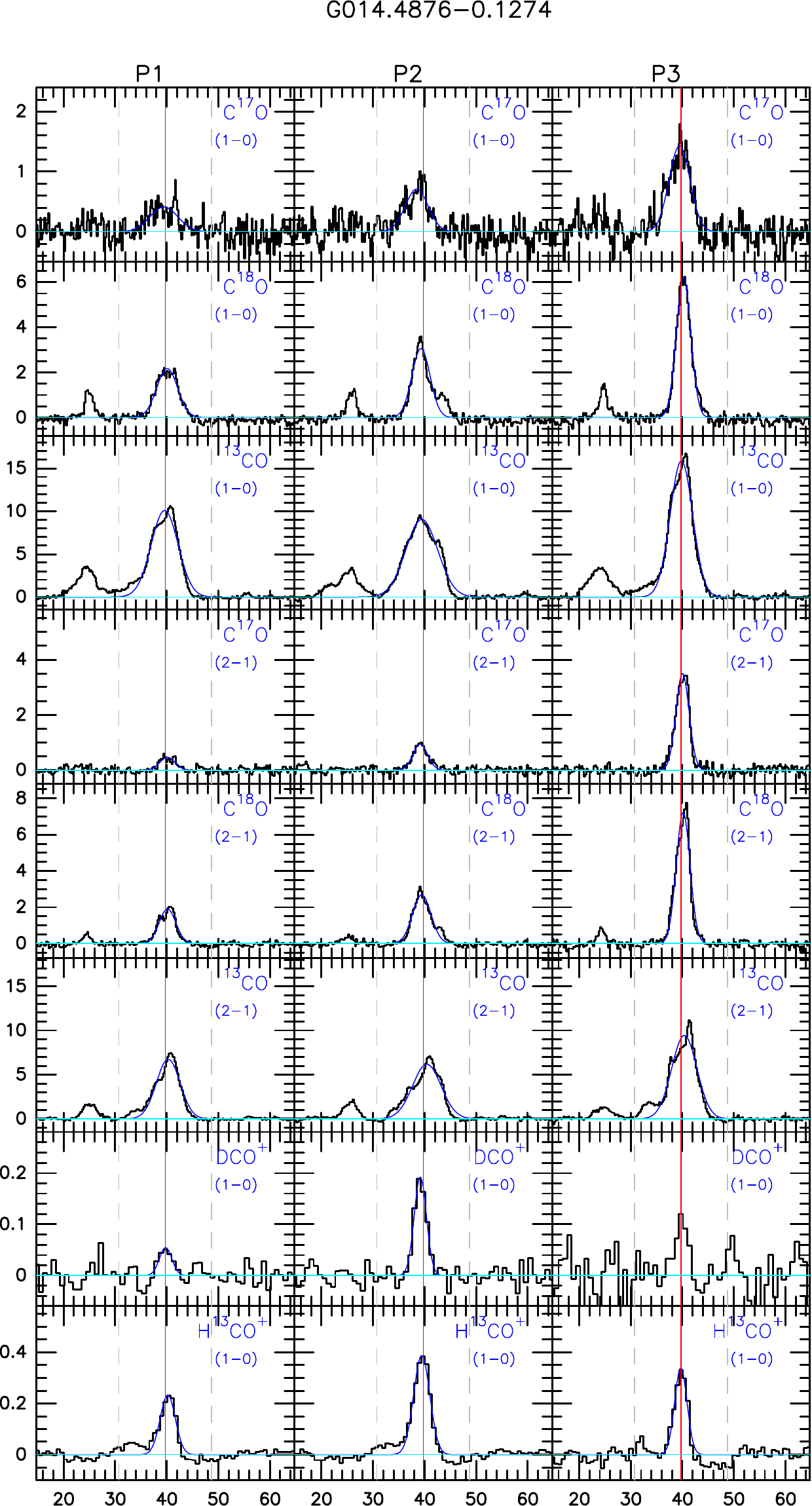}\\
&\multicolumn{2}{c}{\Large$\rm V_{lsr}\,(km\,s^{-1})$}
\end{tabular}
\caption{(continued)}
%\caption{Profiles of the CO and $\rm HCO^+$ isotopologue lines observed using IRAM 30\,m NRO 45\,m, averaged from a beam-sized region with the center  toward P1, P2, and P3 of each source in the plane of the sky. All lines are extracted from images that we regridded to the same pixel size, but whose native angular and velocity resolution we kept as in the observations (see beam information in Table~\ref{tab:lines}). In each panel, two gray dashed vertical lines  indicate the velocity range for which we integrate the intensity; the red  vertical line indicate the $\rm V_{sys}$ of each source. The horizontal cyan line indicates the baseline ($\rm T_{mb}$=0\,K).}\label{spec2}
\end{figure}

\newpage
  \begin{figure}
      \begin{tabular}{p{4.5cm}p{4.5cm}p{4.5cm}p{4.5cm}}
G\,15.22-0.43\\
\includegraphics[align=c,width=4.8cm] {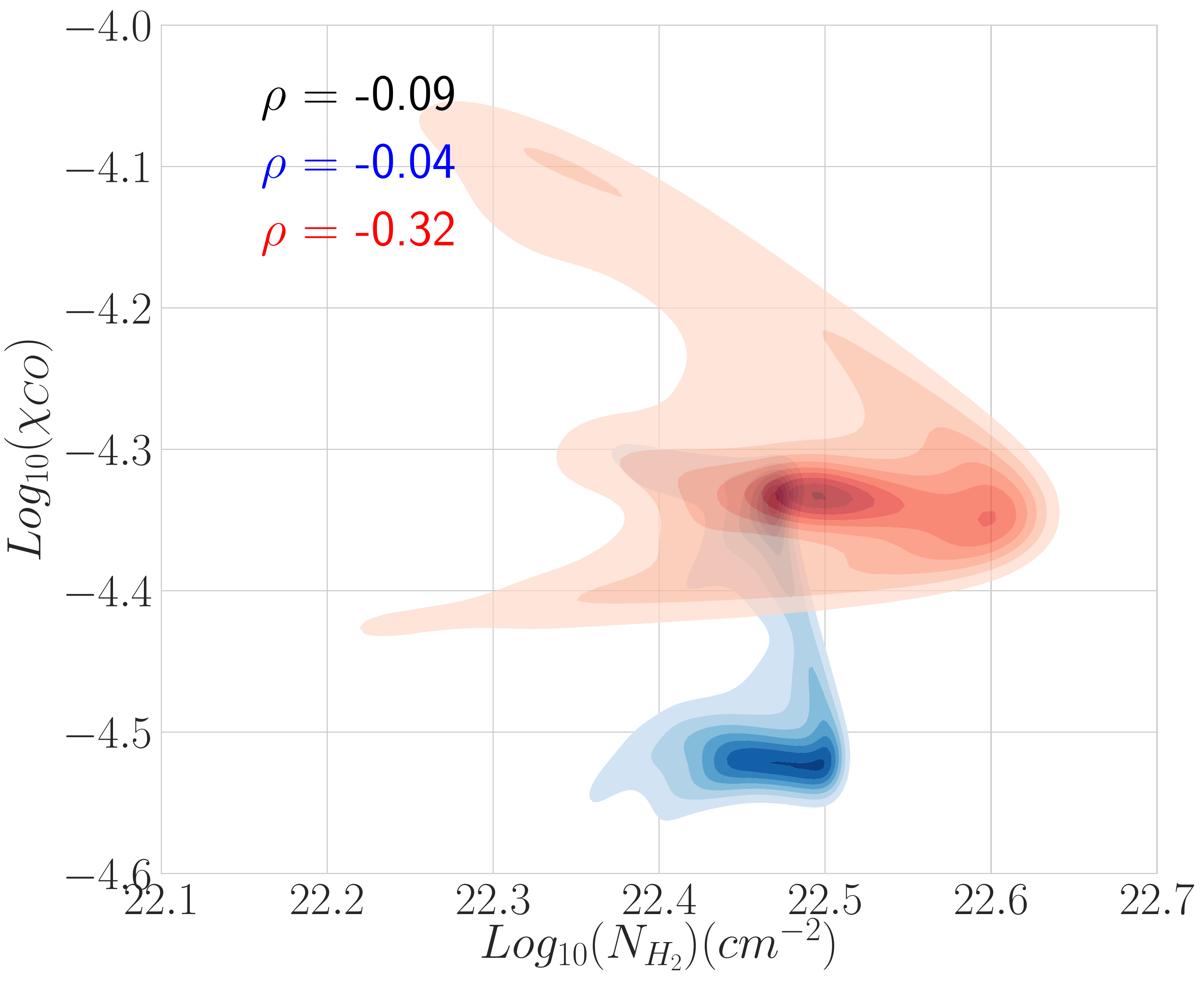}
&\includegraphics[align=c,width=4.8cm] {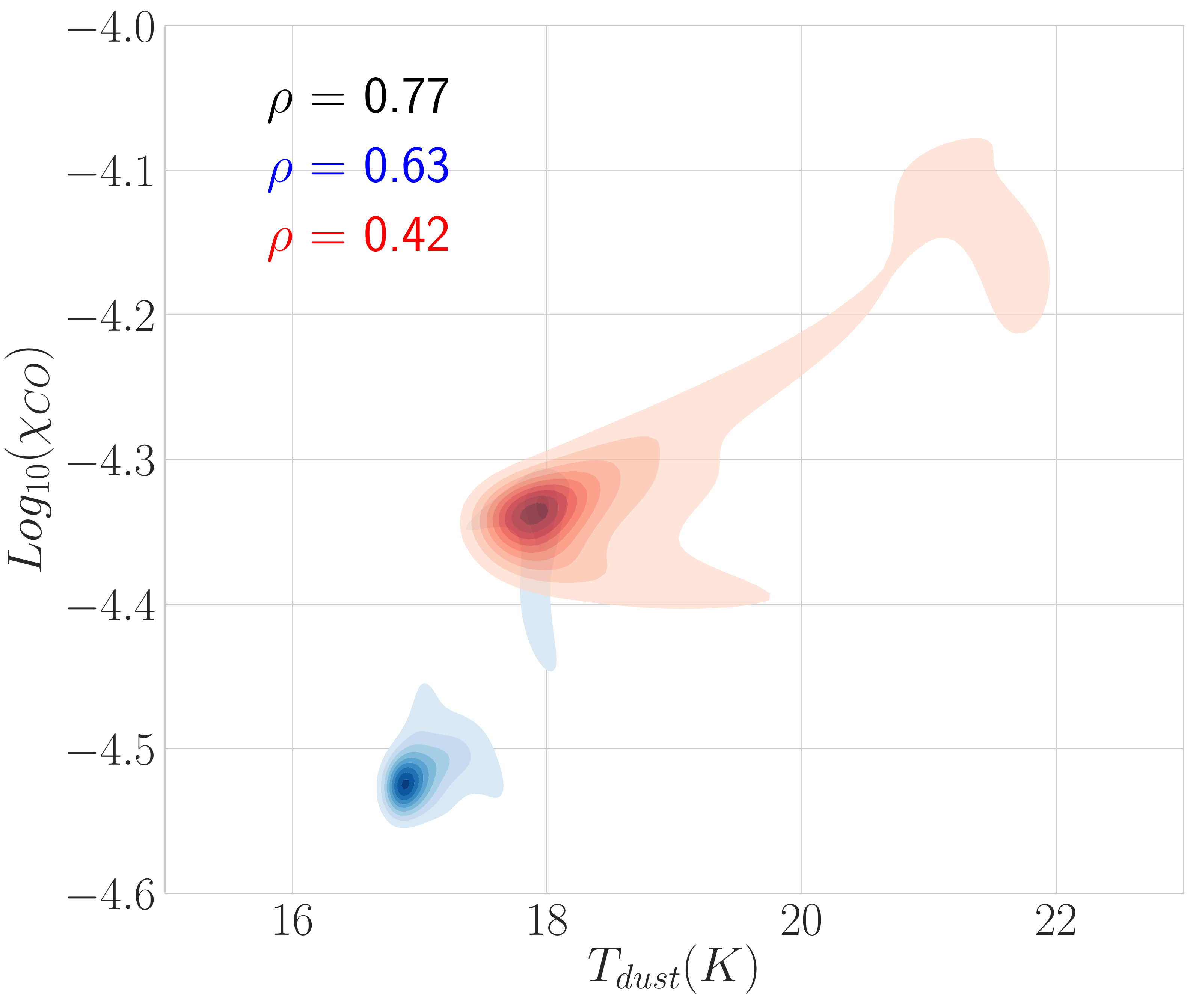}
&\includegraphics[align=c,width=4.8cm] {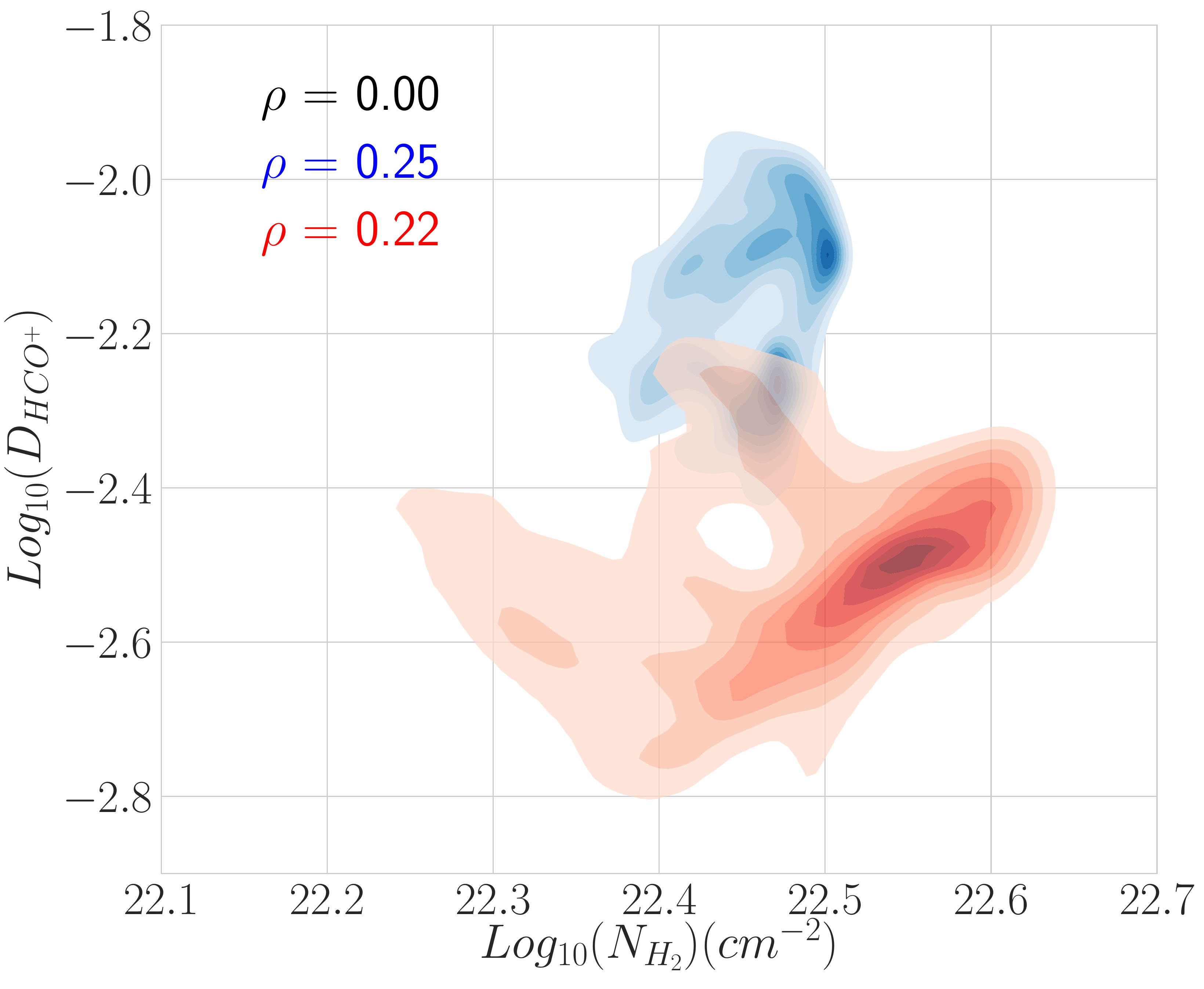}
&\includegraphics[align=c,width=4.8cm] {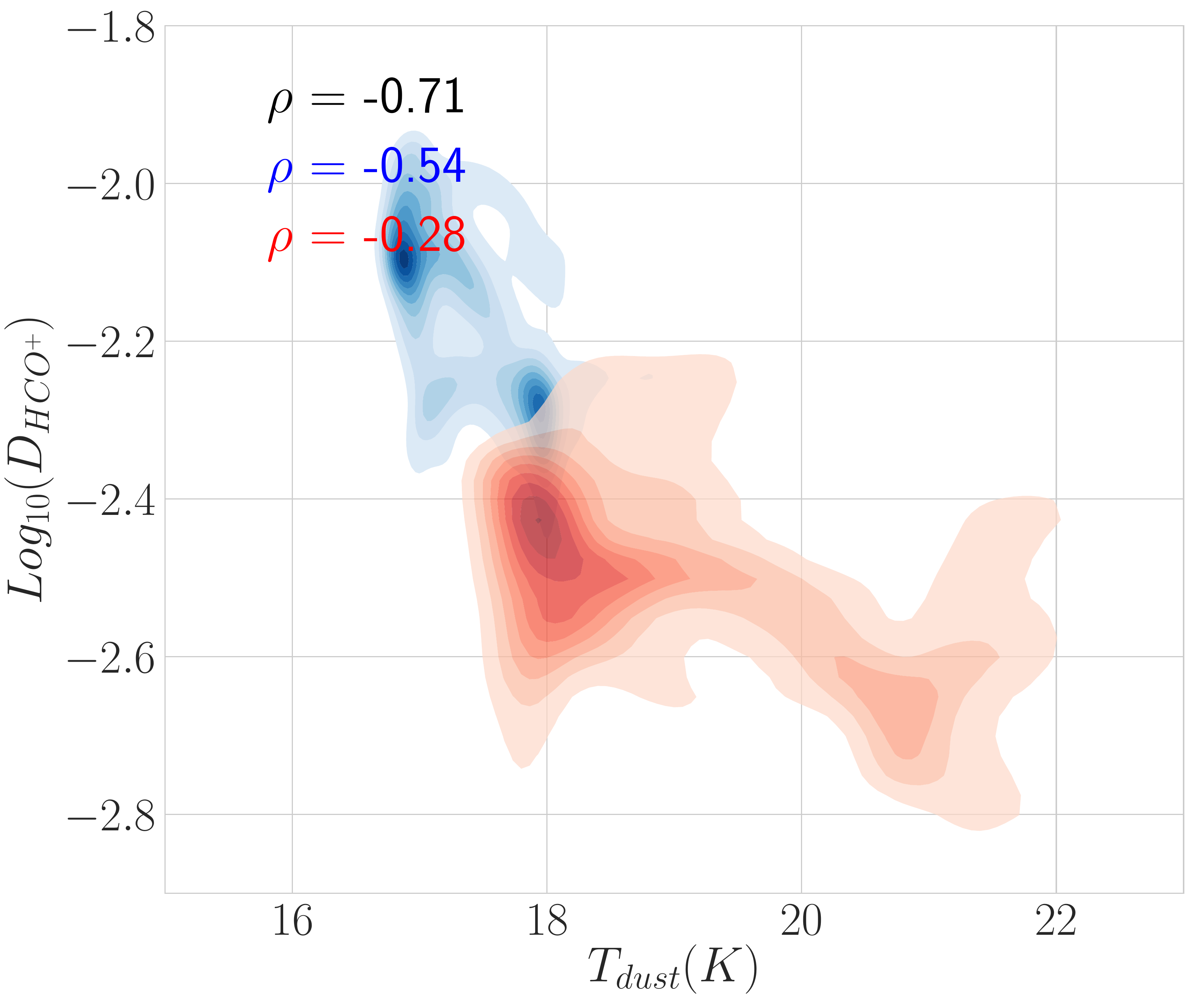}\\[-0.2cm]
G\,11.38+0.81\\
\includegraphics[align=c,width=4.8cm] {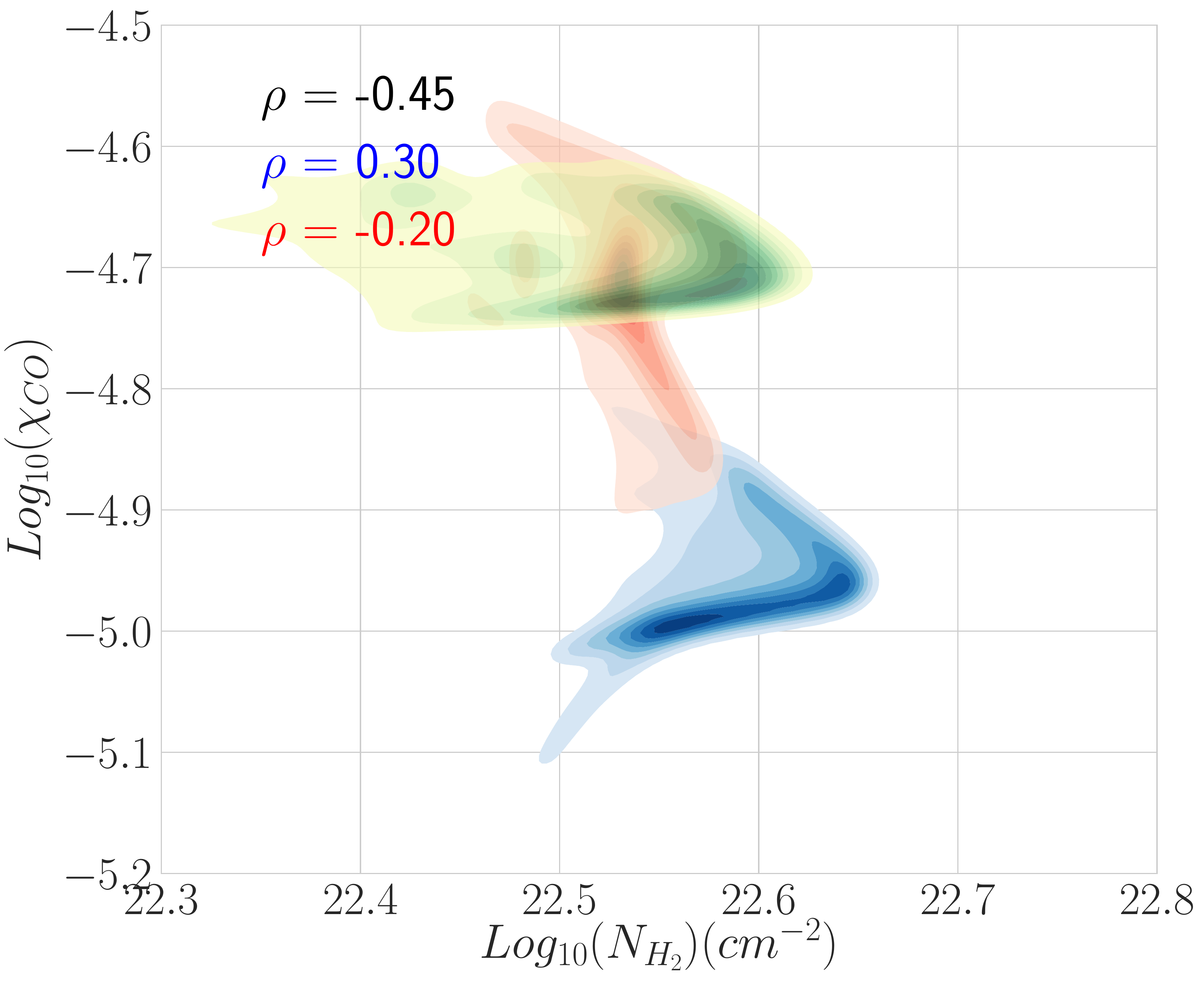}
&\includegraphics[align=c,width=4.8cm] {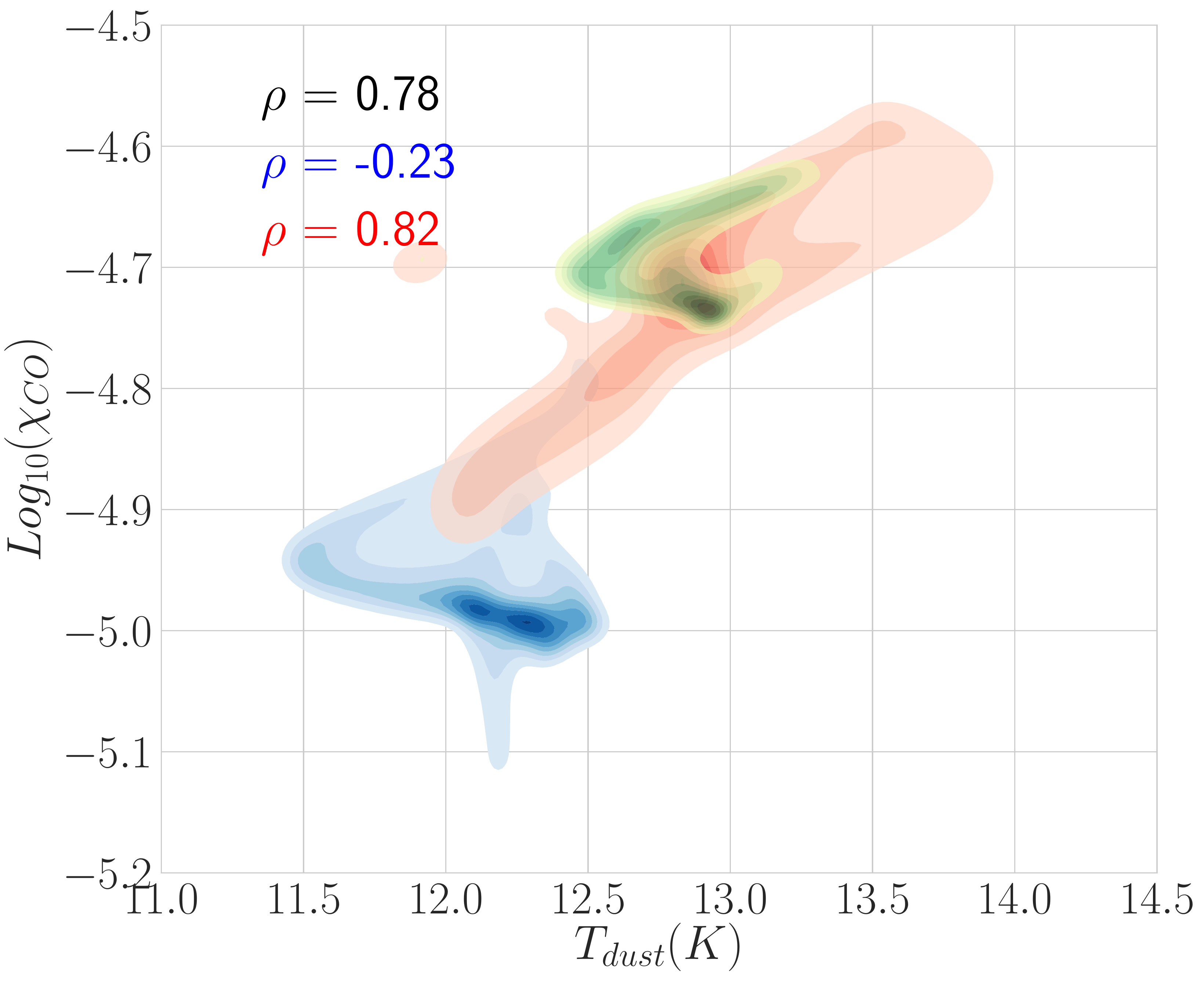}
&\includegraphics[align=c,width=4.8cm] {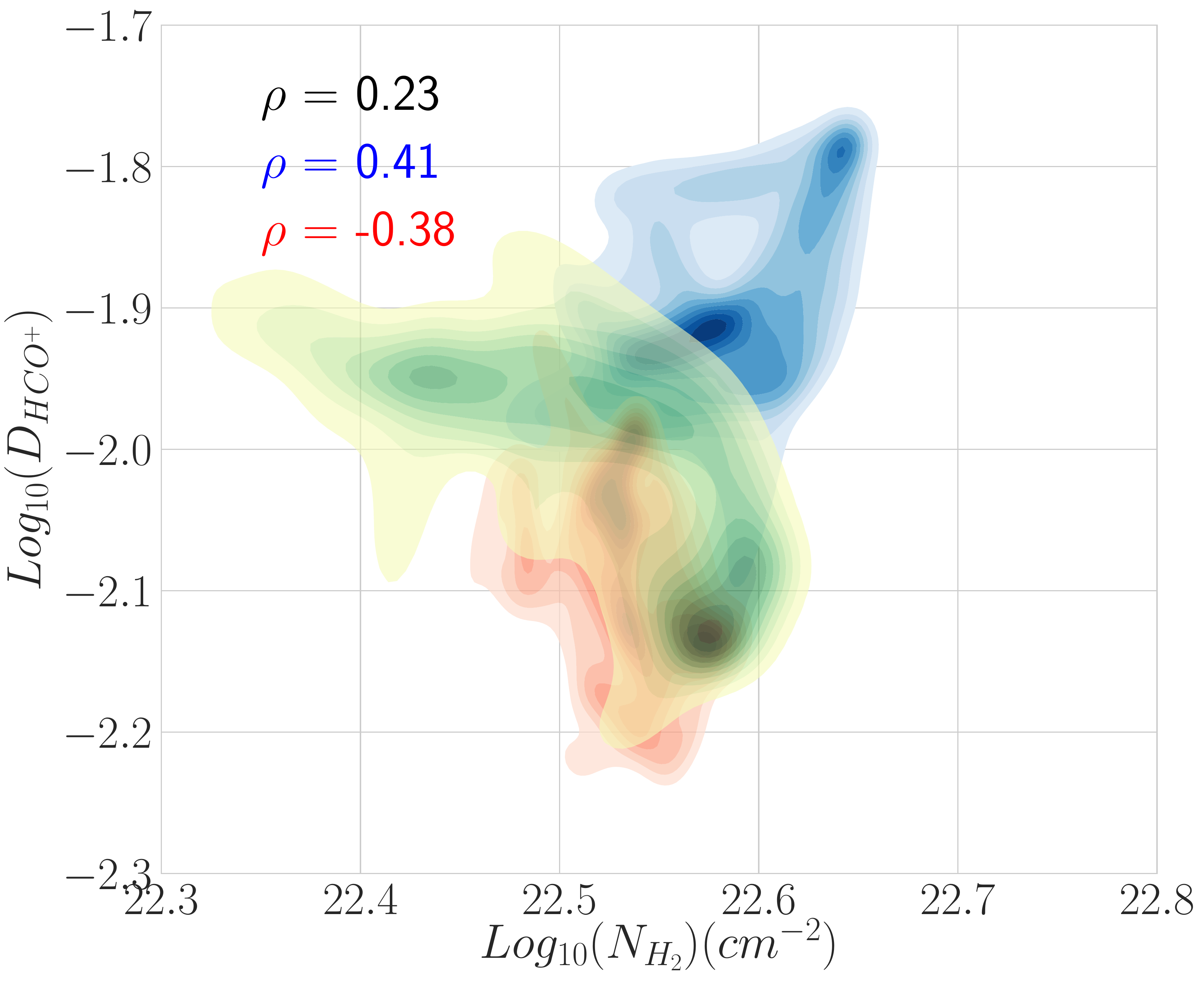}
&\includegraphics[align=c,width=4.8cm] {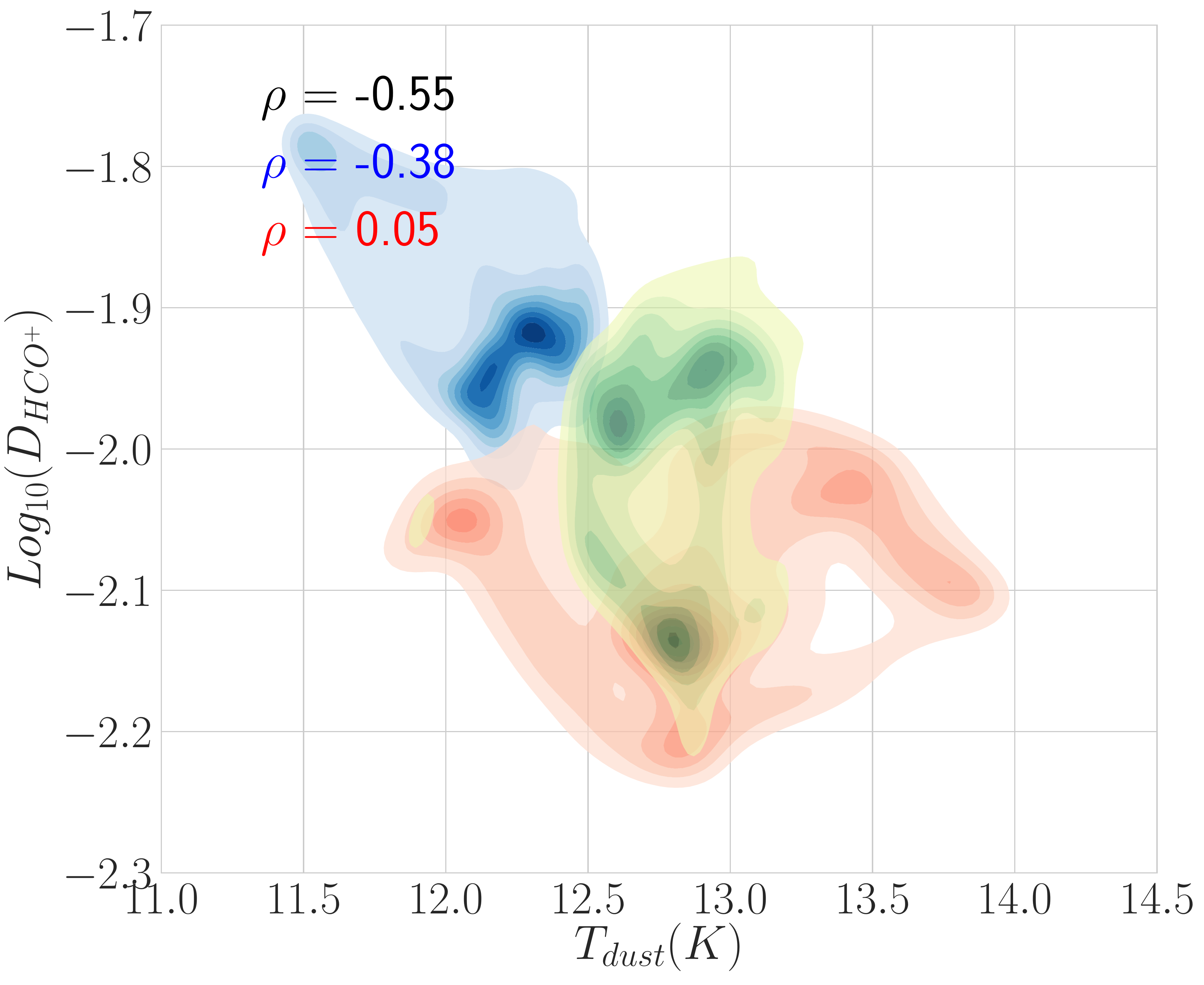}\\[-0.2cm]
G\,14.49-0.13 \\
\includegraphics[align=c,width=4.8cm] {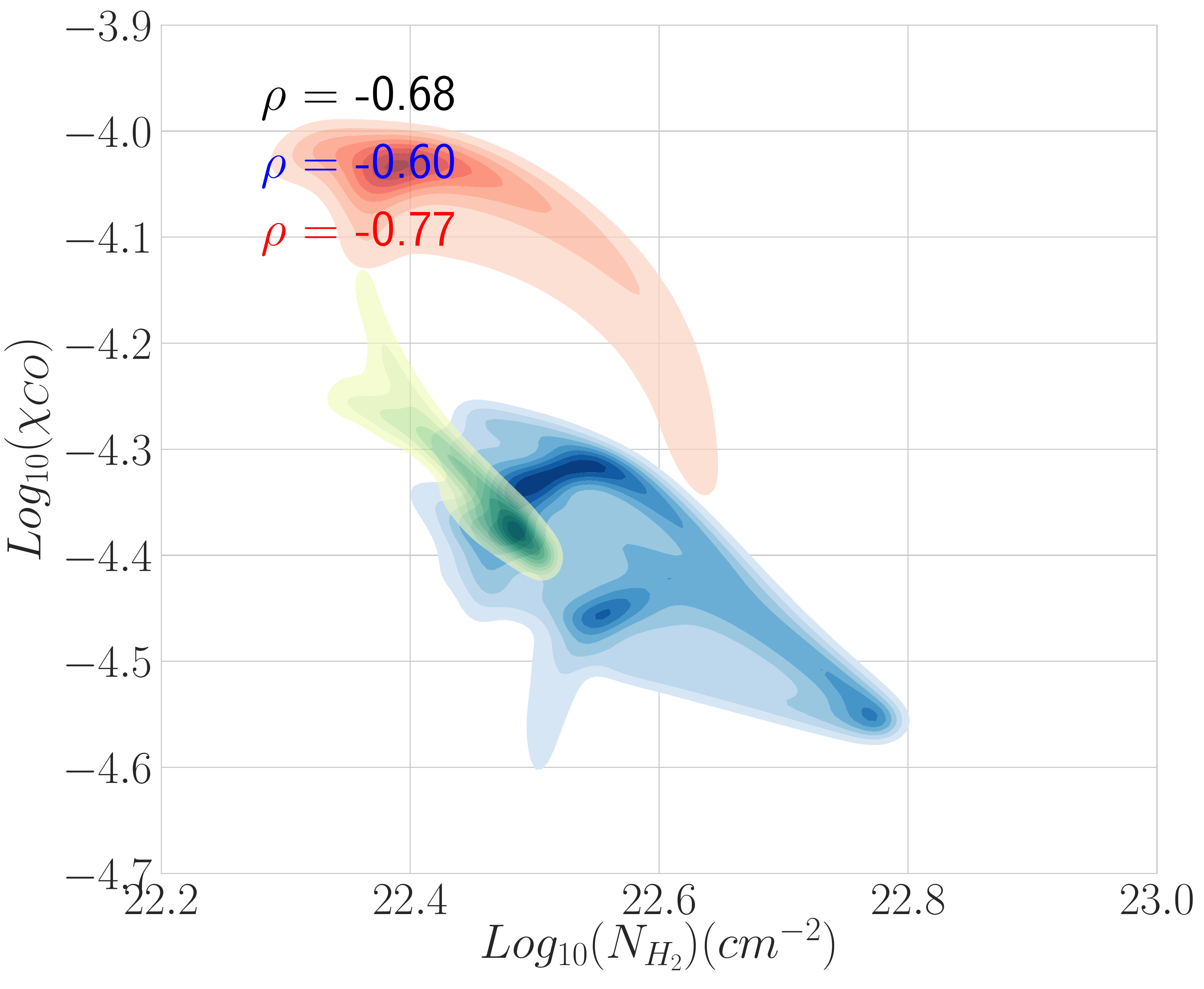}
&\includegraphics[align=c,width=4.8cm] {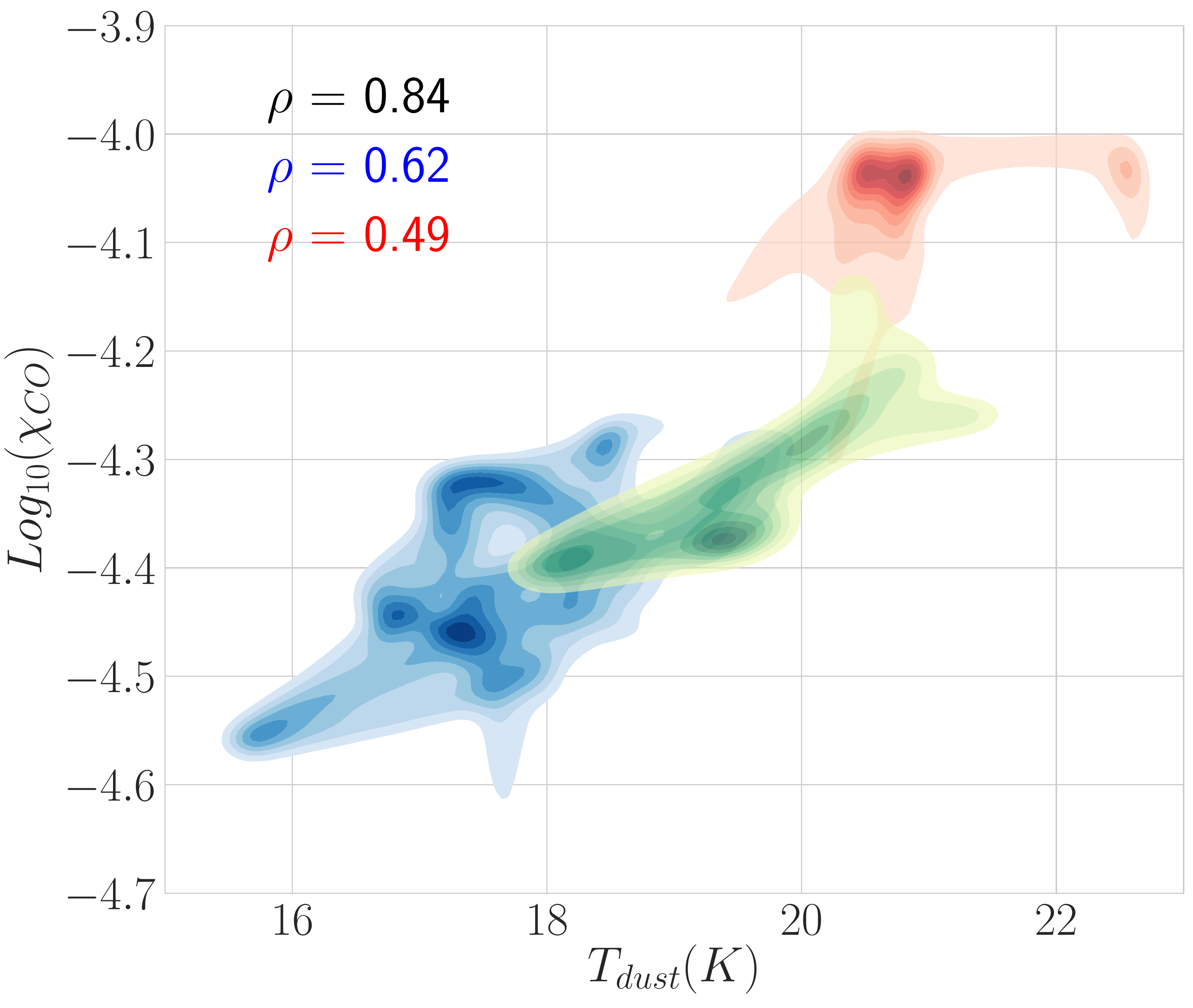}
&\includegraphics[align=c,width=4.8cm] {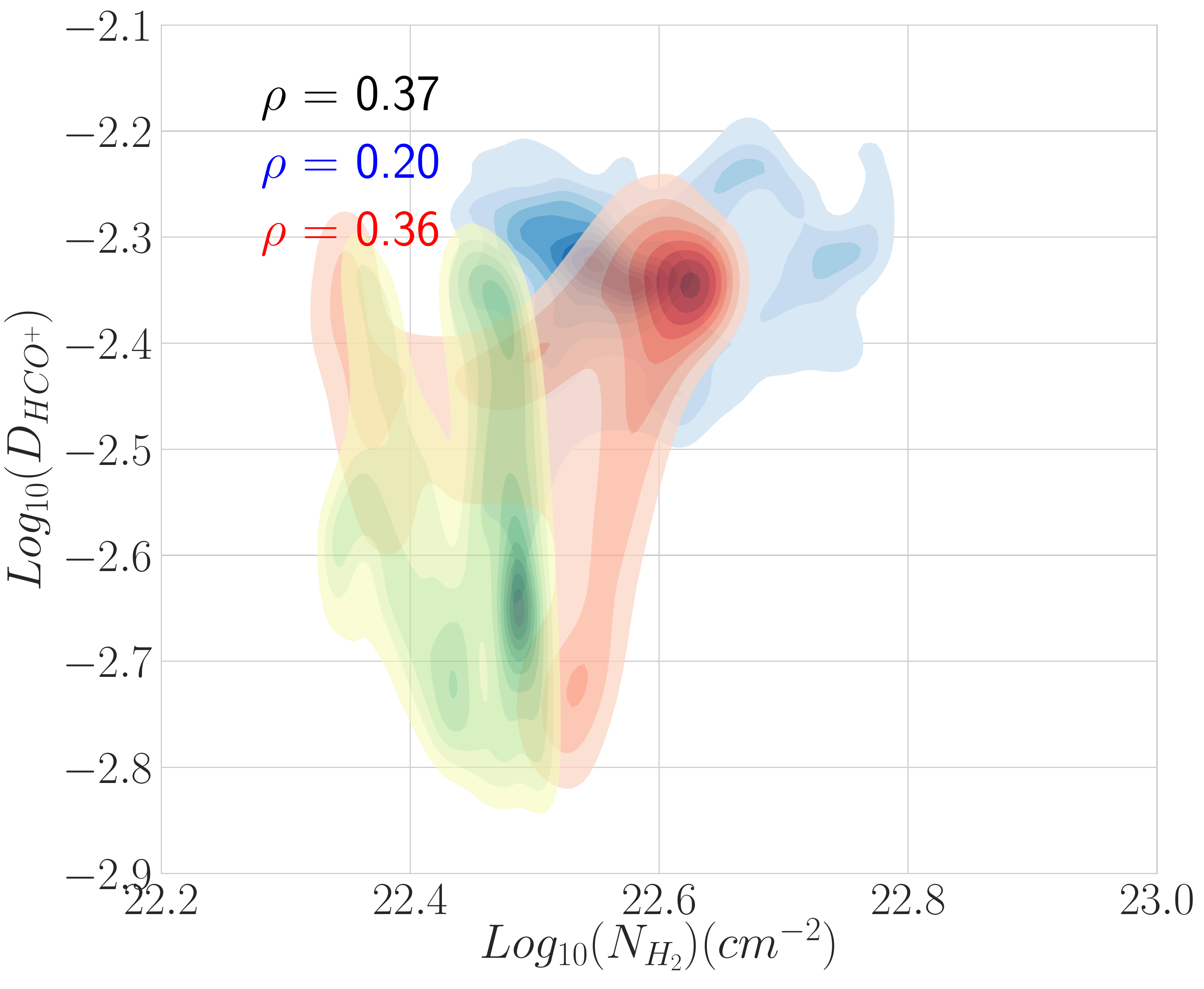}
&\includegraphics[align=c,width=4.8cm] {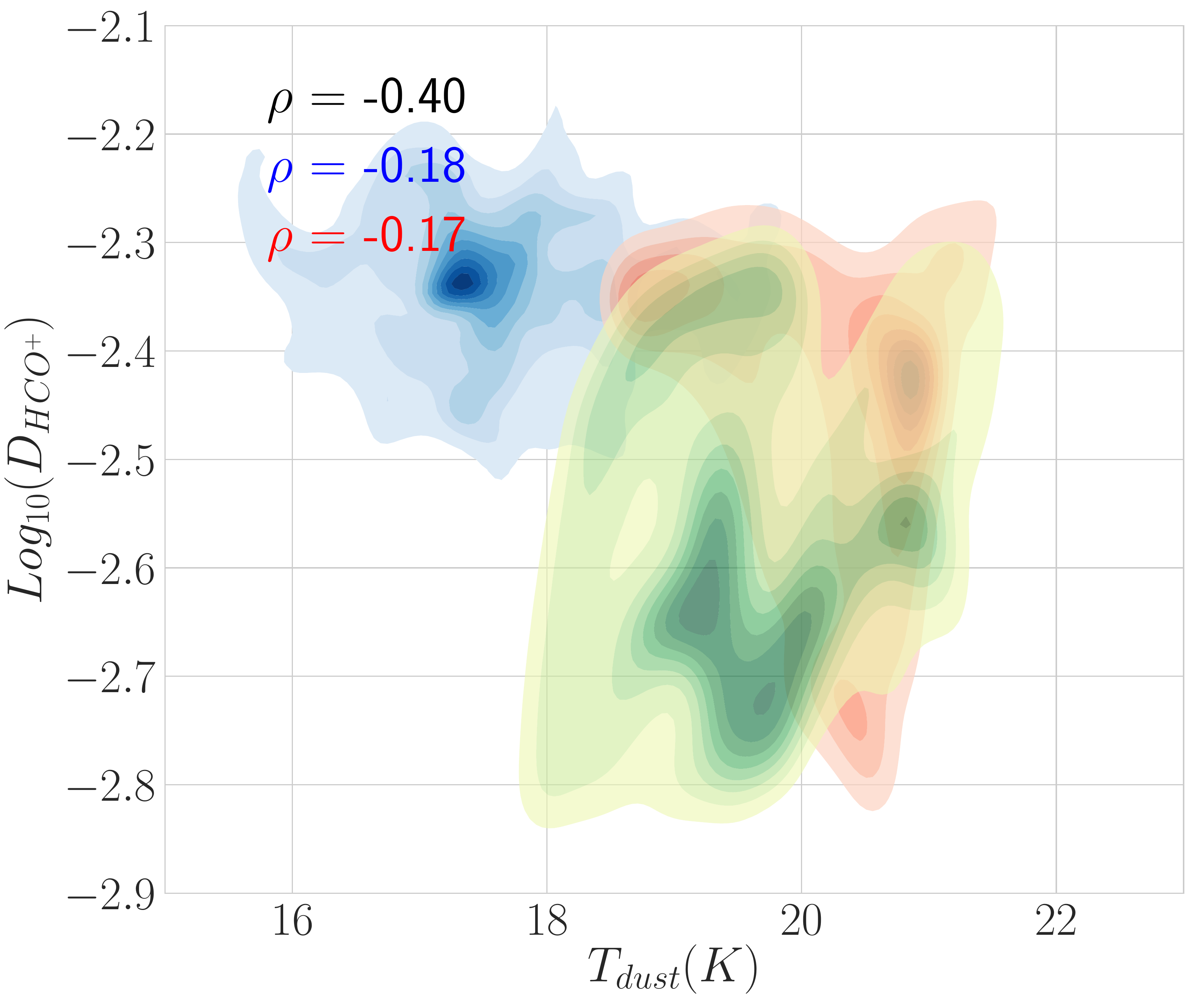}\\[-0.2cm]
  G\,34.74-0.12\\
  \includegraphics[align=c,width=4.8cm] {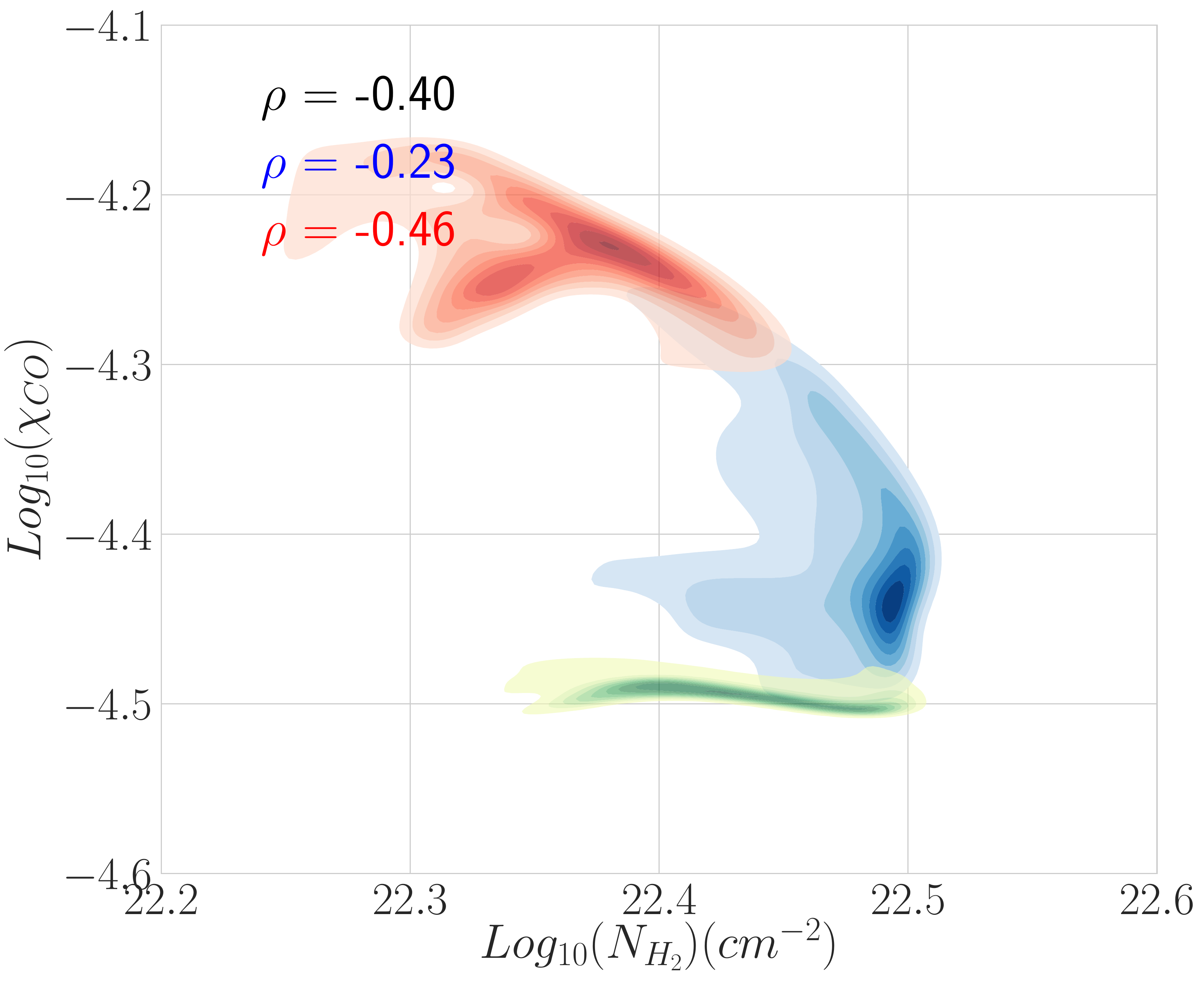}
&\includegraphics[align=c,width=4.8cm] {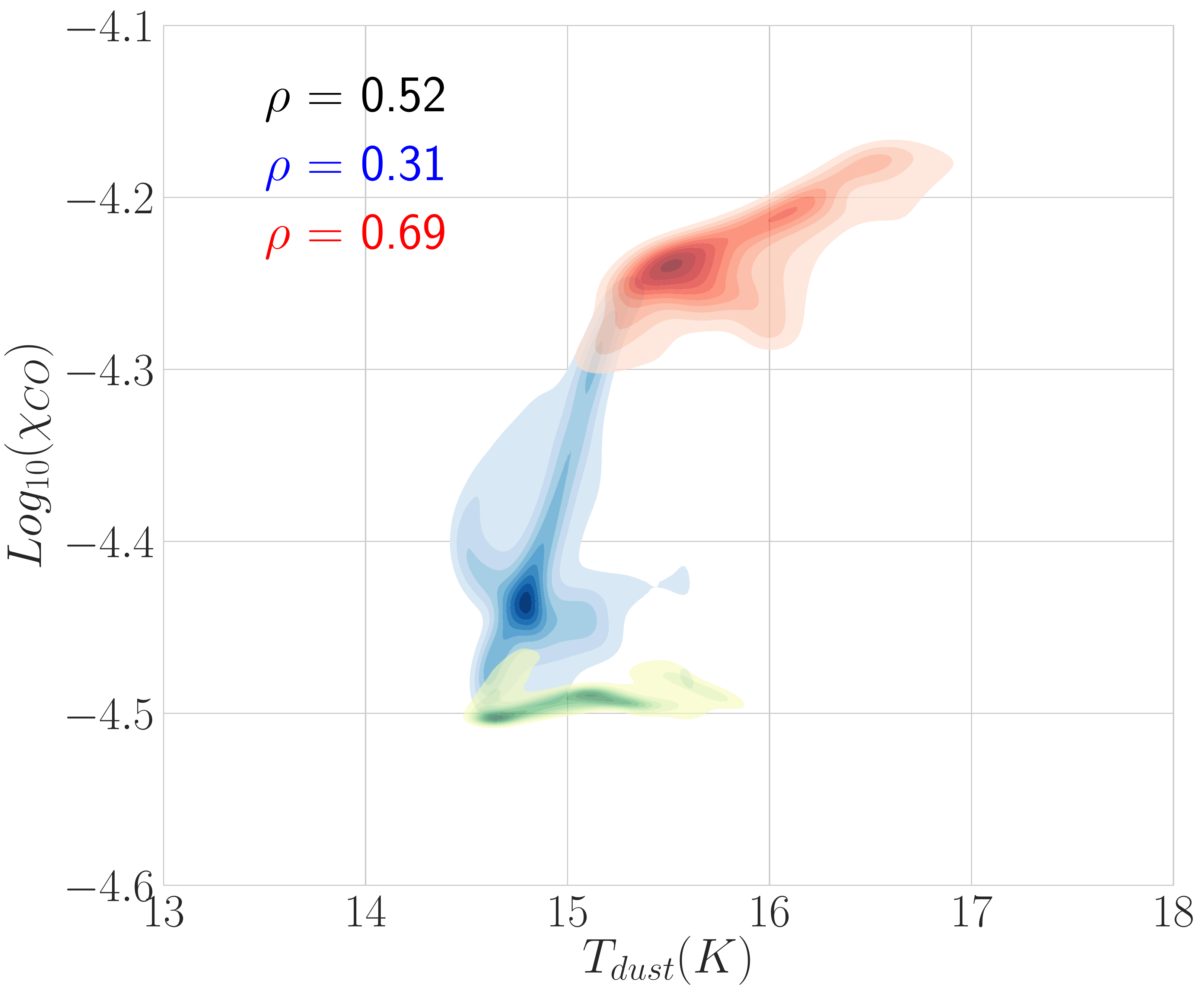}
&\includegraphics[align=c,width=4.8cm] {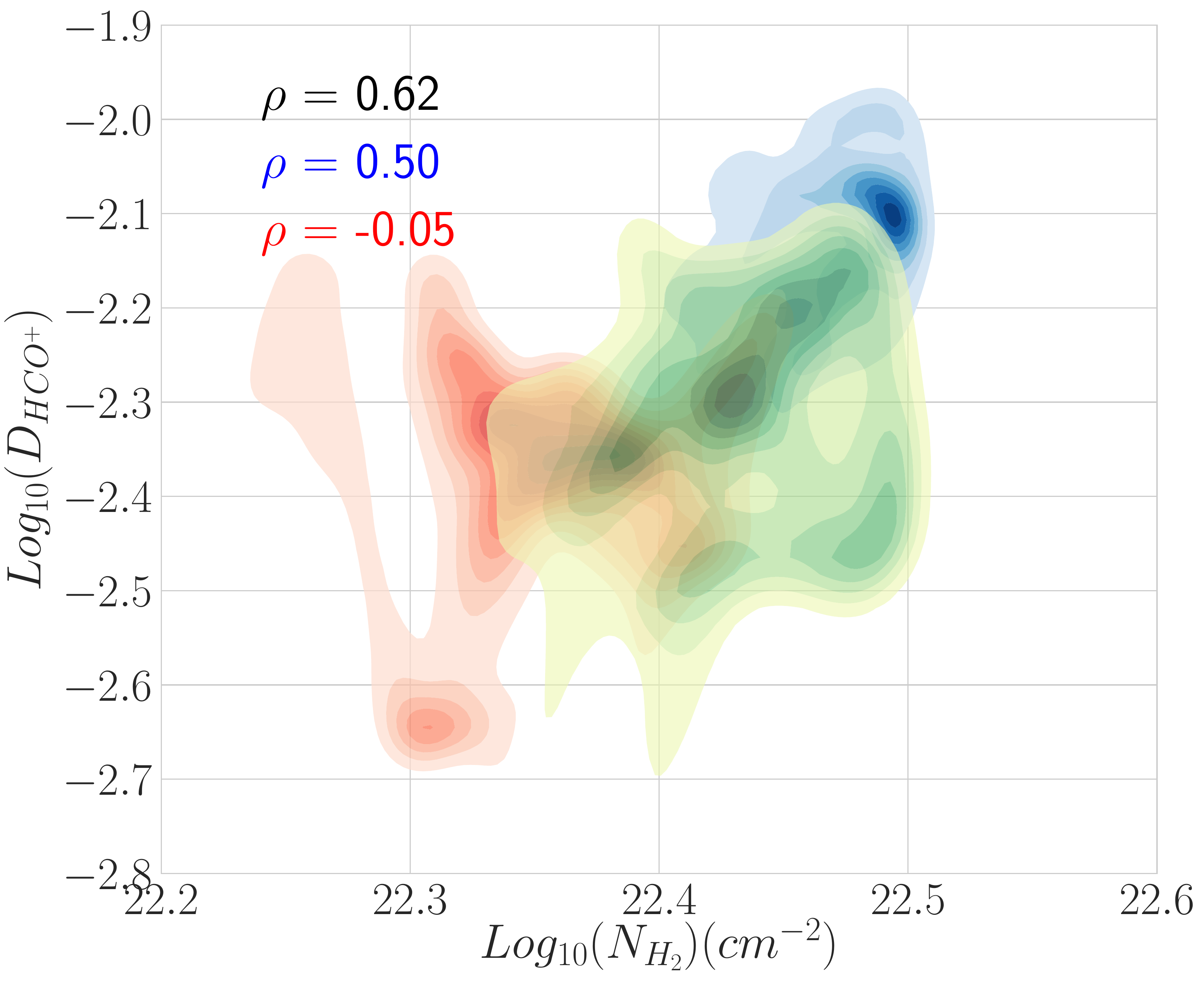}
&\includegraphics[align=c,width=4.8cm] {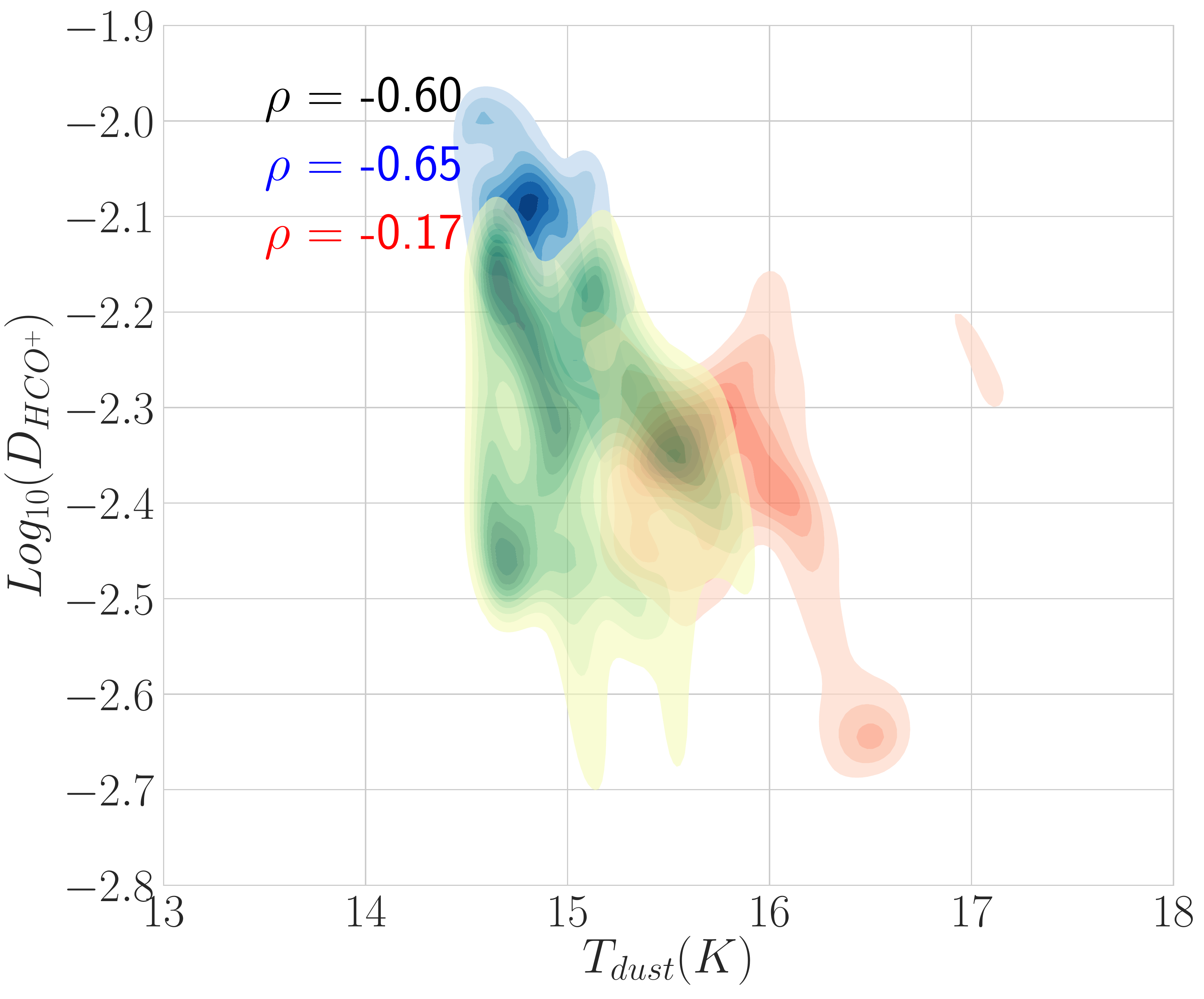}\\[-0.2cm]

\end{tabular}
\caption{Possible correlation between variables and gas density or dust temperature. Values are extracted from pixels after smoothing the parameter maps to the same angular resolution (36\arcsec), and are plotted with a bivariate Gaussian kernel density estimate as contours. {\color{black}The CO- and $\rm DCO^+$-dominant zones are plotted  in red and blue, with the Spearman's rank correlation correlation coefficient $\rho$ given in red and blue, respectively.  The transition zone is plotted in yellowish-green. The  Spearman's rank correlation correlation coefficient $\rho$ of the entire source is given in black. } The pixels where the continuum  at 870\,$\mu$m shows $\rm <5\sigma$ emission or  $\rm DCO^+$\,(1--0)  shows $\rm <3\sigma$ emission are blanked.
%For better comparison with the modeling result (Figure~\ref{fig:model}), we plot parameters  in the same range for all four sources. 
}\label{fig:correlation2}
\end{figure} 

\newpage
\begin{table}
\caption{Sources in our sample and their observation parameters}\label{sourceall}
\scalebox{1.0}{
\begin{tabular}{ccccccccc}
\hline\hline

Source$^a$   & R.A.$^b$      &DEC.$^b$ &$d$$^c$  &$\rm R_{\rm GC}$$^d$ &$\rm V_{sys}$          &$\rm rms_{4.0\,mm}$$^e$          &$\rm rms_{3.4\,mm}$$^f$  &$\rm rms_{1.3\,mm}$$^g$ \\
           &[J2000]  &[J2000]   &(kpc)   &(kpc) &$\rm (km\,s^{-1})$          &(K)                      &(K)                                     &(K)  \\
         \hline
{G\,011.0970-0.1093}    &$\rm 18^h10^m25^s.70$    &$\rm  -19^\circ22^{'}59^{''}.5$    &3.0  &4.9   & 29.8$^h$ &0.05  &0.02 &0.27\\
{G\,011.3811+0.8103}    &$\rm 18^h07^m36^s.41$    &$\rm  -18^\circ41^{'}21^{''}.1$   &2.8   &5.2   & 26.8$^i$  &0.02  &0.02  &0.17\\
%{G\,012.6738-0.0998}    &$\rm 18^h13^m35^s.69$    &$\rm  -17^\circ59^{'}46^{''}.0$   &10.7      & 55.5 \\
{G\,012.9459-0.2488}    &$\rm 18^h14^m41^s.50$    &$\rm  -17^\circ49^{'}41^{''}.9$   &3.0   &5.0   & 34.0$^j$ &0.03 &0.02 &0.28\\
{G\,012.9674-0.2380}    &$\rm 18^h14^m41^s.69$    &$\rm  -17^\circ48^{'}15^{''}.8$   &3.0  &4.9    & 35.0$^k$ &0.04 &0.02 &0.25\\
{G\,014.1842-0.2280}    &$\rm 18^h17^m05^s.14$    &$\rm  -16^\circ43^{'}46^{''}.9$   &3.1   &4.8   & 39.7$^j$ &0.03 &0.04 &0.45\\
{G\,014.2314-0.1758}    &$\rm 18^h16^m59^s.23$    &$\rm  -16^\circ39^{'}47^{''}.9$   &3.0  &4.9    & 37.5$^h$ &0.03 &0.03 &0.23\\
{G\,014.4876-0.1274}     &$\rm 18^h17^m19^s.03$   &$\rm  -16^\circ24^{'}53^{''}.6$   &3.2  &4.9    & 39.7$^h$ &0.03 &0.03  &0.31\\
{G\,014.6858-0.2234}    &$\rm 18^h18^m03^s.67$    &$\rm  -16^\circ17^{'}09^{''}.6$   &3.0 &5.0    & 37.7$^h$ &0.02 &0.02 &0.14 \\
{G\,014.7258-0.2031}    &$\rm 18^h18^m03^s.96$    &$\rm  -16^\circ14^{'}28^{''}.0$   &3.1  &5.0    & 37.5$^h$ &0.04 &0.04 &0.24\\
{G\,015.2169-0.4267}     &$\rm 18^h19^m51^s.19$   &$\rm  -15^\circ54^{'}50^{''}.8$   &1.9 &6.1     & 22.7$^l$ &0.05 &0.04  &0.33\\
{G\,015.5022-0.4201}    &$\rm 18^h20^m23^s.28$    &$\rm  -15^\circ39^{'}33^{''}.8$   &3.2  &5.0    & 39.7$^j$ &0.04 &0.04 &0.37\\
{G\,016.3013-0.5251}    &$\rm 18^h22^m19^s.90$    &$\rm  -15^\circ00^{'}13^{''}.7$   &3.2  &5.2    & 38.3$^h$ &0.05 &0.02 &0.54\\
{G\,018.8008-0.2958}    &$\rm 18^h26^m18^s.94$    &$\rm  -12^\circ41^{'}15^{''}.4$   &5.0  &4.3    & 65.5$^j$ &0.04 &0.02 &0.30 \\
{G\,018.9295-0.0289}     &$\rm 18^h25^m35^s.64$   &$\rm  -12^\circ26^{'}57^{''}.1$   &3.3  &5.2    & 43.6$^m$  &0.06 &0.03 &0.24\\
{G\,022.5309-0.1927}    &$\rm 18^h32^m59^s.64$    &$\rm  -09^\circ20^{'}06^{''}.0$    &5.0 &4.3    & 75.9$^h$ &0.03 &0.03 &0.46\\
{G\,022.6919-0.4519}     &$\rm 18^h34^m13^s.61$   &$\rm  -09^\circ18^{'}42^{''}.5$    &4.9 &4.3    & 76.8$^h$  &0.04 &0.04 &0.26 \\
{G\,022.7215-0.2733}    &$\rm 18^h33^m38^s.38$    &$\rm  -09^\circ12^{'}11^{''}.5$    &4.6  &4.4   & 72.8$^h$ &0.03 &0.02 &0.36\\
{G\,024.5245-0.1397}     &$\rm 18^h36^m30^s.98$   &$\rm  -07^\circ32^{'}28^{''}.0$    &5.7  &4.1   & 90.3$^l$ &0.05 &0.02 &0.23\\
{G\,028.2726-0.1666 }    &$\rm 18^h43^m31^s.18$    &$\rm  -04^\circ13^{'}18^{''}.8$   &4.5  &4.7   & 79.6$^i$   &0.04 &0.03  &0.38\\
{G\,028.3231-0.0676}   &$\rm 18^h42^m46^s.60$    &$\rm  -04^\circ04^{'}11^{''}.9$    &4.6  &4.7    & 79.5$^n$ &0.03 &0.03 &0.11 \\
{G\,028.5246-0.2519}    &$\rm 18^h44^m17^s.14$    &$\rm  -04^\circ02^{'}12^{''}.5$    &4.7 &4.5    & 87.3$^i$ &0.06 &0.03 &0.32 \\
{G\,028.5413-0.2371}    &$\rm 18^h44^m15^s.79$    &$\rm  -04^\circ00^{'}54^{''}.7$    &4.6  &4.6   & 84.3$^i$ &0.04  &0.03 &0.36 \\
%{G\,030.85-0.11}    &$\rm 18^h48^m01^s.85$    &$\rm  -01^\circ54^{'}00^{''}.7$    &6.3      & 99.8 \\
{G\,034.7391-0.1197}     &$\rm 18^h55^m09^s.70$   &$\rm  +01^\circ33^{'}13^{''}.3$   &4.7  &5.2    & 79.0$^h$   &0.02  &0.02 &0.18\\
{G\,034.7798-0.5671}    &$\rm 18^h56^m49^s.73$    &$\rm  +01^\circ23^{'}08^{''}.9$   &2.2 &6.4     & 41.3$^j$ &0.05 &0.03 &0.21\\
\hline\hline

\multicolumn{8}{l}{\color{black} { Note.} $a$. ATLASGAL name.  }\\
\multicolumn{8}{l}{\color{black} ~~~~~~~~~~~$b$. OTF mapping center. }\\
\multicolumn{8}{l}{\color{black} ~~~~~~~~~~~$c$. Kinematic distance from \citet{yuan17}, with an uncertainty of $\rm \pm 0.5$\,kpc. }\\
\multicolumn{8}{l}{\color{black} ~~~~~~~~~~~$d$. Galactocentric distance, calculated by using \citet{wenger18}.}\\
\multicolumn{8}{l}{\color{black} ~~~~~~~~~~~$e$. Measured by IRAM 30\,m in main-beam temperature  $\rm T_{mb}$ (K)  directly from observations without smoothing, }\\
\multicolumn{8}{l}{\color{black} ~~~~~~~~~~~ with an angular resolution of $\sim$36\arcsec~ and velocity resolution of $\sim$0.72\,$\rm km\,s^{-1}$   for 4.0\,mm lines.}\\
\multicolumn{8}{l}{\color{black} ~~~~~~~~~~~$f$. Measured by IRAM 30\,m in main-beam temperature  $\rm T_{mb}$ (K)  directly from observations without smoothing, }\\
\multicolumn{8}{l}{\color{black} ~~~~~~~~~~~  with an angular resolution of $\sim$29\arcsec~ and velocity resolution of $\sim$0.56\,$\rm km\,s^{-1}$   for 3.4\,mm lines.}\\
\multicolumn{8}{l}{\color{black} ~~~~~~~~~~~$g$. Measured by IRAM 30\,m in main-beam temperature  $\rm T_{mb}$ (K) directly from observations without smoothing, }\\
\multicolumn{8}{l}{\color{black} ~~~~~~~~~~~with an angular resolution of $\sim$11\arcsec~ and velocity resolution of $\sim$0.22\,$\rm km\,s^{-1}$   for 1.3\,mm lines.}\\
\multicolumn{8}{l}{\color{black} ~~~~~~~~~~~$h$. \citet{wienen12}.}\\
\multicolumn{8}{l}{\color{black} ~~~~~~~~~~~$i$. \citet{csengeri14}. }\\
\multicolumn{8}{l}{\color{black} ~~~~~~~~~~~$j$. \citet{shirley13}. }\\
\multicolumn{8}{l}{\color{black} ~~~~~~~~~~~$k$. Single-pointed observation using Submillimeter Telescope (SMT) \citep{yuan17}. }\\
\multicolumn{8}{l}{\color{black} ~~~~~~~~~~~$l$. \citet{dempsey13}.}\\
\multicolumn{8}{l}{\color{black} ~~~~~~~~~~~$m$. \citet{purcell12}.}\\
\multicolumn{8}{l}{\color{black} ~~~~~~~~~~~$n$.Pilot study source, with line information given in \citet{feng19a}.}\\

\end{tabular}
}
\end{table}

\begin{table*}[]
%\scalebox{0.9}{
\caption{Targeted lines covered by our IRAM 30\,m and  NRO 45\,m observations
}\label{tab:lines}
\centering
\begin{tabular}{llllllllllll}
%\begin{longtable*}{lllllll}

\hline\hline
Mol. &Freq.  &Transition    &$\rm S\mu^2$$^a$         &$\rm E_{\it u}/k_B$$^a$   &\multicolumn{2}{c}{$n_{crit}^d(\rm cm^{-3})$}  &\multicolumn{3}{c}{$n_{eff}^e(\rm cm^{-3})$} &Telescope &Beam   \\
         &(GHz) &                  &($\rm D^2$)             &  (K)                                                                   &10\,K  &20\,K                                             &10\,K &15\,K &20\,K                                       &    &\\

\hline
HCN   &88.632      &J=1--0$^c$   & 26.8    &4.2                    &4.7E+5 &3.0E+5  &8.4E+3 &5.6E+3 &4.5E+3 &IRAM 30\,m &29.3\\
$\rm H^{13}CN$   &86.340      &J=1--0$^c$   & 26.7    &4.1  &4.3E+5 &2.7E+5   &3.5E+5 &2.2E+5  &1.6E+5 &IRAM 30\,m &30.0\\
$\rm HC^{15}N$   &86.055     &1--0   & 8.9    &4.1                &4.3E+5 &2.7E+5   &$--$&$--$&$--$ &IRAM 30\,m &30.1\\
DCN  &72.415    &J=1--0$^c$    & 26.8    &3.5                      &2.6E+5 &1.6E+5   &$--$&$--$&$--$ &IRAM 30\,m &35.8\\
\hline
HNC &90.664     &1--0   & 9.3    &4.3 				  &1.4E+5 &1.1E+5    &3.7E+3 &2.7E+3 &2.3E+3 &IRAM 30\,m   &28.6\\
$\rm HN^{13}C$   &87.091    &1--0      &7.3    &4.2  		  &9.6E+4 &7.3E+4  &$--$&$--$&$--$ &IRAM 30\,m &29.8\\
$\rm H^{15}NC$  &88.866     &1--0     &7.3    &4.3  		  &1.0E+5 &7.8E+4  &$--$&$--$&$--$ &IRAM 30\,m  &29.2\\
DNC   &76.306     &1--0    & 9.3    &3.7  				  &8.2E+4 &6.3E+4  &$--$&$--$&$--$ &IRAM 30\,m &34.0\\
\hline
$\rm HCO^+$  &89.189    &1--0   &15.2   &4.2  			  &7.0E+4 &4.7E+4  &9.5E+2 &6.4E+2 &5.3E+2 &IRAM 30\,m  &29.1\\
$\rm H^{13}CO^+$  &86.754    &J=1--0$^b$     &15.2   &4.2  &6.2E+4 &4.1E+4 &3.9E+4 &2.7E+4 &2.2E+4 &IRAM 30\,m &29.9\\
$\rm HC^{18}O^+$ &	85.162   &1--0       &15.2   &4.1           &4.2E+4 &2.8E+4  &$--$&$--$&$--$ &IRAM 30\,m  &30.5\\
$\rm DCO^+$  &72.039    &J=1--0$^b$  & 14.5   &3.5          &3.2E+4 &2.1E+4  &$--$&$--$&$--$ &IRAM 30\,m  &36.0\\
\hline
$\rm N_2H^+$ &93.173     &J=1--0$^c$   &104.0     &4.5     &6.1E+4 &4.1E+4  &1.0E+4 &6.7E+3 &5.5E+3 &IRAM 30\,m &27.8\\
$\rm N_2D^+$ &77.109    &J=1--0$^c$  &104.0   &3.7         &5.9E+4 &3.9E+4   &$--$&$--$&$--$ &IRAM 30\,m &33.6\\
\hline

$\rm C^{18}O$    &219.560        &$\rm 2\text{--}1$           &0.02      &15.8     &4.7E+3 &3.8E+3   &$--$&$--$&$--$   &IRAM 30\,m &11.8\\
$\rm ^{13}CO$    &220.400         &$\rm 2\text{--}1$          &0.02      &15.9     &4.8E+3 &3.8E+3   &$--$&$--$&$--$   &IRAM 30\,m &11.8\\
$\rm C^{17}O$    &224.714         &$\rm 2\text{--}1$          &0.02      &16.2     &5.1E+3 &4.1E+3   &$--$&$--$&$--$    &IRAM 30\,m &11.5\\
$\rm C^{18}O$     &109.782       &$\rm 1\text{--}0$           &0.01      &5.3       &7.5E+2 &4.8E+2   &$--$&$--$&$--$  &NRO 45\,m &16.4\\
$\rm ^{13}CO$     &110.201         &$\rm 1\text{--}0$         &0.01      &5.3       &7.6E+2 &4.8E+2  &$--$&$--$&$--$   &NRO 45\,m &16.4\\
$\rm C^{17}O$     &112.359        &$\rm 1\text{--}0$          &0.01      &5.4       &8.2E+2 &5.2E+2   &$--$&$--$&$--$  &NRO 45\,m &16.1\\
\hline
$\rm  H_2CO$  &  72.838 &$\rm 1_{0,1}\text{--}0_{0,0} $  &5.4  &3.5    &4.5E+4  &2.8E+4  &5.0E+4 &3.2E+4 &2.6E+4  &IRAM 30\,m &35.6\\
$\rm  H_2CO$  &  218.222 &$\rm 3_{0,3}\text{--}2_{0,2} $  &16.3   &21.0 &9.7E+5&7.8E+5  &1.5E+5 &8.2E+4 &6.3E+4  &IRAM 30\,m &11.9\\
$\rm  H_2CO$  &  218.476 &$\rm 3_{2,2}\text{--}2_{2,1}  $  &9.1    &68.1 &3.5E+5&3.2E+5 &$--$&$--$&$--$ &IRAM 30\,m &11.9\\
$\rm  H_2CO$  &  218.760 &$\rm 3_{2,1}\text{--}2_{2,0} $   &9.1    &68.1 &3.5E+5&3.2E+5 &$--$&$--$&$--$ &IRAM 30\,m &11.9\\
\hline
$\rm CH_3OH$  &76.510  &$\rm 5_{0 , 5}\text{--}4_{1 , 3}\,E$   &1.9 &47.9   &2.5E+3 &2.1E+3 &$--$&$--$&$--$  &IRAM 30\,m &33.9\\
$\rm CH_3OH$ &84.521   &$\rm 5_{-1, 5}\text{--}4_{0 , 4}\,E$ &3.1    &40.4  &6.1E+3 &5.1E+3 &$--$&$--$&$--$ &IRAM 30\,m  &30.7\\
$\rm CH_3OH$ &218.440  &$\rm 4_{2 , 2}\text{--}3_{1 , 2}\,E$  &3.5 &45.5   &1.6E+5 &1.3E+5 &$--$&$--$&$--$  &IRAM 30\,m &11.9\\

$\rm CH_2DOH$   &89.408    &$\rm 2_{0,2}\text{--}1_{0,1}\,e0$ &1.2  &6.4   &4.1E+4 &3.1E+4  &$--$&$--$&$--$  &IRAM 30\,m &29.0\\
\hline
%%%%%%%%%%%%%%%%%%%%%%%%%%%%%

$\rm OCS$           &85.139        &$\rm 7\text{--}6$                               &3.6          &16.3          &4.1E+3 &3.3E+3 &$--$&$--$&$--$  &IRAM 30\,m &30.5\\
c-$\rm C_3H_2$   &85.339        &$\rm 2_{1,2}\text{--}1_{0,1}$           &48.1        &6.4           &3.2E+5 &1.6E+5 &$--$&$--$&$--$  &IRAM 30\,m &30.4\\                
$\rm CH_3C_2H$   &85.456        &$\rm 5_1\text{--}4_1$                    &1.8         &19.5          &$--$&$--$ &$--$&$--$&$--$  &IRAM 30\,m &30.3\\
$\rm CH_3C_2H$   &85.457        &$\rm 5_0\text{--}4_0$                    &1.9         &12.3          &$--$&$--$ &$--$&$--$&$--$   &IRAM 30\,m &30.3\\        
$\rm NH_2D$  &85.926   	&$\rm 1_{1,1}\,0s\text{--}1_{0,1}\,0a$        &28.6        &20.7         &5.6E+4 &4.8E+4 &$--$&$--$&$--$  &IRAM 30\,m &30.2\\
$\rm SiO$               &86.847        &$\rm 2\text{--}1$                            &19.2        &6.3            &6.7E+4&5.4E+4 &$--$&$--$&$--$  &IRAM 30\,m &29.9\\
$\rm C_2H$       &87.284         &$\rm N=1\text{--}0, J=3/2\text{--}1/2, F=1\text{--}1$        &0.1    &4.2       &4.2E+3 &2.4E+3 &$--$&$--$&$--$  &IRAM 30\,m  &29.7\\
$\rm C_2H$       &87.317         &$\rm N=1\text{--}0, J=3/2\text{--}1/2, F=2\text{--}1$        &1.0    &4.2       &2.7E+4 &1.5E+4 &$--$&$--$&$--$  &IRAM 30\,m   &29.7\\
$\rm C_2H$       &87.329         &$\rm N=1\text{--}0, J=3/2\text{--}1/2, F=1\text{--}0$        &0.5    &4.2       &2.1E+4 &1.2E+4 &$--$&$--$&$--$ &IRAM 30\,m    &29.7\\
$\rm C_2H$       &87.402         &$\rm N=1\text{--}0, J=1/2\text{--}1/2, F=1\text{--}1$        &0.5    &4.2       &1.7E+4 &1.0E+4 &$--$&$--$&$--$   &IRAM 30\,m   &29.7\\
$\rm C_2H$       &87.407         &$\rm N=1\text{--}0, J=1/2\text{--}1/2, F=0\text{--}1$        &0.2    &4.2       &1.8E+4 &1.2E+4 &$--$&$--$&$--$  &IRAM 30\,m   &29.7\\
$\rm C_2H$       &87.446         &$\rm N=1\text{--}0, J=1/2\text{--}1/2, F=1\text{--}0$        &0.1    &4.2        &3.5E+3 &2.1E+3 &$--$&$--$&$--$   &IRAM 30\,m    &29.7\\
$\rm HNCO$       &87.925       &$\rm 4_{0,4}\text{--}3_{0,3}$            &30.8          &10.5               &8.7E+4 &5.3E+4  &$--$&$--$&$--$    &IRAM 30\,m  &29.5\\
$\rm HC_3N$      &90.979         &$\rm 10\text{--}9$                   &138.7          &24.0                  &1.6E+5 &1.2E+5  &4.3E+5 &7.2E+4 &4.3E+4  &IRAM 30\,m  &28.5\\
$\rm ^{13}CS$    &92.494         &$\rm 2\text{--}1$                     &15.3             &6.7                    &3.0E+5 &2.3E+5  &$--$&$--$&$--$   &IRAM 30\,m  &28.0\\

%%%%%%%%%%%%%%%%%%%%%%%%%%%%%

\hline \hline
\multicolumn{12}{l}{{ Note.} $^a$ Line spectroscopic parameters are given according to catalogs including the}\\ 
\multicolumn{12}{l}{~~~ ~~~ ~~JPL \citep[][\url{http://spec.jpl.nasa.gov}]{pickett98} and CDMS \citep[][\url{http://www.astro.uni-koeln.de/cdms/catalog}]{muller05};}\\
\multicolumn{12}{l}{~~~ ~~~ ~~$^b$ Hyperfine splittings are recorded in JPL and CDMS but not resolved in our observations,}\\
\multicolumn{12}{l}{ ~~~ ~~~  ~~~ so the sum of $\rm S\mu^2$ is used for the rotational transitions to calculate the total column density;}\\
\multicolumn{12}{l}{~~~ ~~~ ~~$^c$ Hyperfine splittings are resolved in our observations, and only the sum of $\rm S\mu^2$ is  needed for the  }\\
\multicolumn{12}{l}{ ~~~ ~~~  ~~~ rotational transitions to calculate the total column density;}\\
\multicolumn{12}{l}{~~~ ~~~  ~~$^d$ The critical density of each transition $n_{crit}$ is derived from the Einstein coefficient $A_{ij}$ and the collision rate $C_{ij}$ at 10--20\,K given by }\\
\multicolumn{12}{l}{ ~~~ ~~~  ~~~  LAMDA \citep{schoier05}. We assume that the deuterated lines have the same $C_{ij}$ as their hydrogenated counterparts;}\\
\multicolumn{12}{l}{~~~ ~~~  ~~$^e$ The effective excitation density at kinetic temperatures of 10--20\,K from \citet{shirley15}; ``$--$" indicates an unrecorded value.}\\
%\multicolumn{9}{l}{ ~~~ ~~~  ~~~   ``$--$" labels the non-recorded value;}\\
%\multicolumn{9}{l}{ ~~~ ~~~  ~~~   $^g$ ``N" represents the observations carried by using NRO 45\,m, and  ``I" represents the observations carried by  IRAM 30\,m.}\\
\end{tabular}
%}
%\end{longtable*}
\end{table*}

\begin{table}
\caption{The best-fit parameters (Linewidth $\rm \Delta \upsilon\,(km~s^{-1})$ and integrated intensity $\rm \int T_B(\upsilon)d\upsilon\,(K\, km~s^{-1})$) for lines in Figure \ref{spec1}, given by GILDAS.
}\label{tab:linedeu}
\scalebox{0.8}{
\centering
\begin{tabular}{c|c|c|c|p{1.5cm}p{1.5cm}|p{1.5cm}p{1.5cm}|p{1.5cm}p{1.5cm}|cc}
\hline\hline

\hline
Source
&Line$^a$  &Freq $^b$    &$\theta$$^c$   &\multicolumn{2}{c|}{P1$^d$}   &\multicolumn{2}{c|}{P2$^d$}    &\multicolumn{2}{c|}{P3$^d$}  &Velocity range$^e$    &$\sigma$$^f$\\
&&(GHz)  &(\arcsec)   &$\rm (km~s^{-1})$      &$\rm (K\,km~s^{-1})$  &$\rm (km~s^{-1})$    &$\rm (K\,km~s^{-1})$   &$\rm (km~s^{-1})$    &$\rm (K\,km~s^{-1})$    &$\rm [km~s^{-1},km~s^{-1}]$ &$\rm (K\,km~s^{-1})$\\
\hline
 \hline

 \multirow{5}{*}{G\,15.22-0.43}
&$\rm   ^{13}CO \, (2-1) $ &  220.398   &11.8  &$\rm     2.1 \pm     0.0 $    &$\rm   49.08 \pm    0.16 $   %  &$\rm    20.9 \pm     0.0 $\n
&$\rm     1.5 \pm     0.0 $    &$\rm   23.72 \pm    0.27 $   %  &$\rm    22.0 \pm     0.0 $\n
 &$\rm     1.7 \pm     0.0 $    &$\rm   47.65 \pm    0.21 $   %  &$\rm    20.2 \pm     0.0 $\n
 &[14, 32]  &1.41\\ %  &$\rm    80.0 \pm     0.0 $\n
&$\rm   C^{18}O \, (2-1) $ &  219.560  &11.8 &$\rm     1.6 \pm     0.0 $    &$\rm   10.95 \pm    0.11 $   %  &$\rm    20.9 \pm     0.0 $\n
&$\rm     0.9 \pm     0.0 $    &$\rm    4.41 \pm    0.09 $   %  &$\rm    22.0 \pm     0.0 $\n
&$\rm     1.3 \pm     0.0 $    &$\rm    8.88 \pm    0.16 $   %  &$\rm    20.2 \pm     0.0 $\n
 &[14, 32]  &1.16\\%  &$\rm    79.8 \pm     0.1 $\n
&$\rm  C^{17}O \, (2-1) $ &   224.714  &11.5  &$\rm     2.1 \pm     0.1 $    &$\rm    3.39 \pm    0.10 $   %  &$\rm    20.9 \pm     0.0 $\n
&$\rm     1.6 \pm     0.2 $    &$\rm    1.28 \pm    0.10 $   %  &$\rm    22.0 \pm     0.1 $\n
&$\rm     1.8 \pm     0.1 $    &$\rm    2.62 \pm    0.15 $   %  &$\rm    20.3 \pm     0.0 $\n
 &[14, 32]  &0.94\\%  &$\rm    80.1 \pm     0.0 $\n
%$\rm  HCO^+ \, (1-0) $ &  89.189   &$\rm     3.6 \pm     0.1 $    &$\rm    8.56 \pm    0.20 $   %  &$\rm    77.3 \pm     0.0 $\n
%&$\rm     2.8 \pm     0.1 $    &$\rm    4.62 \pm    0.13 $  &[74, 83] \\%  &$\rm    77.8 \pm     0.0 $\n
&$\rm  H^{13}CO^+ \, (1-0) $ &  86.754  &29.9  &$\rm     1.8 \pm     0.1 $    &$\rm    1.01 \pm    0.03 $   %  &$\rm    20.9 \pm     0.0 $\n
&$\rm     1.3 \pm     0.1 $    &$\rm    0.65 \pm    0.03 $   %  &$\rm    21.9 \pm     0.0 $\n
&$\rm     1.7 \pm     0.1 $    &$\rm    0.73 \pm    0.04 $   %  &$\rm    20.4 \pm     0.1 $\n
 &[14, 32]  &0.35\\%  &$\rm    79.5 \pm     0.0 $\n
&$\rm  DCO^+ \, (1-0) $ &72.039 &36.0   &$\rm     1.9 \pm     0.3 $    &$\rm    0.32 \pm    0.04 $   %  &$\rm    21.1 \pm     0.1 $\n
 &$\rm     1.3 \pm     0.1 $    &$\rm    0.38 \pm    0.04 $   %  &$\rm    22.0 \pm     0.1 $\n
 &$\rm     -- $$^g$    &$\rm   \sigma=0.04 $$^g$   %  &$\rm    27.4 \pm     0.0 $\n
 &[14, 32]   &0.09\\%  &$\rm    79.7 \pm     0.0 $\n
  &$\rm   ^{13}CO \, (1-0) $ &110.201 &16.4  &$\rm     2.1 \pm     0.0 $    &$\rm   32.14 \pm    0.26 $   %  &$\rm    20.5 \pm     0.0 $\n
  &$\rm     1.6 \pm     0.0 $    &$\rm   19.04 \pm    0.26 $   %  &$\rm    21.5 \pm     0.0 $\n
  &$\rm     1.8 \pm     0.0 $    &$\rm   23.67 \pm    0.27 $   %  &$\rm    19.9 \pm     0.0 $\n
  &[14, 32] &8.57 \\
 &$\rm   C^{18}O \, (1-0) $ &109.783  &16.4   &$\rm     1.9 \pm     0.0 $    &$\rm    4.39 \pm    0.10 $   %  &$\rm    21.2 \pm     0.0 $\n
  &$\rm     1.1 \pm     0.0 $    &$\rm    2.33 \pm    0.08 $   %  &$\rm    22.3 \pm     0.0 $\n
  &$\rm     1.7 \pm     0.1 $    &$\rm    2.54 \pm    0.10 $   %  &$\rm    20.6 \pm     0.0 $\n
  &[14, 32]  &0.67\\
 &$\rm   C^{17}O \, (1-0) $ &112.359  &16.0 &$\rm     1.8 \pm     0.6 $    &$\rm    0.99 \pm    0.19 $   %  &$\rm    21.6 \pm     0.1 $\n
 &$\rm     0.9 \pm     0.3 $    &$\rm    0.31 \pm    0.10 $   %  &$\rm     3.5 \pm     0.1 $\n
 &$\rm     4.4 \pm     1.1 $    &$\rm    0.79 \pm    0.18 $   %  &$\rm    18.7 \pm     0.5 $\n
  &[14, 32]  &1.43\\
 \hline

 \multirow{5}{*}{G\,11.38+0.81}
&$\rm   ^{13}CO \, (2-1) $ &  220.398   &11.8  &$\rm     3.9 \pm     0.1 $    &$\rm   11.28 \pm    0.16 $   %  &$\rm    27.6 \pm     0.0 $\n
&$\rm     3.6 \pm     0.1 $    &$\rm    9.14 \pm    0.15 $   %  &$\rm    26.8 \pm     0.0 $\n  
&$\rm     3.4 \pm     0.1 $    &$\rm    9.77 \pm    0.14 $   %  &$\rm    27.7 \pm     0.0 $\n
 &[21, 33]  &0.55\\ %  &$\rm    80.0 \pm     0.0 $\n
&$\rm   C^{18}O \, (2-1) $ &  219.560 &11.8  &$\rm     1.9 \pm     0.0 $    &$\rm    6.02 \pm    0.10 $   %  &$\rm    28.0 \pm     0.0 $\n
&$\rm     1.8 \pm     0.1 $    &$\rm    3.59 \pm    0.11 $   %  &$\rm    27.2 \pm     0.0 $\n
&$\rm     1.9 \pm     0.1 $    &$\rm    3.86 \pm    0.10 $   %  &$\rm    27.6 \pm     0.0 $\n
 &[21, 33]    &0.59\\%  &$\rm    79.8 \pm     0.1 $\n
&$\rm  C^{17}O \, (2-1) $ &   224.714 &11.5  &$\rm     2.3 \pm     0.1 $    &$\rm    2.13 \pm    0.10 $   %  &$\rm    27.9 \pm     0.1 $\n
&$\rm     2.1 \pm     0.2 $    &$\rm    1.13 \pm    0.08 $   %  &$\rm    27.2 \pm     0.1 $\n
&$\rm     2.4 \pm     0.3 $    &$\rm    1.07 \pm    0.10 $   %  &$\rm    27.4 \pm     0.1 $\n
 &[21, 33]    &0.53\\%  &$\rm    80.1 \pm     0.0 $\n
%$\rm  HCO^+ \, (1-0) $ &  89.189   &$\rm     3.6 \pm     0.1 $    &$\rm    8.56 \pm    0.20 $   %  &$\rm    77.3 \pm     0.0 $\n
%&$\rm     2.8 \pm     0.1 $    &$\rm    4.62 \pm    0.13 $  &[74, 83] \\%  &$\rm    77.8 \pm     0.0 $\n
&$\rm  H^{13}CO^+ \, (1-0) $ &  86.754  &29.9  &$\rm     1.8 \pm     0.1 $    &$\rm    0.74 \pm    0.03 $   %  &$\rm    27.9 \pm     0.0 $\n
&$\rm     2.0 \pm     0.1 $    &$\rm    0.57 \pm    0.02 $   %  &$\rm    27.0 \pm     0.0 $\n
&$\rm     1.9 \pm     0.1 $    &$\rm    0.54 \pm    0.02 $   %  &$\rm    27.5 \pm     0.0 $\n
 &[21, 33]    &0.13\\%  &$\rm    79.5 \pm     0.0 $\n
&$\rm  DCO^+ \, (1-0) $ &72.039  &36.0  &$\rm     2.0 \pm     0.1 $    &$\rm    0.60 \pm    0.04 $   %  &$\rm    27.7 \pm     0.1 $\n
&$\rm     1.6 \pm     0.1 $    &$\rm    0.41 \pm    0.02 $   %  &$\rm    27.1 \pm     0.0 $\n
 &$\rm     1.7 \pm     0.1 $    &$\rm    0.36 \pm    0.02 $   %  &$\rm    27.4 \pm     0.0 $\n
 &[21, 33]     &0.08\\%  &$\rm    79.7 \pm     0.0 $\n
  &$\rm   ^{13}CO \, (1-0) $ &110.201 &16.4 &$\rm     3.5 \pm     0.0 $    &$\rm   17.71 \pm    0.12 $   %  &$\rm    27.3 \pm     0.0 $\n
  &$\rm     3.1 \pm     0.0 $    &$\rm   13.45 \pm    0.13 $   %  &$\rm    26.8 \pm     0.0 $\n
  &$\rm     3.4 \pm     0.0 $    &$\rm   14.97 \pm    0.11 $   %  &$\rm    27.1 \pm     0.0 $\n
  &[21, 33] &11.18 \\
 &$\rm   C^{18}O \, (1-0) $ &109.783  &16.4  &$\rm     2.0 \pm     0.0 $    &$\rm    7.31 \pm    0.10 $   %  &$\rm    28.2 \pm     0.0 $\n
 &$\rm     1.7 \pm     0.0 $    &$\rm    4.60 \pm    0.08 $   %  &$\rm    27.5 \pm     0.0 $\n
 &$\rm     1.9 \pm     0.0 $    &$\rm    4.79 \pm    0.09 $   %  &$\rm    27.7 \pm     0.0 $\n
  &[21, 33] &1.27\\
 &$\rm   C^{17}O \, (1-0) $ &112.359  &16.0 &$\rm     3.6 \pm     0.4 $    &$\rm    2.70 \pm    0.20 $   %  &$\rm    27.9 \pm     0.1 $\n
 &$\rm     1.7 \pm     0.2 $    &$\rm    1.18 \pm    0.12 $   %  &$\rm    27.4 \pm     0.1 $\n
 &$\rm     2.0 \pm     0.2 $    &$\rm    1.22 \pm    0.12 $   %  &$\rm    27.8 \pm     0.1 $\n
  &[21, 33] &1.16\\
 \hline

 \multirow{5}{*}{G\,14.49-0.13}
&$\rm   ^{13}CO \, (2-1) $ &  220.398   &11.8  &$\rm     5.4 \pm     0.1 $    &$\rm   38.74 \pm    0.87 $   %  &$\rm    40.3 \pm     0.1 $\n
&$\rm     6.5 \pm     0.2 $    &$\rm   42.87 \pm    1.06 $   %  &$\rm    40.4 \pm     0.1 $\n
&$\rm     5.8 \pm     0.1 $    &$\rm   57.89 \pm    0.97 $   %  &$\rm    40.4 \pm     0.0 $\n
 &[31, 49]   &1.86\\ %  &$\rm    80.0 \pm     0.0 $\n
&$\rm   C^{18}O \, (2-1) $ &  219.560 &11.8  &$\rm     3.8 \pm     0.1 $    &$\rm    7.69 \pm    0.21 $   %  &$\rm    40.1 \pm     0.1 $\n
&$\rm     4.2 \pm     0.1 $    &$\rm   12.04 \pm    0.22 $   %  &$\rm    39.4 \pm     0.0 $\n
&$\rm     3.1 \pm     0.0 $    &$\rm   24.20 \pm    0.29 $   %  &$\rm    40.1 \pm     0.0 $\n
 &[31, 49]  &1.20\\%  &$\rm    79.8 \pm     0.1 $\n
&$\rm  C^{17}O \, (2-1) $ &   224.714 &11.5  &$\rm     3.5 \pm     0.3 $    &$\rm    1.69 \pm    0.13 $   %  &$\rm    40.1 \pm     0.1 $\n
&$\rm     3.3 \pm     0.2 $    &$\rm    3.34 \pm    0.15 $   %  &$\rm    39.1 \pm     0.1 $\n
&$\rm     3.0 \pm     0.1 $    &$\rm   11.17 \pm    0.17 $   %  &$\rm    40.0 \pm     0.0 $\n
 &[31, 49] &0.96\\%  &$\rm    80.1 \pm     0.0 $\n
%$\rm  HCO^+ \, (1-0) $ &  89.189   &$\rm     3.6 \pm     0.1 $    &$\rm    8.56 \pm    0.20 $   %  &$\rm    77.3 \pm     0.0 $\n
%&$\rm     2.8 \pm     0.1 $    &$\rm    4.62 \pm    0.13 $  &[74, 83] \\%  &$\rm    77.8 \pm     0.0 $\n
&$\rm  H^{13}CO^+ \, (1-0) $ &  86.754  &29.9  &$\rm     3.5 \pm     0.2 $    &$\rm    0.86 \pm    0.05 $   %  &$\rm    40.2 \pm     0.1 $\n
&$\rm     3.7 \pm     0.1 $    &$\rm    1.53 \pm    0.05 $   %  &$\rm    39.4 \pm     0.1 $\n
 &$\rm     2.8 \pm     0.2 $    &$\rm    1.02 \pm    0.06 $   %  &$\rm    39.7 \pm     0.1 $\n
 &[31, 49]  &0.28\\%  &$\rm    79.5 \pm     0.0 $\n
&$\rm  DCO^+ \, (1-0) $ &72.039 &36.0    &$\rm     2.9 \pm     0.7 $    &$\rm    0.17 \pm    0.04 $   %  &$\rm    39.8 \pm     0.4 $\n
&$\rm     2.8 \pm     0.2 $    &$\rm    0.57 \pm    0.04 $   %  &$\rm    39.2 \pm     0.1 $\n
&$\rm     2.6 \pm     0.8 $    &$\rm    0.32 \pm    0.09 $   %  &$\rm    39.8 \pm     0.4 $\n
 &[31, 49]   &0.12\\%  &$\rm    79.7 \pm     0.0 $\n
  &$\rm   ^{13}CO \, (1-0) $ &110.201 &16.4 &$\rm     6.2 \pm     0.2 $    &$\rm   66.23 \pm    1.76 $   %  &$\rm    39.6 \pm     0.1 $\n
  &$\rm     7.3 \pm     0.2 $    &$\rm   70.33 \pm    1.82 $   %  &$\rm    39.4 \pm     0.1 $\n
  &$\rm     5.1 \pm     0.1 $    &$\rm   86.92 \pm    1.68 $   %  &$\rm    39.8 \pm     0.0 $\n
  &[31, 49] &46.70\\
 &$\rm   C^{18}O \, (1-0) $ &109.783  &16.4  &$\rm     4.8 \pm     0.2 $    &$\rm   11.10 \pm    0.39 $   %  &$\rm    40.1 \pm     0.1 $\n
 &$\rm     4.2 \pm     0.2 $    &$\rm   13.56 \pm    0.46 $   %  &$\rm    39.3 \pm     0.1 $\n
 &$\rm     3.3 \pm     0.1 $    &$\rm   21.86 \pm    0.38 $   %  &$\rm    40.2 \pm     0.0 $\n
  &[31, 49] &5.68\\
 &$\rm   C^{17}O \, (1-0) $ &112.359  &16.0  &$\rm     6.4 \pm     0.7 $    &$\rm    2.79 \pm    0.28 $   %  &$\rm    39.4 \pm     0.3 $\n
 &$\rm     5.2 \pm     0.4 $    &$\rm    3.95 \pm    0.28 $   %  &$\rm    38.5 \pm     0.2 $\n
 &$\rm     4.8 \pm     0.2 $    &$\rm    7.26 \pm    0.30 $   %  &$\rm    39.4 \pm     0.1 $\n
  &[31, 49] &2.19\\
\hline

 \multirow{5}{*}{G\,34.74-0.12}
&$\rm   ^{13}CO \, (2-1) $ &  220.398  &11.8   &$\rm     4.6 \pm     0.1 $    &$\rm   16.70 \pm    0.19 $   %  &$\rm    78.5 \pm     0.0 $\n
&$\rm     4.7 \pm     0.1 $    &$\rm   16.22 \pm    0.22 $   %  &$\rm    78.5 \pm     0.0 $\n  
&$\rm     5.4 \pm     0.1 $    &$\rm   21.63 \pm    0.34 $   %  &$\rm    77.5 \pm     0.0 $\n
 &[72, 85]  &0.68\\ %  &$\rm    80.0 \pm     0.0 $\n
&$\rm   C^{18}O \, (2-1) $ &  219.560  &11.8 &$\rm     2.2 \pm     0.1 $    &$\rm    5.21 \pm    0.11 $   %  &$\rm    78.7 \pm     0.0 $\n
&$\rm     2.4 \pm     0.0 $    &$\rm    6.50 \pm    0.11 $   %  &$\rm    78.6 \pm     0.0 $\n
&$\rm     2.2 \pm     0.1 $    &$\rm    6.40 \pm    0.14 $   %  &$\rm    77.7 \pm     0.0 $\n
 &[72, 85]  &0.61\\%  &$\rm    79.8 \pm     0.1 $\n
&$\rm  C^{17}O \, (2-1) $ &   224.714 &11.5  &$\rm     2.4 \pm     0.2 $    &$\rm    1.42 \pm    0.12 $   %  &$\rm    78.8 \pm     0.1 $\n
&$\rm     1.9 \pm     0.1 $    &$\rm    2.03 \pm    0.10 $   %  &$\rm    78.5 \pm     0.0 $\n
&$\rm     2.4 \pm     0.2 $    &$\rm    1.88 \pm    0.10 $   %  &$\rm    77.7 \pm     0.1 $\n
 &[72, 85]  &0.56\\%  &$\rm    80.1 \pm     0.0 $\n
%$\rm  HCO^+ \, (1-0) $ &  89.189   &$\rm     3.6 \pm     0.1 $    &$\rm    8.56 \pm    0.20 $   %  &$\rm    77.3 \pm     0.0 $\n
%&$\rm     2.8 \pm     0.1 $    &$\rm    4.62 \pm    0.13 $  &[74, 83] \\%  &$\rm    77.8 \pm     0.0 $\n
&$\rm  H^{13}CO^+ \, (1-0) $ &  86.754  &29.9  &$\rm     2.3 \pm     0.1 $    &$\rm    0.87 \pm    0.03 $   %  &$\rm    78.8 \pm     0.0 $\n
&$\rm     2.5 \pm     0.1 $    &$\rm    0.92 \pm    0.02 $   %  &$\rm    78.6 \pm     0.0 $\n 
&$\rm     2.0 \pm     0.1 $    &$\rm    0.71 \pm    0.02 $   %  &$\rm    77.8 \pm     0.0 $\n
 &[72, 85]  &0.18\\%  &$\rm    79.5 \pm     0.0 $\n
&$\rm  DCO^+ \, (1-0) $ &72.039 &36.0 &$\rm     2.0 \pm     0.3 $    &$\rm    0.34 \pm    0.04 $   %  &$\rm    78.6 \pm     0.1 $\n
&$\rm     2.3 \pm     0.2 $    &$\rm    0.53 \pm    0.04 $   %  &$\rm    78.5 \pm     0.1 $\n
&$\rm     1.7 \pm     0.2 $    &$\rm    0.31 \pm    0.04 $   %  &$\rm    77.8 \pm     0.1 $\n
 &[72, 85]   &0.19\\%  &$\rm    79.7 \pm     0.0 $\n
  &$\rm   ^{13}CO \, (1-0) $ &110.201 &16.4 &$\rm     5.6 \pm     0.1 $    &$\rm   27.72 \pm    0.20 $   %  &$\rm    78.1 \pm     0.0 $\n
 &$\rm     4.8 \pm     0.0 $    &$\rm   25.95 \pm    0.17 $   %  &$\rm    78.1 \pm     0.0 $\n 
 &$\rm     5.6 \pm     0.1 $    &$\rm   36.06 \pm    0.36 $   %  &$\rm    77.4 \pm     0.0 $\n
  &[72, 85] &14.26\\
 &$\rm   C^{18}O \, (1-0) $ &109.783  &16.4  &$\rm     2.4 \pm     0.1 $    &$\rm    5.98 \pm    0.13 $   %  &$\rm    78.9 \pm     0.0 $\n
 &$\rm     2.5 \pm     0.0 $    &$\rm    8.43 \pm    0.11 $   %  &$\rm    78.8 \pm     0.0 $\n
 &$\rm     2.6 \pm     0.1 $    &$\rm    8.44 \pm    0.13 $   %  &$\rm    78.3 \pm     0.0 $\n
  &[72, 85] &1.59\\
 &$\rm   C^{17}O \, (1-0) $ &112.359  &16.0 &$\rm     1.9 \pm     0.4 $    &$\rm    1.22 \pm    0.17 $   %  &$\rm    78.8 \pm     0.1 $\n
 &$\rm     4.7 \pm     0.4 $    &$\rm    3.03 \pm    0.23 $   %  &$\rm    78.1 \pm     0.2 $\n
 &$\rm     4.6 \pm     0.4 $    &$\rm    3.32 \pm    0.22 $      %  &$\rm    77.6 \pm     0.1 $\n
 &[72, 85] &1.57\\
 \hline

\hline\hline

  \multicolumn{10}{l}{{ Note.} $a$. Lines are extracted from images by averaging a beam-sized region centered at P1, P2, and P3  in the plane of the sky. }\\
  \multicolumn{10}{l}{~~~~~~~~~~~~~~ All  line images have the same pixel size, but whose native angular and
velocity resolution we kept as in the observations;}\\
\multicolumn{10}{l}{~~~~~~~~~~$b$.  Rest frequency is given from  the main line of the hyperfine splittings;}\\
\multicolumn{10}{l}{~~~~~~~~~~$c$. Angular resolution from observations;}\\
\multicolumn{10}{l}{~~~~~~~~~~$d$. Uncertainties on the measured intensities are typically $\le10\%$;}\\
% \multicolumn{7}{l}{~~~~~~~~~~$f$. Hyperfine splitting of HCN\,(1--0) is not well fitted by one velocity component,}\\
%  \multicolumn{7}{l}{~~~~~~~~~~~~~~ see the discussion in \citet[][]{feng16a} and \citet{beuther07c};}\\
\multicolumn{10}{l}{~~~~~~~~~~$e$. The velocity range we integrate for individual lines to obtain their intensity maps is in Figure~\ref{spec1};}\\
\multicolumn{10}{l}{~~~~~~~~~~$f$. The $\sigma$ rms on the molecular line intensity maps in is Figure~\ref{spec1}.}\\
\multicolumn{10}{l}{~~~~~~~~~~$g$. Line emission $\rm <3\sigma$ rms.}
\end{tabular}
}
\end{table}

\newpage
\begin{table}[htbp]
\caption{Gas parameters of our targets}\label{tab:gaspara}
\scalebox{0.63}{

\begin{tabular}{p{1.3cm}p{1.6cm}|p{0.8cm}|ccc|ccc|ccc|ccc}
\hline\hline

Properties & &Source                    &\multicolumn{3}{c|}{G\,15.22-0.43}                  &\multicolumn{3}{c|}{G\,11.38+0.81}                                                                   &\multicolumn{3}{c|}{G\,14.49-0.13}     &\multicolumn{3}{c}{G\,34.74-0.12} 
   \\
   \hline
                                                 \multicolumn{2}{c}{$\theta\rm=16\arcsec-20\arcsec$}           &&CaseA$^a$         &Case B$^b$            &          
                                                                                              &CaseA$^a$	      &Case B$^b$            &          
                                                                                              &CaseA$^a$         &Case B$^b$            &           
                                                                                              &CaseA$^a$          &Case B$^b$           &          \\
   \hline                                                             
\multirow{3}{*} {$\rm N_{C^{18}O}$}  &\multirow{3}{*} {($\rm \times10^{15}\,cm^{-2}$)}  
   &P1 &$\rm  3.1\pm 0.6$       &$\rm  2.6\pm 0.3$   &
           &$\rm  3.4\pm 1.7$	       &$\rm  10.9\pm 4.6$      &
           &$\rm  12.4\pm 6.5$      &$\rm  22.8\pm 4.5$      &
           &$\rm  6.7\pm 2.9$       &$\rm  9.7\pm 2.6$      &\\
&&P2 &$\rm  6.1\pm 0.1$       &$\rm  5.5\pm 0.4$        &  
          &$\rm  5.0\pm 2.5$	      &$\rm  13.7\pm 5.0$        & 
          &$\rm  10.5\pm 6.3$      &$\rm  22.5\pm 4.6$    &
          &$\rm  5.3\pm 2.2$       &$\rm  6.6\pm 1.3$    &\\
&&P3 &$\rm  5.5\pm 0.2$       &$\rm  4.8\pm 0.3$     &
          &$\rm  3.2\pm 1.3$	      &$\rm  4.4\pm 0.8$      &     
          &$\rm  21.7\pm 10.5$   &$\rm  26.6\pm 3.2$     &
          &$\rm  7.8\pm 4.0$       &$\rm  15.6\pm 4.7$     &\\                                                                                                                                                                                                                    
   \hline     
   \multirow{3}{*} {$f_D(\rm C^{18}O)$$^{g}$} &\multirow{3}{*} {} 
   &P1        &$\rm  4.1\pm 0.9$    & &
           	 &$\rm  11.9\pm 6.2$    & &
           	 &$\rm 4.8\pm 2.5 $      & &
                 &$\rm 3.1\pm 1.3 $         & &\\
&&P2        &$\rm  3.1\pm0.1$      & & 
          	 &$\rm  6.7\pm3.3$        & &
          	 &$\rm 3.3\pm 2.0 $         & &
                 &$\rm 3.7\pm 1.6 $          & &\\
&&P3        &$\rm 1.3\pm0.1 $         & &
          	 &$\rm 7.9\pm 3.4 $      &  &   
          	 &$\rm 1.7\pm 0.9 $         & &
                 &$\rm 1.2\pm 0.6$         & &\\   
        
 \hline       
\hline
                                                 \multicolumn{2}{c}{$\theta\rm\sim35\arcsec$}           &&CaseC$^c$         &Case D$^d$                      &Case E$^e$
                                                                                              &CaseC$^c$	      &Case D$^d$                      &Case E$^e$
                                                                                              &CaseC$^c$         &Case D$^d$                      &Case E$^e$
                                                                                              &CaseC$^c$          &Case D$^d$                      &Case E$^e$\\
   \hline                                                             
\multirow{3}{*} {$\rm N_{C^{18}O}$}  &\multirow{3}{*} {($\rm \times10^{15}\,cm^{-2}$)}  
   &P1  &$\rm 2.5\pm 0.1$ &--$^f$			&$\rm 3.5\pm 2.7$
           &$\rm 1.5\pm 0.3$ &--$^f$			&$\rm 1.4\pm 0.9$   
           &$\rm 5.2\pm 0.9$ &$\rm 5.2\pm 0.5$       &$\rm 5.3\pm 3.8$
            &$\rm 3.6\pm 0.1$ &$\rm 3.4\pm 0.4$      	&$\rm  3.8\pm 2.6$\\
&&P2  &$\rm 4.9\pm 0.1$  &--$^f$			&$\rm 5.3\pm 2.8$ 
          &$\rm 2.2\pm 0.2$  &--$^f$		&$\rm 1.8\pm 1.1$  
          &$\rm 4.0\pm 0.9$ &$\rm 4.0\pm 1.5$        &$\rm 4.3\pm3.5 $
           &$\rm 2.8\pm0.1$ &$\rm 2.7\pm0.3$         &$\rm  3.0\pm2.4$\\
&&P3  &$\rm 3.8\pm 0.1$ &--$^f$			&$\rm 4.7\pm 3.1$   
           &$\rm 1.7\pm 0.3$ &--$^f$		&$\rm 1.7\pm 1.2$              
          &$\rm 7.0\pm 0.9$ &$\rm  7.7\pm 3.3$ 	&$\rm 7.9\pm 4.9$
          &$\rm 3.4\pm 0.1$ &$\rm 3.3\pm 1.1$        &$\rm 3.9\pm 2.9 $  \\                                                                                                                                                                                                                    
   \hline                                                                               
\multirow{3}{*} {$f_D(\rm C^{18}O)$$^{g}$} &\multirow{3}{*} {} 
   &P1 &$\rm 4.1\pm 0.1 $ &--$^f$			&$\rm 2.9\pm 2.1 $      
          &$\rm 12.8\pm2.3$ &--$^f$			&$\rm  13.6\pm8.9$       
          &$\rm 5.4\pm 1.0 $ &$\rm 8.3\pm 0.8 $    &$\rm 5.3\pm3.7$
          &$\rm 3.8\pm 0.1 $ &$\rm 5.0\pm 0.7$    	&$\rm 3.6\pm 2.4$\\
&&P2  &$\rm 2.7\pm 0.1 $ &--$^f$			&$\rm 2.5\pm 1.3$  
           &$\rm  7.6\pm0.9$ &--$^f$			&  $\rm 9.0\pm 5.3$                    
           &$\rm 3.5\pm 0.8 $ &$\rm 5.3\pm 1.9 $ 	&$\rm 3.3\pm 2.3 $
           &$\rm 4.5\pm 0.1 $  &$\rm 5.9\pm0.7$    &$\rm 4.1\pm 2.9$  \\
 &&P3  &$\rm 1.5\pm 0.1 $ &--$^f$			&$\rm 1.2\pm 0.8$   
            &$\rm  9.2\pm1.6$ &--$^f$			&$\rm 8.9\pm 6.2$                        
            &$\rm 2.8\pm 0.4 $ &$\rm  3.5\pm 1.5 $  &$\rm 2.5\pm 1.5 $
            &$\rm 2.1\pm 0.1 $ &$\rm 2.0\pm 0.7 $     &$\rm 1.9\pm 1.3$\\     
                                                                                                                     
  \hline                                                                                
\multirow{3}{*} {$\rm N_{H^{13}CO^+}$}   &\multirow{3}{*} { ($\rm \times10^{12}\,cm^{-2}$)}  
  &P1    &$\rm 1.4\pm 0.1 $  &--$^f$			&$\rm 2.5\pm 1.9 $    
            &$\rm 0.7\pm 0.3$   &--$^f$			&$\rm 1.3\pm 0.9   $                 
            &$\rm 2.6\pm 0.1$  &$\rm 2.7\pm 0.3$ 	&$\rm 3.3\pm 1.3$
            &$\rm  1.4\pm 0.1$ &$\rm  1.4\pm 0.3$     &$\rm 2.3\pm 1.2$  \\
&&P2   &$\rm 2.3\pm 0.1 $ &--$^f$			&$\rm 3.1\pm 1.5  $
           &$\rm 1.1\pm 0.3$  &--$^f$			&$\rm 2.0\pm 1.1 $                   
           &$\rm  1.5\pm 0.1$  &$\rm 1.5\pm 0.5 $ 	&$\rm 1.9\pm1.0 $
           &$\rm  1.5\pm 0.1$ &$\rm  1.4\pm0.2$    	&$\rm  2.8\pm1.6$    \\
&&P3  &$\rm 1.4\pm 0.1 $ &--$^f$			&$\rm 2.0\pm 1.3 $      
           &$\rm 0.7\pm 0.3$  &--$^f$			&$\rm 1.6\pm 1.1 $                 
           &$\rm  1.8\pm 0.1$  &$\rm 1.6\pm 0.6 $ 	&$\rm 2.1\pm 1.0  $
           &$\rm  1.4\pm0.1$  &$\rm 1.3\pm 0.4 $   	&$\rm 2.2\pm 1.3$\\                                                                               
   \hline                                                                               
\multirow{3}{*} {$\rm N_{DCO^+}$}    &\multirow{3}{*} {($\rm \times10^{11}\,cm^{-2}$)}  
   &P1   &$\rm 5.8\pm 0.6$  &--$^f$			&$\rm 12.7\pm 9.6$                     
            &$\rm 4.9\pm 1.3$ &--$^f$			&$\rm 10.6\pm 6.2$                    
            &$\rm 6.1\pm 0.5$ &$\rm 6.2\pm 1.5$ 	&$\rm 7.6\pm 2.9$
            &$\rm  5.9\pm 0.5$ &$\rm  6.0\pm 1.4$     &$\rm 10.3\pm 5.7 $  \\
&&P2   &$\rm 4.1\pm 0.6$ &--$^f$			&$\rm   5.7\pm 2.9$                                           
            &$\rm 5.0\pm 1.4$ &--$^f$			&$\rm 9.5\pm 5.2 $               
            &$\rm <1.2$$^i$ &$\rm <0.8$$^i$         &$\rm <1.1$$^i$
            &$\rm <2.4$$^i$ &$\rm <2.3$$^i$          &$\rm <3.5$$^i$   \\
&&P3   &$\rm <2.1 $$^i$  &--$^f$			 &$\rm <3.4 $$^i$           
           &$\rm 2.4\pm 1.4$ &--$^f$                        &$\rm <5.5$              
           &$\rm 3.6\pm 0.5$ &$\rm  2.9\pm 1.7$      &$\rm 3.9\pm1.8$
           &$\rm <3.3$$^i$ &$\rm <2.8 $$^i$         &$\rm < 4.1$$^i$  \\   
   \hline                                                                               
\multirow{3}{*} {$D$$\rm _{HCO^+}$$^{h}$} &\multirow{3}{*} {} 
&P1  &$\rm 1.0\%\pm 0.1\% $   &--$^f$				    &$\rm 1.0\%\pm 0.8\% $                              
         &$\rm 1.6\%\pm 0.8\% $  &--$^f$       		            &$\rm 1.7\%\pm 1.3\% $              
         &$\rm 0.5\%\pm 0.1\% $ &$\rm 0.5\%\pm 0.1\% $   &$\rm 0.5\%\pm 0.3\% $
         &$\rm 1.0\%\pm 0.1\%$ &$\rm  0.9\%\pm 0.3\%$   &$\rm 1.0\%\pm 0.7\%$   \\
&&P2    &$\rm 0.4\%\pm 0.1\% $  &--$^f$ 			     &$\rm 0.4\%\pm 0.2\% $    
            &$\rm 1.0\%\pm 0.3\% $ &--$^f$			     &$\rm 1.0\%\pm 0.8\% $                
            &$\rm <0.1\% $$^i$  &$\rm <0.1\% $$^i$ 	     &$\rm <0.1\% $$^i$
             &$\rm  <0.4\%$$^i$  &$\rm  <0.4\%$$^i$   	    &$\rm <0.4\%$$^i$   \\
&&P3  &$\rm <0.4\%$$^i$  &--$^f$			     &$\rm <0.4\%$$^i$                            
           &$\rm < 0.7\%$$^i$  &--$^f$			     &$\rm < 0.7\%$$^i$                               
           &$\rm 0.4\%\pm0.1\% $ &$\rm  0.4\%\pm 0.2\%$  &$\rm 0.4\%\pm0.3\% $
           &$\rm <0.5 \%$$^i$ &$\rm <0.5\% $$^i$         &$\rm < 0.6\%$$^i$     \\

%%%%%%%%%%%%%%%%%%%%%%%%%%%%%

\hline \hline
\multicolumn{15}{l}{\color{black} { Note.} Here P1, P2, and P3 denote the $\rm DCO^+$-dominant,   transition, and the CO-dominant  zones, respectively. }\\
\multicolumn{15}{l}{\color{black} ~~~~~~~~~ $a$. Use $T_{\rm dust}$ and derived from  (2-1)/(1-0) lines of $\rm C^{18}O$, at an angular resolution of 18\arcsec or 20\arcsec. }\\
\multicolumn{15}{l}{\color{black} ~~~~~~~~~ $b$. Use the best fit of (2-1)/(1-0) lines of $\rm C^{18}O$ from RADEX at an angular resolution of 16.4\arcsec. }\\
\multicolumn{15}{l}{\color{black} ~~~~~~~~~ $c$. Use $T_{\rm dust}$ and derived from the $\rm C^{18}O$\,(2-1) line at an angular resolution of 34.7\arcsec. }\\
\multicolumn{15}{l}{\color{black} ~~~~~~~~~ $d$. Use $T_{kin}$($p$-$\rm NH_3$) at an angular resolution of 34.7\arcsec. }\\
\multicolumn{15}{l}{\color{black} ~~~~~~~~~ $e$.  Use $\rm T_{rot}$($p$-$\rm H_2CO$) at an angular resolution of 35.6\arcsec. }\\
\multicolumn{15}{l}{\color{black} ~~~~~~~~~ $f$. Here ``--" indicates the location where we do not have  $\rm NH_3$ observations.}\\
\multicolumn{15}{l}{\color{black} ~~~~~~~~~ $g$. The $\rm C^{18}O$ depletion is derived by assuming the expected abundance with respect to $\rm H_2$  as}\\
\multicolumn{15}{l}{~~~~~~~~~~~~ %$\rm  9.50\times10^{-5} {\it exp}(1.11-0.13{\it D}_{GC,kpc})/ (58.80{\it D}_{GC,kpc}+37.10)$,      
Eqns.~\ref{eq:nc18oex}--\ref{eq:dc18o} and assuming a gas-to-dust mass ratio $log(\gamma)=\rm 0.087 {\it R_{\rm GC}}(kpc)+1.44$  \citep{draine11,giannetti17b}.}\\
\multicolumn{15}{l}{\color{black} ~~~~~~~~~$h$. D-fraction is derived from the  $\rm DCO^+$\,(1--0) and $\rm H^{13}CO^+$\,(1--0) lines by assuming that they are optically thin,}\\
\multicolumn{15}{l}{~~~~~~~~~~~~   have the same beam filling toward each pixel, and have a constant fraction of $\mathcal{R}_{\rm ^{12}C/^{13}C}\sim  6.1{\it R}_{\rm GC}({\rm kpc})+ 14.3 $ \citep{giannetti14};}\\
\multicolumn{11}{l}{\color{black} ~~~~~~~~~ $i$. An upper limit is given when the detected $\rm DCO^+$\,(1--0) shows $\rm <3\sigma$ emission.}\\

\end{tabular}

}
\end{table}

\end{document}